\documentclass[aps,prd,twocolumn,nofootinbib,showpacs,preprintnumbers,superscriptaddress]{revtex4-2}
\usepackage[pdftex]{graphicx}
\usepackage[usenames]{color}
\usepackage[dvips]{epsfig}
\usepackage{hyperref}
\usepackage{bbm}
\usepackage{marvosym}
\usepackage{verbatim}
\usepackage{psfrag}
\usepackage{slashed}
\usepackage{float}
\usepackage{amsmath}
\usepackage{dcolumn}
\usepackage{amssymb}
\usepackage[normalem]{ulem}
\usepackage{wasysym}
\usepackage{multirow}
\usepackage{enumerate}
\usepackage{adjustbox}
\usepackage{mathtools}
\usepackage{diagbox}
\usepackage{array} 
\usepackage{ bbold }
\usepackage[table,x11names]{xcolor}
\usepackage{array,booktabs}
\usepackage{amsfonts}
\usepackage[caption=false]{subfig}
\DeclareMathOperator{\sech}{sech}

\begin{document}

\title{Spinodal-assisted nucleation in the two-dimensional $q-$state  Potts model with short-to-long range interactions}

\author{G.~Gagliardi}
\email{giuseppe.gagliardi@roma3.infn.it}
\affiliation{Istituto Nazionale di Fisica Nucleare, Sezione di Roma Tre, Via della Vasca Navale 84, I-00146 Rome, Italy}
\author{F.~Macheda}
\email{francesco.macheda@kcl.ac.uk}
\affiliation{Department of Physics, King’s College London,Strand, London WC2R 2LS, United Kingdom}
\affiliation{Istituto Italiano di Tecnologia, Graphene Labs, Via Morego 30, I-16163 Genova, Italy}

\pacs{64.60.De, 05.50.+q,05.10.Ln,64.60.Qb, 64.60.qe}

\begin{abstract}
We study homogeneous nucleation in the two-dimensional $q-$state Potts model for $q=3,5,10,20$ and ferromagnetic couplings $J_{ij} \propto \Theta (R - |i-j|)$, by means of Monte Carlo simulations employing heat bath dynamics. Metastability is induced in the low temperature phase through an instantaneous quench of the magnetic field coupled to one of the $q$ spin states. The quench depth is adjusted, depending on the value of temperature $T$, interaction range $R$, and number of states $q$, in such a way that a constant nucleation time is always obtained. In this setup we analyze the crossover between the classical compact droplet regime occurring in presence of short range interactions $R \sim 1$, and the long-range regime $R\gg 1$ where the properties of nucleation are influenced by the presence of a mean-field spinodal singularity. We evaluate the metastable susceptibility of the order parameter as well as various critical droplet properties, which along with the evolution of the quench depth as a function of $q,T$ and $R$, are then compared with the field theoretical predictions valid in the large $R$ limit in order to find the onset of spinodal-assisted nucleation.  We find that, with a mild dependence on the values of $q$  and $T$ considered, spinodal scaling holds for interaction ranges $R\gtrsim 8-10$,  and that signatures of the presence of a pseudo-spinodal are already visible for remarkably small interaction ranges $R\sim 4-5$. The influence of spinodal singularities on the occurrence of multi-step nucleation is also discussed.
\end{abstract}
\maketitle

\section{Introduction}
\label{Intro}
Nucleation phenomena have been the subject of numerous studies in the last century. When a modification in the external conditions of a thermally equilibrated medium occurs, the system can be brought into a region where a different phase would exist in thermal equilibrium. It may then happen that the relaxation towards the new equilibrium state is preceded by the formation of one or more metastable states, whose decays are hindered by the presence of free energy barriers. The nucleation of a given metastable state occurs, according to classical nucleation theory (CNT)~\cite{10.1002/andp.19354160806,10008767133,9780691085951}, with the appearence of a fluctuation-induced critical droplet configuration, whose subsequent growth is favored by a decrease of the free energy. Examples of metastable states are common in nature and can be observed in very different contexts---from supercooled vapors and liquids~\cite{10.1038/35065704}, to the false vacuum associated to the electroweak transition~\cite{10.1103/PhysRevD.16.1762}, and in two-dimensional superfluids~\cite{10.1088/0022-3719/5/11/002}. Many properties of nucleation, like metastable lifetimes, critical droplet profile, and the cascade of decays through which equilibration happens, depend on the specific type of modification of the external parameters that brings the system out of equilibrium, as well as on the microscopic details of the interactions. In this work we are concerned with the physics of spinodal-assisted nucleation in systems with short-to-long range interactions, studying as a toy-model the two-dimensional $q-$state  Potts model.\\
\indent The spinodal limit is essentially a mean-field concept and defines the absolute limit of metastability, where a local minimum of the free energy turns into an inflection point. In the mean-field limit the spinodal points sharply separate the region where a metastable state is infinitely long lived from the one where metastability does not occur at all; hence they are singularity points where the susceptibility $\chi_{\phi}$ of the order parameter diverges~\cite{10.1088/0034-4885/50/7/001,10.1063/1.4959235}.  Observations of spinodal-like singularities are usually hindered by thermal fluctuations, and in systems where mean-field is a bad approximation, the concept of a spinodal singularity lacks any basis. However, by studying simple models via Monte Carlo simulations, it is possible to monitor the evolution of the nucleation pattern as a function of the interaction range, finding the onset of spinodal-assisted nucleation. This idea was pursued for the first time in Refs.~\cite{10.1103/PhysRevLett.49.1262,10.1103/PhysRevB.38.11607,Monette1992} in the case of the Ising model, where metastable states can be created e.g. with a quench of the magnetic field in the ordered phase. For large values of the interaction range, their findings were in agreement with the analytical predictions based on a semi-classical treatment of the scalar $\phi^{3}$ theory~\cite{10.1103/PhysRevB.29.2698,10.1103/PhysRevB.31.6127,10.1103/PhysRevB.28.445}, which extended previous works on nucleation close to the condensation point~\cite{10.1016/0003-4916(67)90200-X,10.1088/0305-4470/13/5/034}. It was found that, as opposite to the case of short range interactions where the critical droplet is a compact object, for long range interactions and close to the spinodal point the critical droplet consists of a small-amplitude extended fluctuation having the properties of a percolating cluster, which grows by compactification around its center.\\
\indent In this work we perform a systematic study of nucleation in the two-dimensional $q-$state  Potts model with ferromagnetic couplings of the form $J_{ij}(R)\propto \Theta(R- |i-j|)$. Our main goal is to describe the crossover between nucleation in presence of short range interactions (SRI, $R\sim 1$) and long range interactions (LRI, $R\gg 1$) for different values of $q$, mapping out the region of the parameter space exhibiting spinodal nucleation. Since the existence of a sharp spinodal point is often assumed in the discussion of nucleation in real systems, where this might not always be well justified, investigating to what extent it can be observed in simple systems depending on the length scale of the interactions, can give an insight on the reliability of such assumption. Moreover, in the $q$-state Potts model the complex free energy landscape enriches the phenomenology since many saddle points exist and a given metastable state can decay following different channels. In this case, the presence of the spinodal singularity may not only affect the shape of the nucleating droplet, but modifies the decay chain through which thermalization is achieved. \\
\indent We concentrate on metastable states created from a completely ordered phase with all spins aligned in a given direction by instantaneously turning on a magnetic field that disfavours such phase. In this setup, there exists for all $q$ a critical value of the magnetic field $h=h_{\text{sp}}(q)$ that gives rise to a spinodal instability. By getting closer to $h_{\text{sp}}(q)$, the mean-field free energy curvature of the metastable state gets flatter and flatter along one specific direction in the order parameter space; hence, invoking the Ostwald step-rule~\cite{ostwald1897file}, for sufficiently large $R$ nucleation must take place along this direction. For $q\geq 3$, this channel is different from the dominant nucleation channel for shallow quenches, where a fluctuation of \emph{one} of the $q-1$ equilibrium phases initiate the decay. The critical droplet associated to the spinodal point is instead expected to be a delocalized fluctuation where  all the $q-1$ spins that are not coupled to the magnetic field \emph{locally coexist}. This regime is not to be confused with what happens in the case of shallow quenches in the so-called coalescence regime, where many droplets of different spins are created in \emph{different} regions of space, if the volume is sufficiently large~\cite{10.1142/S0129183102003255,corberi2021phases}. After the decay, the system retains a residual $Z_{q-1}$ symmetry for a certain amount of time and, depending on the temperature value, a second metastable state can form. \\
\indent We restrict ourselves to the so-called single droplet regime, and choose the lattice volume $V=L\times L$ so as to prevent the contemporary appearance of multiple droplets. We focus on two temperatures: $T= 0.5T_{c}$ and $T=0.8T_{c}$, where $T_{c}$ is the mean-field melting temperature given by
\begin{align}
\label{mean_field_critical_coupling}
T_{c}^{-1} \equiv J_{c} = \frac{q-1}{2(q-2)}\log{(q-1)}\, ,
\end{align}
and where $2J/V$ is the all-to-all mean-field ferromagnetic coupling. 
We use heat bath dynamics to evolve the system, and for each simulation point we analyze up to $2\cdot 10^{2}$ nucleation events in order to evaluate several observables such as average nucleation time, metastable susceptibility and droplet properties. The latter are determined exploiting the formalism of the Coniglio-Klein clusters~\cite{10.1088/0305-4470/13/8/025,10.1007/978-0-387-30440-3_104}. We then compare our numerical results with the theoretical predictions obtained through a semi-classical expansion around the mean-field limit.\\
\indent The paper is organized as follows: in Section~\ref{I} we introduce the $q-$state  Potts model, focusing on its mean-field limit, the related spinodals and the phase diagram in the $T-h$ plane. In Section~\ref{II} we generalize the semi-classical analysis of Ref.~\cite{10.1103/PhysRevB.29.2698} to the Potts model, finding the $q-$dependence of cluster properties and nucleation times. In Section~\ref{III} we describe the numerical methods and in Section~\ref{IV} we present our results. Finally in Section~\ref{V} we draw our conclusions.   
\section{The model}
\label{I}
\begin{figure*}[t!]
\includegraphics[scale=0.42]{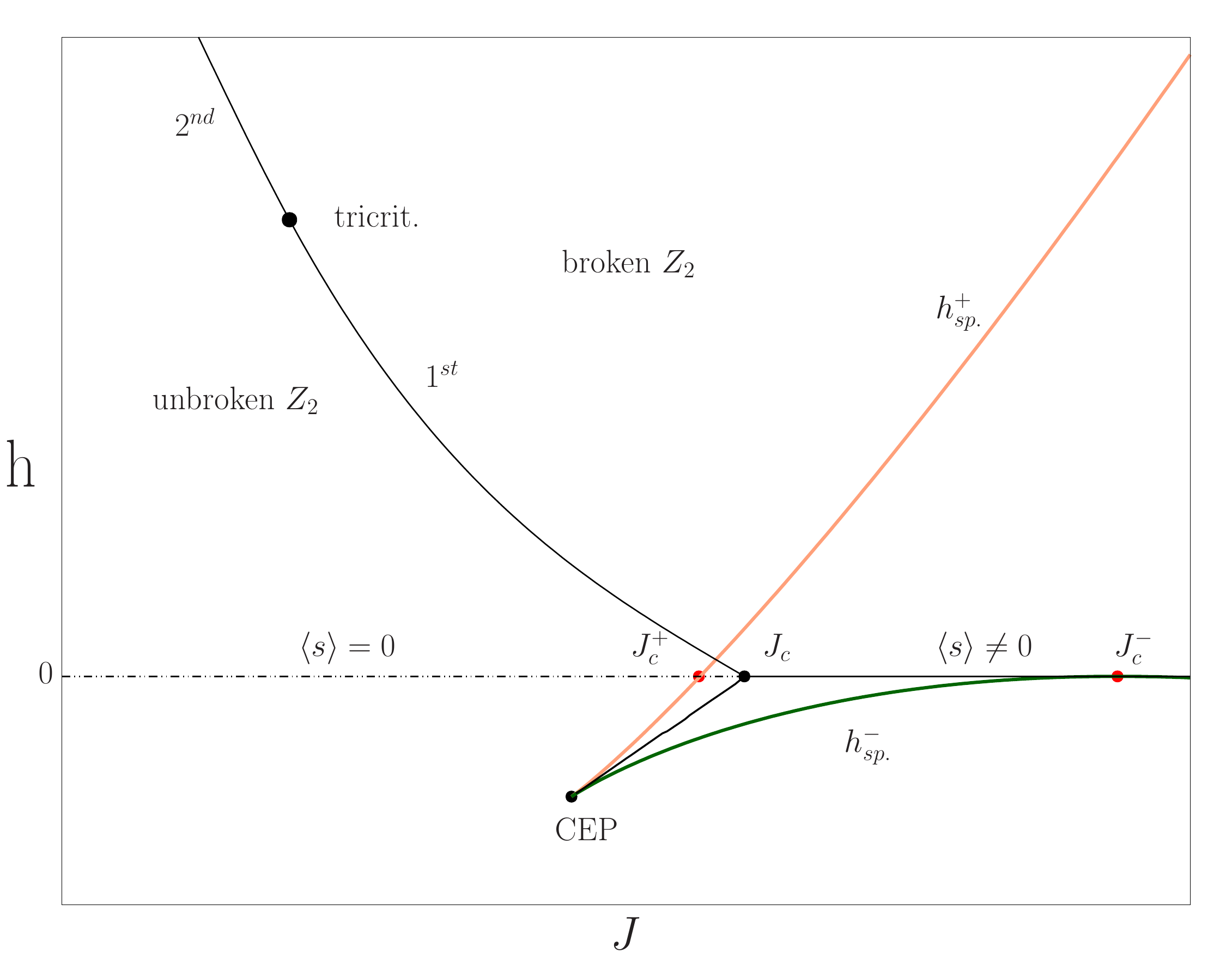}
\caption{\small \it Sketch of the phase diagram of the three-state Potts model in the mean-field limit. At $h=0$ the system undergoes a first order phase transition at $J_c$ , while for $h \neq 0$
the transition point moves in the $h-J$ plane, defining two first order lines lying respectively in the upper and lower half-plane. For $h>0$ the first order transition line culminates in a tricritical point, followed by a line of second order transitions that converges to the Ising critical point at $J=0.5$ in the limit $h\to +\infty$. Instead, for $h<0$ the line of first order phase transitions terminates in a Critical End Point (CEP) beyond which there are no true phase transitions, and the high and low temperature phase are smoothly connected (crossover). The CEP corresponds to the intersection point of the two spinodal lines $h_{sp}^{+}$ and $h_{sp}^{-}$. }
\label{fig:Spinodal_lines}
\end{figure*}

The Hamiltonian of the $q$-state Potts model in presence of external magnetic fields $h_{\alpha}$ is given by~(\cite{10.1103/RevModPhys.54.235})
\begin{align}
H=-\sum_{ij} J_{ij} \delta_{\sigma_i \sigma_j} -\sum_i\sum_{\alpha=0}^{q-1} h_{\alpha}\delta_{\alpha \sigma_i}\, ,
\label{eq:PottsModel}
\end{align}
where $i,j$ are lattice sites, $\sigma_{i}\in \{0,\ldots, q-1\}$ is the spin field, and $J_{ij}$ is the ferromagnetic coupling
\begin{align}
\label{ferro_coupling}
J_{ij}= \begin{cases}
 dJ/A(R) > 0\quad & \textrm{if} \quad |i-j|\leq R\\
 0 & \textrm{otherwise} 
\end{cases}
\end{align}
where $d$ is the dimensionality of the system and $A(R)$ is the number of spins that interact with any fixed $\sigma_{i}$, i.e.
\begin{align}
{\displaystyle \sum_{j}}\, J_{ij} = Jd\, .
\end{align}
\begin{figure*}%
\centering
     \subfloat[\label{fig:ContourPloth0}]{%
       \includegraphics[scale=0.68]{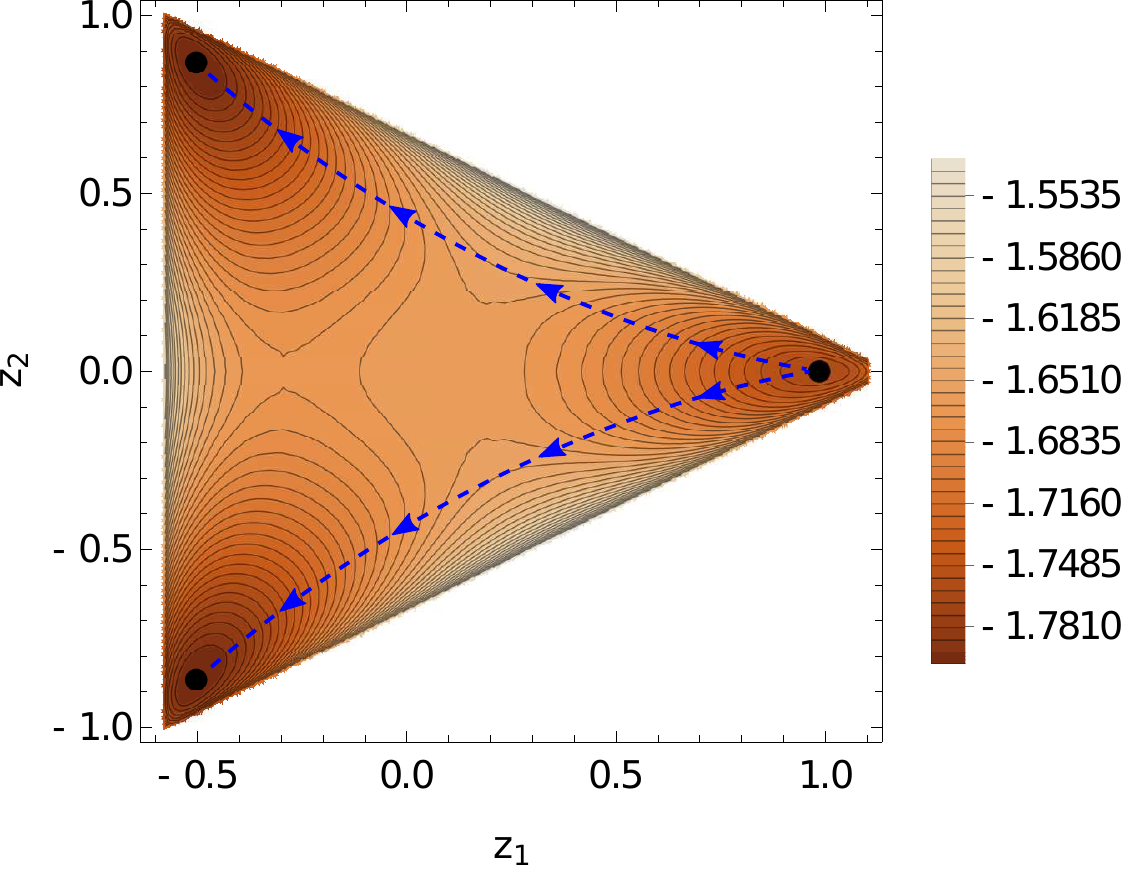}}
       \hspace{5mm}
     \subfloat[\label{fig:ContourPlothsp}]{%
       \includegraphics[scale=0.68]{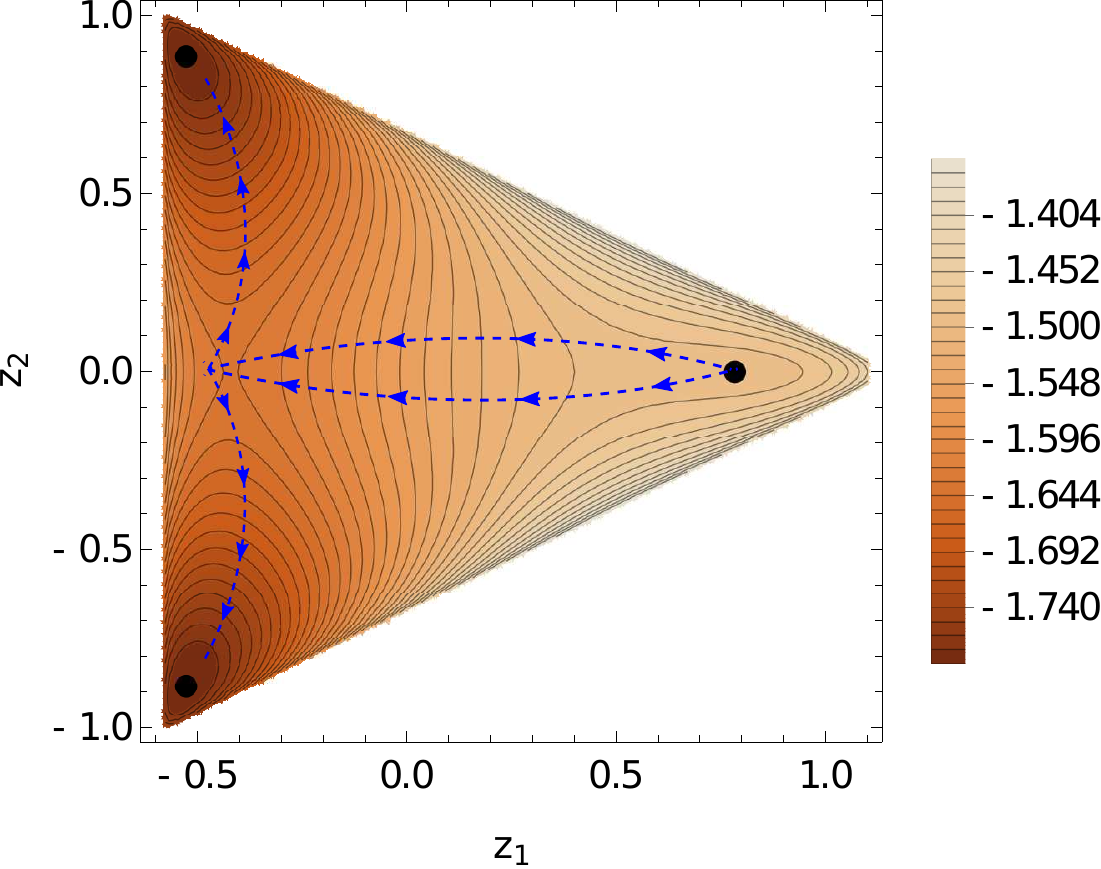}}
\caption{\small \it Contour plot of the free energy iso-surfaces in the $z_{1}-z_{2}$ plane for the three-state Potts model at $T=0.8T_{c}$ for (a) a shallow magnetic field $h=0.05$ and (b) a strong magnetic field $h=0.36$. The black points correspond to the positions of the three minima of the free-energy. The oriented blue lines display the most probable nucleation paths.}
\label{fig:ContourPlot}
\end{figure*}
In the long range limit $R\to\infty$, the free energy density $\mathcal{F}/V$ of the system can be computed exactly using standard combinatorial analysis~\cite{10.1103/RevModPhys.54.235}. As a function of the spin occupation numbers $n_\alpha =\langle \delta_{\alpha\sigma_{i}} \rangle$ one has
\begin{equation}
\label{mean_field_free_energy}
\frac{\mathcal{F}(J,h,\lbrace n \rbrace )}{V}=\sum_{\alpha=0}^{q-1} n_\alpha\log{n_\alpha } -Jd n_{\alpha}^2 -h_{\alpha}n_{\alpha}
\end{equation}
where $\sum_{\alpha=0}^{q-1}n_{\alpha}=1$. As it is well known, in the mean-field limit and at zero external field, the system undergoes a continuous phase transition for $q = 2$ and a discontinuous one for $q\geq 3$~(\cite{10.1088/0034-4885/50/7/001}) at a critical coupling $J_{c}$ given by Eq.~[\ref{mean_field_critical_coupling}]. Instead, in the case of next-neighbor interactions in two dimensions, the transition is first order only for $q\geq 5$.\\
\indent As already outlined in the introduction, in the ordered phase $J>J_{c}$ metastable states can be created by e.g. letting the system thermalize in one of the $q$ degenerate minima and then applying a magnetic field anti-parallel to it. Without loss of generality we take this state to be the minimum rich of $0-$th spins, i.e.
\begin{align}
n_{0} > n_{1} = n_{2} = \ldots = n_{q-1}\, ,
\end{align}
and turn on instantaneously a magnetic field $h_{\alpha} = -h\cdot\delta_{\alpha 0}$ with $h > 0$. After the quench, the system relaxes into a metastable minimum that according to the mean-field free energy of Eq.~[\ref{mean_field_free_energy}] corresponds to the following values of the occupation numbers
\begin{align}
&n_0^{ms}=\frac{1}{q}\left[1+(q-1)s\right]\, , \nonumber \\
&n_1^{ms}=\ldots=n_{q-1}^{ms}=\frac{1}{q}(1-s)\, ,
\label{eq:occupations}
\end{align}
and where the value of the scalar order parameter $-\frac{1}{q-1}\leq s\leq 1$ is obtained from 
\begin{align}
\frac{\partial (\mathcal{F}/V)}{\partial s} \propto  \log\left[ \frac{1+ (q-1)s}{1-s} \right]-2dJs + h = 0\, ,
\end{align}
with $\partial^{2}\mathcal{F} \geq 0$, and $s> 0$.
Depending on the magnitude of $h$, different nucleation paths can be followed by the system. An illustrative example is provided in Fig.~[\ref{fig:ContourPlot}], where the mean-field free energy is plotted for $q=3$ in the $z_{1}-z_{2}$ plane, with
\begin{align}
z_{1} = \frac{2n_{0}- n_{1}-n_{2}}{\sqrt{3}}\, ,\qquad  z_{2} = n_{1} - n_{2}\, . 
\end{align}
\begin{figure*}
\hspace{-0.3cm}
\subfloat[\label{fig:free_spinodal}]{%
       \includegraphics[scale=0.44]{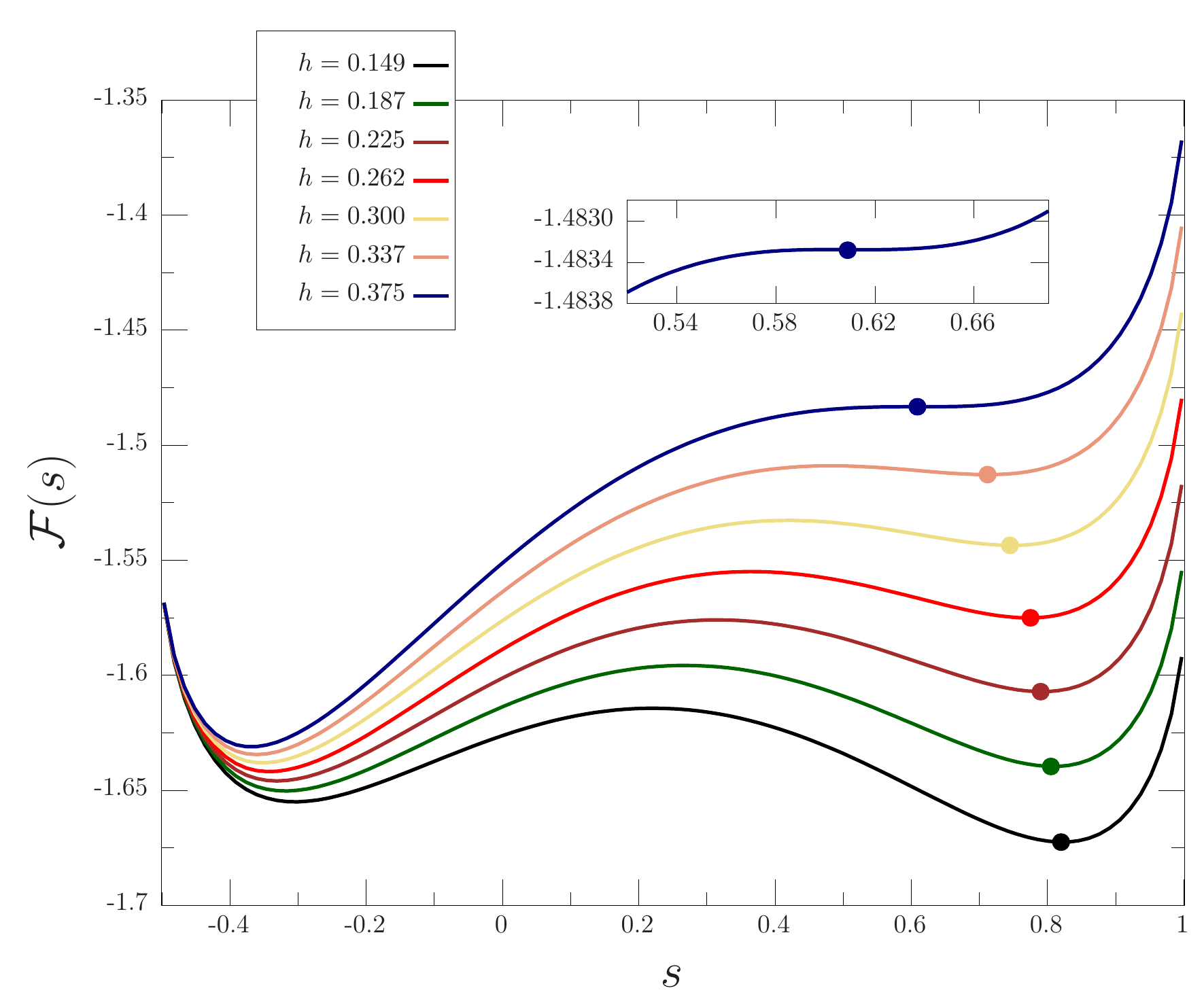}}
       \hspace{1.7cm}
     \subfloat[\label{free_non_spinodal}]{%
       \includegraphics[scale=0.44]{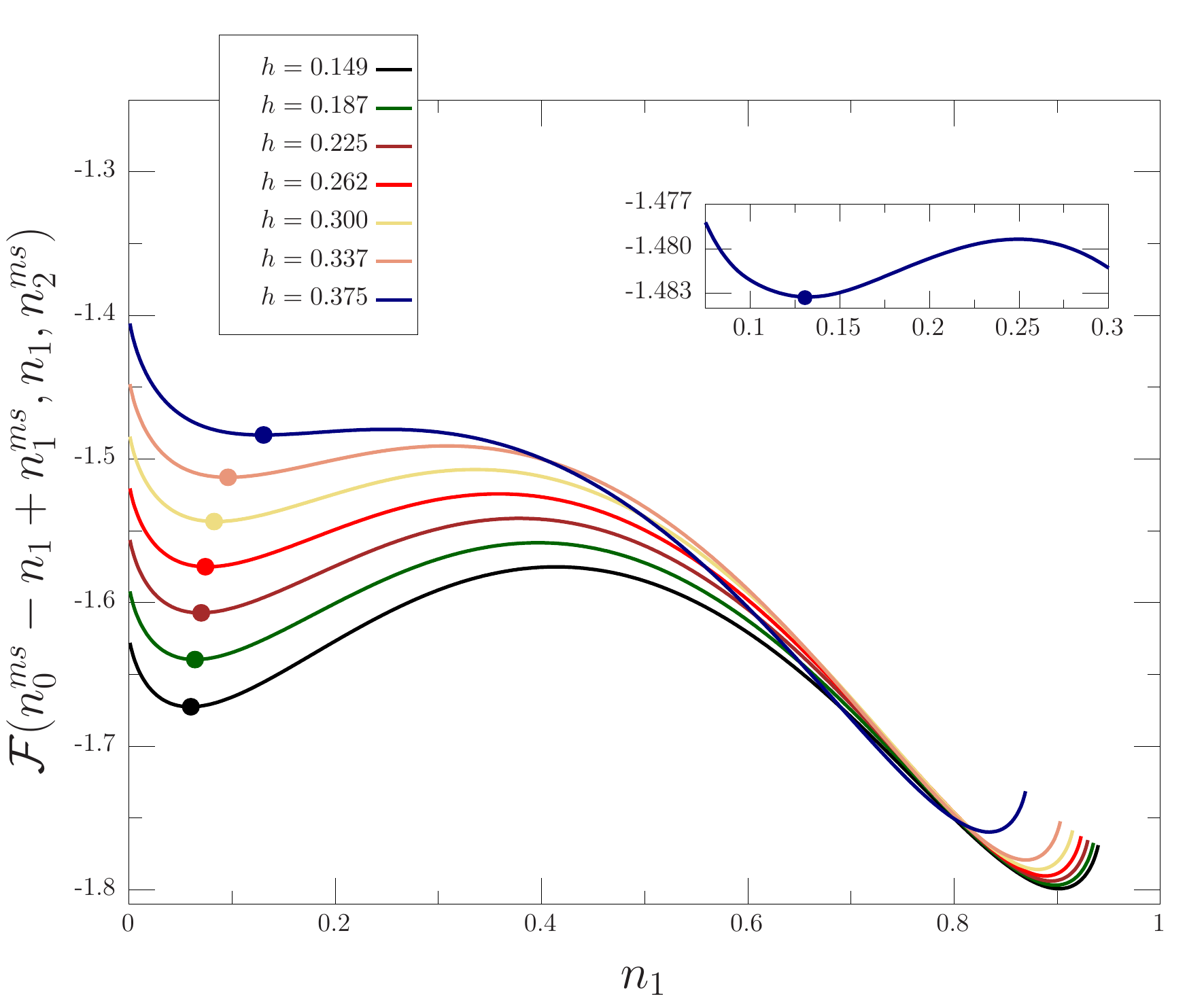}}
\caption{\small\it Mean-field free energy of the three-state Potts model for $T=0.8T_{c}$ along the symmetric channel (a) and  as a function of the occupation number $n_{1}$ keeping $n_{2}$ fixed to its metastable value $n_{2}^{ms}$ (b). The solid points indicate the position of the metastable state as a function of the magnetic field $h$. By increasing the value of $h$, the free energy along the symmetric channel develops an inflection point at $h=0.375$, while in the other direction a non-zero free energy barrier still separates the metastable state from the stable one. }
\label{fig:free_energy_slices}
\end{figure*}
The triangular symmetry of the free energy is broken by the presence of a non-zero $h$. For shallow quenches (Fig.~[\ref{fig:ContourPloth0}]) there are only two sectors in the $z_1-z_2$ plane where nucleation is expected to take place. These sectors host paths connecting the metastable state with one of the two degenerate global minima of the free energy at $z_{2} \ne 0$. In this situation nucleation happens through the formation of a single droplet of either spins $\sigma=1$ or $\sigma=2$. On the contrary, for deep quenches an additional nucleation channel along the line $z_{2}=0$ opens up (Fig.~[\ref{fig:ContourPlothsp}]). On this line, the system still possesses a $Z_{2}$ symmetry corresponding to the exchange of the occupation numbers $n_{1}$ and $n_{2}$. In this case the system gets into a state that does not correspond to an absolute minimum of the free energy. Eventually, a second decay along one of the two blue lines depicted in Fig~\ref{fig:ContourPlothsp} occurs, and thermalization is achieved. This picture applies to higher $q$ as well: for $h> h_{thr}(q)$, a nucleation channel where $n_{\alpha}=n_{\beta}$ for $\alpha,\beta>0$ is always present. We will refer to this nucleation path as the \emph{symmetric channel}. A relevant question is then whether the symmetric channel is the favored one. According to the Ostwald step-rule, the system nucleates more often into the state separated from the metastable one by the smallest free energy barrier. By increasing the magnitude of the magnetic field $h$, the curvature of the free energy density in the metastable minimum becomes smaller and smaller in the direction of the symmetric channel (see Fig.~[\ref{fig:free_energy_slices}]), becoming exactly zero at a spinodal point obtained by solving
\begin{align}
\label{def_spinodal}
\frac{\partial \mathcal{F}(s)}{\partial s} = \frac{\partial^{2} \mathcal{F}(s)}{\partial s^{2}} = 0 \, ,
\end{align}
with $\mathcal{F}$ given by Eq.~[\ref{mean_field_free_energy}]. The solution of the linear system determines two spinodal lines $h^{\pm}_{sp}(J,q)$, where
\begin{align}
&h^{\pm}_{sp}(J,q)= -\log\left[ \frac{1+(q-1)s^{\pm}_{sp}}{1-s^{\pm}_{sp} }\right] + 2dJs^{\pm}_{sp}\, ,  \\[6pt]
&s^{\pm}_{sp}(J,q)=\frac{ (q-2)\pm \sqrt{(q-2)^2+ 4(q-1)(1-\frac{q}{2dJ})}}{2(q-1)} \, .
\label{eq:Pottspinodals}
\end{align} 
The spinodal  $h^{-}_{sp}$ corresponds to the instability point of the metastable minimum rich of the $q-1$ spins $\sigma \in \{1,\ldots,q-1\}$, since $h^{-}_{sp} \leq 0$. The second solution $h^{+}_{sp}$ is instead the relevant one to our discussion, and corresponds to the absence of a free energy barrier along the symmetric channel. In the following sections, we will often use $h_{sp}$ and $s_{sp}$ in place of $h_{sp}^{+}$ and $s^{+}_{sp}$.\\
\indent The two spinodal lines $h^{\pm}_{sp}(J)$ are shown in Fig.~[\ref{fig:Spinodal_lines}] together with a sketch of the phase diagram, for $q=3$. They represent second order lines characterized by the divergence of the susceptibility
\begin{align}
\chi_{s}(J) = \frac{1}{V}\frac{\partial^{2} \mathcal{F}(J,h)}{\partial h^{2}}\big|_{h=h_{sp}^{\pm}}\, ,
\end{align}
measured in the metastable state. For all $q> 2$, the two lines $h^{\pm}_{sp}(J)$ intersect the horizontal axis $h=0$ in exactly two points $J_{c}^{+}$ and $J_{c}^{-}$ with $J_{c}^{+}<J_{c}<J_{c}^{-}$, which correspond respectively to the limit of metastability of the high and low temperature phases. Below $J_{c}^{+}$ the spinodal magnetic field $h_{sp}^{+}(J)$ turns negative, and the two spinodal lines meet in a critical end-point (CEP) at
\begin{align}
J_{CEP} &= \frac{2}{d}\frac{(q-1)}{q}\, ,\\[8pt]
h_{CEP} &=  \frac{2(q-2)}{q}-\log(q-1)\, ,
\end{align}   
which represents the end-point of the first order phase transition line departing from the horizontal axis at $J_{c}$.
\section{Semi-classical expansion around the mean-field limit}
\label{II}
In the large $R$ limit and for deep quenches close to the spinodal point, the decay of the metastable state can be described analytically, following the field-theoretical approach firstly developed by Klein and Unger in Refs.~\cite{10.1103/PhysRevB.29.2698, 10.1103/PhysRevB.31.6127, 10.1103/PhysRevB.28.445}. The starting point is the partition function of the $q-$state  Potts model with  arbitrary ferromagnetic couplings $J_{ij}$ and magnetic fields $h_{\alpha} = -h\,\delta_{\alpha 0}$, which we conveniently rewrite in terms of clock spins $\{\lambda_{i}\}$ as  
\begin{align}
\label{clock_potts_partition_function}
\mathcal{Z}(J ,h) = {\displaystyle \sum_{\{\lambda_{i}\}}}\exp{\bigg\{ \sum_{ij}\sum_{k=1}^{q-1} \frac{J_{ij}}{q} \lambda_{i}^k \lambda_{j}^{-k} -\frac{h}{q}\sum_{i}\sum_{k=1}^{q-1} \lambda_i^k\bigg\}}\, ,
\end{align}     
where $\lambda_{i} \in \{1,\omega, \omega^{2},\ldots, \omega^{q-1}\}$ and  $\omega = e^{2\pi i/q}$. The equivalence between the previous partition function and the one obtained from the Hamiltonian of Eq.~[\ref{eq:PottsModel}] follows from the identity
\begin{align}
\sum_{k=1}^{q} \lambda^{k}_{i} \, \lambda^{-k}_{j} = q\,\delta_{\lambda_{i} \lambda_{j}}\, .
\end{align}
Field theoretical methods can be applied after performing an Hubbard-Stratonovich transformation \cite{PhysRevLett.3.77,1957SPhD....2..416S} via 
\begin{align}
{\displaystyle\prod_{ij}\prod_{k=1}^{q-1}}\exp{\bigg\{ \frac{J_{ij}}{q} \lambda_{i}^k \lambda_{j}^{-k}\bigg\}} &\propto \int \left[ d\phi\right]  \textrm{exp}\bigg\{ {\displaystyle\sum_{ij}\sum_{k=1}^{q-1}} \phi_{k,i}\lambda_{i}^{k}\nonumber \\[8pt]
&+\phi^{*}_{k,i}\lambda_{i}^{-k}-q\phi_{k,i}^{*}J^{-1}_{ij}\phi_{k,j}\bigg\}\,,
\end{align}
where we introduced the auxiliary complex fields $\phi_{k}$ with $k \in \{1,\ldots, q-1\}$ and $\phi_{k}^{*} = \phi_{q-k}$. Inserting this equality back into Eq.~[\ref{clock_potts_partition_function}] and performing the sum over the original clock spins $\{\lambda_{i}\}$, one gets 
\begin{align}
\label{H-S_Potts1}
&\mathcal{Z}(J, h) \propto \int \left[ d\phi \right]\,  e^{-\mathcal{S}(\{\phi_{k,i}\})}   \, ,  
\end{align}
with
\begin{align}
&\mathcal{S}(\{ \phi_{k,i}\}) = {\displaystyle \sum_{k}} {\displaystyle \sum_{ij}}\, q\phi^{*}_{k,i}\, J^{-1}_{ij}\, \phi_{k,j} -{\displaystyle \sum_{i}}\, V(\{\phi_{k,i}\})~,
\label{H-S_Potts2}
\end{align}
and where the local potential $V(\{\phi_{k,i}\})$ is given by
\begin{align}
\label{eq:potential_HS}
V\left(\{\phi_{k,i}\}\right) = \log{ \bigg( \,\,{\displaystyle \sum_{\lambda=1}^{\omega^{q-1}}}\,{\displaystyle \prod_{k=1}^{q-1}}\,\, e^{\phi^{*}_{k,i}\lambda^{-k} + \phi_{k,i}\lambda^{k} - \frac{h}{q}\lambda^{k}}\bigg) }\, .
\end{align}
The kinetic term in the r.h.s. of Eq.~[\ref{H-S_Potts2}] can be expanded in powers of the interaction range $R$. Specializing to the functional form $J_{ij}$ of Eq.~[\ref{ferro_coupling}], the first two terms of the expansion are given by
\begin{align}
\label{eq:gradient_expansion}
q\,{\displaystyle \sum_{ij}}\,\, \phi^{*}_{k,i}\,\, J^{-1}_{i j}\,\,\phi_{k,j} &=  \frac{q}{dJ}\sum_{i}\bigg\{ I_{2}^{d}(R) \frac{R^{2}}{2}| \nabla^{L}\phi_{k,i}|^2  \nonumber \\[8pt]
&+ |\phi_{k,i}|^2  + \mathcal{O}(R^{4}) \bigg\}\, ,
\end{align}
where $\nabla^{L}$ is a lattice discretization of the derivative operator, $d$ is the dimensionality of the system, and
\begin{align}
\label{eq:I2}
I_{2}^{d}(R) = \frac{1}{R^{2}}\frac{ \sum_{j} |j-i|^{2} J_{ij}}{\sum_{j} J_{ij}}~.
\end{align}
This number quickly converges to its asymptotic value $\lim_{R\to\infty}I_{2}^{d}(R) = d/(d+2)$. 
When the interaction range $R$ becomes infinite, the term $\big|\nabla^{L}\phi_{k,i}\big|^2$ must be zero for configurations having a non-zero statistical weight, and one recovers the mean-field result discussed in the previous section. However, at finite $R$ this term allows for instanton configurations, i.e. non-uniform finite energy solutions of the equation of motion, which climb the energy barrier between the metastable minimum and states having lower free energy. Within non-classical nucleation theory, such field configurations are identified with the critical droplets~\cite{10.1016/0003-4916(67)90200-X,10.1088/0305-4470/13/5/034}. \\
\indent Before turning into the computation of the critical droplet, we discuss the interpretation of the Hubbard fields $\phi_{k}$. To this purpose, we perform a field transformation defining the Hubbard occupation numbers $\tilde{n}_{\alpha}\in\mathbb{R}$ via
\begin{align}
\phi_{k} = \frac{Jd}{q}\sum_{\alpha=0}^{q-1} \tilde{n}_{\alpha}e^{-2\pi k\alpha i/q}\, ,\quad \sum_{\alpha}\tilde{n}_{\alpha} = 1~.
\end{align}
In terms of the $\{ \tilde{n}_{\alpha}\}$ the action takes the form
\begin{widetext}
\begin{align}
\label{eq:lagrangian_tilde_n}
\mathcal{S}\left(\{\tilde{n}_{\alpha,i}\}\right) &= \sum_{i} \frac{Jd}{q}I_{2}^{d}(R)\frac{R^{2}}{2}\left[ (q-1) \sum_{\alpha=0}^{q-1}\nabla^{L}\tilde{n}_{\alpha,i}\nabla^{L} \tilde{n}_{\alpha,i} - \sum_{\alpha \ne \beta}\nabla^{L}\tilde{n}_{\alpha,i}\nabla^{L}\tilde{n}_{\beta,i}\right] - \tilde{V}(\left\{\tilde{n}_{\alpha,i}\right\})\, , \nonumber \\[8pt]
\tilde{V}(\left\{\tilde{n}_{\alpha,i}\right\}) &= \log{ \left( {\displaystyle \sum_{\alpha=0}^{q-1}} \exp{ \left\{ 2dJ\, \tilde{n}_{\alpha,i} -h\, \delta_{\alpha 0}\right\}}   \right)} -\frac{Jd}{q}\left[ (q-1)\sum_{\alpha=0}^{q-1} \tilde{n}_{\alpha,i}^{2} - {\displaystyle \sum_{\alpha \ne \beta}} \tilde{n}_{\alpha,i}\tilde{n}_{\beta,i}    \right]~. 
\end{align}
\end{widetext}
In the mean-field limit, the Hubbard occupation numbers $\tilde{n}_{\alpha}$ correspond to the occupation numbers defined in Sec.~[\ref{I}]. This can be easily checked comparing the stationary points of the potential in Eq.~[\ref{mean_field_free_energy}] with the ones obtained from the potential $\tilde{V}(\{\tilde{n}_{\alpha,i}\})$ in Eq.~[\ref{eq:lagrangian_tilde_n}]. However, away from the mean-field limit this does not hold true, and the Hubbard fields play the role of effective magnetic fields. In the following, we will make the assumption that close enough to the spinodal point the quasi-equilibrium properties of the spins are well described by the Lagrangian $\mathcal{L}(\{n_{\alpha,i}\})$ in a neighborhood of the metastable minimum, identifying the Hubbard occupation numbers $\tilde{n}_{\alpha}$ with the spin occupation numbers $n_{\alpha}$.\\
\indent Making use of the action $\mathcal{S}$, we can evaluate the profile of the critical droplet. To do so, we first replace in Eq.~[\ref{eq:lagrangian_tilde_n}] the sum over the lattice volume $\sum_{i}$ with a continuum integral $\int d^{d}\mathbf{r}$, and $\nabla^{L}\to \nabla_{\mathbf{r}}$. This is justified under the assumption of slowly varying fields, which holds true if $R$ is large enough.   
The critical droplets describing the decay of the metastable state in Eq.~[\ref{eq:occupations}] are non-uniform configurations of the form $n_{\alpha}(\mathbf{r}) = n_{\alpha}^{ms} + n_{\alpha}^{cr}(\mathbf{r})$ that correspond to saddle points of the functional integral, and satisfy the boundary conditions
\begin{align}
\label{eq:boundary_conditions}
\lim_{\mathbf{r}\to\infty} n_{\alpha}^{cr}(\mathbf{r}) =0\, , \qquad \nabla_{\mathbf{r}} n_{\alpha}^{cr}(\mathbf{r})\bigg|_{\mathbf{r}=0} = 0\, . 
\end{align}
As usual these are obtained by solving the equations of motion
\begin{align}
\nabla_{\mathbf{r}}\frac{\partial\mathcal{L}}{\partial \nabla_{\mathbf{r}}n_{\alpha}(\mathbf{r})} &= \frac{\partial\mathcal{L}}{\partial n_{\alpha}(\mathbf{r})}\, ,\nonumber \\[8pt]
\quad S\left(n_{\alpha}(\mathbf{r})\right) &= \int d^{d} \mathbf{r}\,\mathcal{L}\left(n_{\alpha}(\mathbf{r})\right)~.
\end{align}
\indent For general $q$, there are many directions in the space of the occupation numbers $n_{\alpha}$ for which a saddle point solution exists.
Here, we mainly concentrate on the direction that allegedly represents the dominant nucleation path in the case of deep quenches close to the mean-field spinodal line, i.e. the symmetric channel discussed in Sec.~[\ref{I}]. Along this path the occupation numbers $n_{\alpha}$ with $\alpha \geq 1$ are all equals; hence we seek for a solution of the equations of motion of the form
\begin{align}
n_{\alpha}(\mathbf{r})= \begin{cases}
\frac{1}{q}[1 + (q-1)\cdot (s^{ms}+s^{cr}(\mathbf{r}))] & \alpha = 0 \\
\frac{1}{q}[1 - (s^{ms} +s^{cr}(\mathbf{r}))] & \alpha > 0
\end{cases}  \, .
\end{align} 
The Lagrangian $\mathcal{L}( n_{\alpha}(\mathbf{r}))$, when restricted to the symmetric path, takes the form
\begin{align}
\label{eq:symmetric_potential}
\mathcal{L}\left(s(\mathbf{r})\right) &= \frac{Jd(q-1)}{q}~I_{2}^{d}(R)\frac{R^{2}}{2}|\nabla_{\mathbf{r}} s(\mathbf{r})|^{2} - \tilde{V}(s(\mathbf{r}))\, ,\nonumber \\[8pt]
\tilde{V}(s(\mathbf{r})) &=\log \left[ e^{ (q-1)\cdot\left(\frac{2dJ}{q}s(\mathbf{\mathbf{r}})-\frac{h}{q}\right)}+(q-1)e^{ -\frac{2dJ}{q}s(\mathbf{\mathbf{r}})+\frac{h}{q} }\right] \nonumber \\[8pt]
&- Jd\frac{(q-1)}{q}s^{2}(\mathbf{r})\, ,
\end{align}
up to irrelevant additive constants. 
For $h\lesssim h_{sp}(J,q)$ the potential is almost flat in the region around the metastable minimum $s \sim s^{ms}$; thus a Taylor expansion in $s(\mathbf{r})$ up to the cubic term provides a good approximation. This gives
\begin{align}
\tilde{V}(s(\mathbf{r})) \sim \textrm{const.} + a(q,J)s^{3}(\mathbf{r}) + b(q,h)\,s(\mathbf{r})\, , 
\label{eq:Vcubic}
\end{align}  
where
\begin{align}
\label{eq:coefficient}
a(q, J) &= \frac{2 (dJ)^2 (q-1) \sqrt{\frac{q (q(dJ-2)+2)}{dJ}}}{3 q^2}\, ,  \nonumber \\[8pt]
b(q,h)&=\frac{q-1}{q}(h-h_{sp})\, ,
\end{align}
and order $\mathcal{O}\left(s^{2}\cdot  (h-h_{sp})\right)$ terms have been also neglected. The $\mathrm{SO}(d)$-symmetric solution to the corresponding Euler-Lagrange equation
\begin{align}
\label{eq:E-L_equation}
&R^{2}\left[\partial^{2}_{r}s^{cr}(r) + \frac{d-1}{r}\partial_{r}s^{cr}(r)\right]= \frac{1}{K}\frac{\partial\tilde{V}(s^{cr}(r))}{\partial s^{cr}(r)}\, , \nonumber \\[8pt]
 &K = I_{2}^{d}(R)\frac{Jd(q-1)}{q}\, , 
\end{align}
with the boundary conditions of Eq.~[\ref{eq:boundary_conditions}] and for arbitrary coefficients, is thoroughly discussed in Refs.~\cite{10.1103/PhysRevB.28.445,10.1103/PhysRevB.29.2698}. The equation describes the one-dimensional motion of a particle subject to the potential $-\tilde{V}$, and to a time-dependent friction force. By neglecting the radial first derivative term, Eq.~[\ref{eq:E-L_equation}] can be solved analytically, giving
\begin{align}
\label{eq:saddle_point}
s^{cr}(r) &= -\sqrt{\frac{3|b(q,h)|}{a(q,J)}}\sech^{2}{\left(\frac{r-r_{0}}{\xi_{cr}} \right)} \, , \\[8pt]
\xi_{cr} /R &= \sqrt{2K} \left(3|b(q,h)|a(q,J)\right)^{-\frac{1}{4}}\, ,
\end{align}
where the center of the droplet $r_{0}$ is a free parameter. 
The critical droplet described by Eq.~[\ref{eq:saddle_point}] is a  diffuse field configuration having an amplitude proportional to $\sqrt{h-h_{sp}}$ and extension $\xi_{cr} \propto R\,(h-h_{sp})^{-1/4}$. The role of the neglected first derivative term is to enhance the amplitude of the critical fluctuation at small $r \lesssim \xi_{cr}$. 
Numerical solutions to the exact Eq.~[\ref{eq:E-L_equation}] are easily obtained using shooting methods. 

The energy associated to the critical fluctuation is instead obtained by evaluating the functional integral Eq.~[\ref{H-S_Potts1}] on the previous solution. Making use of the approximate solution of Eq.~[\ref{eq:saddle_point}] one gets
\begin{align}
\label{eq:critical_droplet_energy}
E_{\textrm{drop}}(J,q,h,R) = \Omega(d)\theta\,  R^{d} K^{\frac{d}{2}} \frac{\big|b(q,h)\big|^{\frac{6-d}{4}}}{\left(a(q,J)\right)^{\frac{d+2}{4}}}\, \,
\end{align}
where $\theta \sim 1.3090$, and $\Omega(d)$ is the solid angle in $d$ dimensions. If one uses the exact solution of Eq.~[\ref{eq:E-L_equation}], the energy of the critical droplet gets enhanced by a constant factor 1.2564 for all values of the parameters.\\
\indent After the formation of the critical droplet, the early stages of nucleation are described, within a quasi-static approximation, through a semi-classical expansion of the functional integral around the saddle point $s^{cr}(r)$, via
\begin{widetext}
\begin{align}
\mathcal{S}(s^{cr}(r) + \delta s(\mathbf{r})) &\simeq \mathcal{S}(s^{cr}(r)) + \frac{1}{2}\int d\mathbf{r}\, d\mathbf{r'}\,\, \delta s(\mathbf{r})\frac{\partial^{2} \mathcal{S}}{\partial s(\mathbf{r}) \partial s(\mathbf{r'})}\bigg|_{s=s^{cr} }\!\!\!\!\!\!\!\!\!\delta s(\mathbf{r'}) ~,\nonumber \\[8pt]  
M(\mathbf{r},\mathbf{r'}) &= \frac{\partial^{2} \mathcal{S}}{\partial s(\mathbf{r}) \partial s(\mathbf{r'})}\bigg|_{s=s^{cr} }= \left[-KR^{2}\nabla^{2}_{\mathbf{r}} + \sqrt{12|a(q,J)b(q,h)|} + +6\alpha(q,J)s^{cr}(r)\right]\delta(\mathbf{r},\mathbf{r'})~. 
\end{align}
\end{widetext}
The quadratic form $M(\mathbf{r},\mathbf{r'})$ possesses an \emph{unique} negative eigenvalue $\lambda_{-}(J,q,h)$ with corresponding eigenfunction $\psi_{\lambda_{-}}(r)$,  called the \emph{growth mode} (see e.g. Ref.~\cite{10.1016/0003-4916(67)90200-X}). Distorsions of the droplet profile along this eigenmode grow exponentially over time since they correspond to a decrease in the free energy. For $t\gtrsim t_{nucl}$, where $t_{nucl}$ is the time where the critical fluctuation appears in the system, one expects the profile of the critical droplet to evolve according to
\begin{align}
s(r,t) - s^{ms} = s^{cr}(r) -\delta e^{\kappa|\lambda_{-}|t} \psi_{\lambda_{-}}(r)\, ,
\end{align}
where $\delta$ is some positive constant reflecting the instability that initiate the decay, while $\kappa$ is a  positive coefficient which depends on the microscopic dynamic of the system~\cite{10.1016/0003-4916(69)90153-5}. In the case of the $q-$state  Potts model, using the coefficients $a(q,J)$ and $b(q,h)$ in Eq.~[\ref{eq:coefficient}], and the result of Ref.~\cite{10.1103/PhysRevB.28.445}, one gets
\begin{align}
\label{eq:growth_mode}
\lambda_{-}(J,q,h) &= -\frac{5\sqrt{3}}{4}\sqrt{a(q,J)\, |b(q,h)|}\, , \\
\psi_{-}(r) &= \sech^{3}{\left(\frac{ r - r_{0}}{\xi_{cr}}\right)}\, . 
\end{align} 
Thus, as usual for spinodal nucleation, during the early stages of growth one expects that the critical droplet compactifies around its center, since the eigenfunction $\psi_{-}(r)$ drops to zero faster than $s^{cr}(r)$. After the saturation of the center, the droplet starts to expand, as already observed in numerical simulations of the two-dimensional Ising model in Ref.~\cite{10.1103/PhysRevLett.68.2336}.\\
\indent Concerning the average nucleation time $\tau$, it can be related to the energy of the critical droplet via~\cite{10.1016/0003-4916(67)90200-X, 10.1016/0003-4916(69)90153-5}
\begin{align}
\tau = \frac{1}{V}\exp{\left(E_{drop} + f_{1-loop} + f_{dyn}\right)}\, , 
\label{eq:tauexp}
\end{align}
where $f_{1-loop}$ is a contribution coming from the so-called capillary wave approximation ~\cite{10.1063/1.4959235}, i.e. obtained by performing a semi-classical expansion around both the metastable minimum and the saddle point solution of Eq.~[\ref{eq:saddle_point}], while $f_{dyn}$ is the so-called dynamical prefactor~\cite{10.1016/0003-4916(69)90153-5}. Both contributions enter as (slowly varying) logarithmic corrections of the form $f_{1-loop}\sim f_{dyn} \propto -\log{|h-h_{sp}|}$, that, for any fixed non-zero value of $h-h_{sp}$, are subleading with respect to $E_{drop}$ at sufficiently large $R$. In Sec.~[\ref{IV}] we will compare the scaling of the nucleation times obtained from Monte Carlo simulations with the theoretical prediction of Eq.~[\ref{eq:tauexp}].\\
\subsection{Validity of the mean-field approach}
\label{sub:validity}
\indent In order to compare the analytical prediction with the result of numerical simulations at finite $R$, it is important to identify the range of validity of the mean-field approach. In thermal equilibrium this is provided by the Ginzburg criterion which states that, for a general system with order parameter $\phi$, the corrections to mean-field behavior close to a critical point are small if
\begin{align}
\langle (\delta \phi)^{2} \rangle_{T} \ll \langle \phi \rangle^{2}_{T} \implies \chi_{\phi} \ll \xi^{d}_{\phi} \langle \phi \rangle ^{2}_{T}~,
\end{align}
where $\chi_{\phi}$ and $ \xi_{\phi}$ are respectively the susceptibility and correlation length of the order parameter. The concept expressed by the Ginzburg criterion in the previous equation can be generalized to describe the onset of mean-field-like behavior close to the spinodal point~\cite{10.1103/PhysRevA.29.341}. A necessary condition for the validity of the mean-field equations is that the typical mean-square fluctuations of the order parameter must be much smaller than the square of the amplitude of the critical fluctuation, i.e.
\begin{align}
\langle (\delta \phi(r))^{2} \rangle_{ms, \ell^{d}} \ll (\phi^{cr}(r=0) - \phi^{cr}(r=\infty))^{2}~,
\end{align}
where $\langle . \rangle_{ms}$ denotes the average in the metastable state, and the l.h.s. is averaged over a volume $V_{\ell} = \ell^{d}$. By taking $\ell = \xi^{ms}_{\phi}$ one gets  
\begin{align}
\label{eq:ginzburg_spinodal}
\chi^{ms}_{\phi} \ll \left(\xi^{ms}_{\phi}\right)^{d}\cdot\left(\phi^{cr}(r=0) - \phi^{cr}(r=\infty)\right)^{2}~,
\end{align}
where $\chi^{ms}_{\phi}$ and $\xi^{ms}_{\phi}$ are the susceptibility and correlation length of the order parameter, as measured in the metastable state. Close to the spinodal point, the fluctuations of the field $s$ around its metastable value $s^{ms}$ can be computed from Eqs.~[\ref{eq:symmetric_potential}],~[\ref{eq:Vcubic}]. Performing a Gaussian approximation around $s=s^{ms}$, the two point function in momentum space takes the usual Ornstein-Zernike form 
\begin{align}
\label{eq:2pt}
\langle \delta s(p) \delta s(-p)\rangle_{ms} = \frac{1}{KR^{2}}\frac{1}{p^{2} + \left(\frac{12|a(J,q)b(h,q)|}{K^{2}R^{4}}\right)^{1/2}}~,    
\end{align}
where $K$ has been defined in Eq.~[\ref{eq:E-L_equation}], and $\delta s = s-s^{ms}$. We thus have
\begin{align}
\label{eq:chi_xi_mf}
\chi_{s}^{ms} &= (12|a(J,q)b(q,h)|)^{-1/2}\, ,  \\[8pt]
\label{eq:xi_mf}
\quad  \xi_{s}^{ms} &= R\frac{\sqrt{K}}{(12|a(J,q)b(h,q)|)^{1/4}}~.
\end{align}
Substituting back the values of $\chi_{s}^{ms}$ and $\xi_{s}^{ms}$ into Eq.~[\ref{eq:ginzburg_spinodal}], and making use of Eq.~[\ref{eq:saddle_point}], one gets the following condition for the validity of mean-field 
\begin{align}
\label{eq:cond_mean_field}
G \,\stackrel{def}{=}\, \frac{3}{12^{(d-2)/4}} R^{d} K^{d/2} \frac{|b(q,h)|^{(6-d)/4}}{a(J,q)^{(d+2)/4}} \gg\, 1~.  
\end{align}
Thus, for large interaction ranges $R$, for $d<6$ mean-field provides a good description with the exception of a narrow region around the spinodal point where $b(q,h) \to0$. The condition expressed by Eq.~[\ref{eq:cond_mean_field}] can be equivalently written in terms of the energy of the critical droplet since
\begin{align}
\frac{E_{drop}}{G} = (12)^{(d-2)/4}\Omega(d)\,\frac{\theta}{3}\,\stackrel{d=2}{\approx}\, 2.7.
\end{align}
Hence, having a sufficiently large energy barrier is a necessary condition to observe mean-field like behavior.\\ 
\indent The emergence of the \emph{spinodal} Ginzubrg parameter can be also understood through the following argument. The renormalization group analysis of Ref.~\cite{10.1103/PhysRevB.18.6199} showed that in the exact mean-field limit, there is a fixed point associated to the spinodal singularity. 
The spinodal fixed point is approached sending to zero an unique relevant scaling variable $u_{h}$, that in our case is proportional to $h- h_{sp}$ (at small $h-h_{sp}$) ,  i.e. to the distance from the spinodal point. In Ref.~\cite{10.1103/PhysRevB.18.6199} the scaling variable $u_{h}$ is associated to a critical exponent $y_{u_{h}} = \frac{2}{3}d$. This fact implies that in a neighborhood of the spinodal fixed point, the singular part of the (metastable) free energy density $f^{ms}_s(u_h)$ scales as
\begin{align}
\label{scaling_RG}
f^{ms}_{s}(u_h) = b^{-nd} f^{ms}_{s}\left( b^{ny_{u_{h}}} u_{h} \right)~.
\end{align}
In the previous expression we iterated $n$ times the renormalization group transformation using a coarse graining parameter $b$. Eq.~[\ref{scaling_RG}] implies that close to the spinodal point $f_{s}(u_{h}) \propto u_{h}^{3/2} \sim (h-h_{sp})^{3/2}$ in agreement with Eq.~[\ref{eq:chi_xi_mf}] since $s^{ms}\sim (h-h_{sp})^{1/2}$ and
\begin{align}
f^{ms}_s(s^{ms}) = \tilde{V}(s^{ms}) \propto (h-h_{sp})^{3/2}~.    
\end{align}
The spinodal fixed point turns out to be unstable below six dimensions if a gradient term $u_{R}^{-1}\nabla^{2}$, i.e. a finite range interaction, is present in the starting Hamiltonian. This reflects the fact that for $d<6$, as $h\to h_{sp}$ at fixed $R$, a well defined spinodal does not exist since the free energy cost associated to the critical fluctuation goes to zero (Eq.~[\ref{eq:critical_droplet_energy}]). In Ref.~\cite{10.1103/PhysRevB.18.6199} it was found that the coupling $u_{R}$ of the gradient term scales under a renormalization group transformation as $b^{y_{u_{R}}}$ with $y_{u_{R}} = 2- d/3$, and for $u_{R} \ne 0$ and $d<6$ the RG flow moves away from the spinodal fixed point. The crossover between mean-field-like and non-mean-field behavior can be discussed assuming that close to the spinodal fixed point $u_{R}= u_{h} = 0$, the singular part of the metastable free energy density satisfies a scaling relation of the form
\begin{align}
\label{eq:two_scaling_RG}
f^{ms}_{s}(u_{h}, u_{R}) = b^{-nd}f^{ms}_{s}\left( b^{ny_{u_{h}}}u_{h}, b^{ny_{u_{R}}}u_{R}\right)~.
\end{align}
By choosing $\big|b^{ny_{u_{h}}}u_{h}\big| = 1$, one gets
\begin{align}
f^{ms}_{s}(u_{h}, u_{R}) &= \big| u_{h} \big|^{3/2}f^{ms}_{s}\left(1,
u_{R}/u_{h}^{(y_{u_{R}}/y_{u_{h}})}\right)\nonumber \\[8pt]
&= \big |u_{h}\big|^{3/2}\Psi\left(u_{R}/u_{h}^{(y_{u_{R}}/y_{u_{h})}} \right)~,
\end{align}
where $\Psi$ is a universal scaling function and $y_{u_{R}}/y_{u_{h}}$ is the crossover exponent \cite{9780521499590}. Close to the spinodal,  $u_{h} \propto h-h_{sp}$, and $u_{R} \propto R^{-2}$, hence
\begin{align}
\label{eq:final_Ginz_cross}
\Psi\left(u_{R}/u_{h}^{(y_{u_{R}}/y_{u_{h})}} \right) &= \hat{\Psi} \left(R^{-2}/(h-h_{sp})^{(y_{u_{R}}/y_{u_{h})}}\right) \nonumber \\[8pt]
&= \tilde{\Psi}\left( G^{-1} \right)~,
\end{align}
where in the last line we have used $y_{u_{R}}/y_{u_{h}} = 3/d - 1/2$ and Eq.~[\ref{eq:cond_mean_field}], and we dropped the dependence of $G$ on $J$ and $q$. The previous expression shows that, as expected, the crossover region corresponds to $G \gg 1$. Approaching the spinodal point keeping $G$ fixed, allows to be sensible to the mean-field spinodal critical exponents. Moreover when $G\gg 1$ also the mean lifetime of the metastable state is very long (Eqs.~[\ref{eq:ginzburg_spinodal}],~[\ref{eq:tauexp}]). This justifies the underlying assumption made in writing Eq.~[\ref{eq:two_scaling_RG}], i.e. the possibility of treating the metastable state as an equilibrium state with a restricted partition function consisting of all microstates in the neighborhood of the metastable minimum.

\section{Numerical methods}
\label{III}
The simulations have been carried out on a $L\times L$ square lattice employing periodic boundary conditions in all directions. We considered four different values of $q=3,5,10,20$, interaction ranges $R\in     \left\{2,\ldots, 15\right\}$ and two values of temperature: $T=0.5T_{c}$ and $T=0.8T_{c}$, where $T_{c}$ is the mean-field melting temperature defined in Eq.~[\ref{mean_field_critical_coupling}]. The heat bath algorithm has been used to evolve the system, i.e. local spin updates $(\sigma_{i}=\mu)\to (\sigma_{i}=\nu)$ are accepted with probability
\begin{align}
P(\mu\to\nu) = \frac{e^{-H_{\nu}}}{{\displaystyle \sum_{\rho=0}^{q-1}}e^{-H_{\rho}}}\, , \qquad \mu,\nu \in \{0,q-1\}\, .
\end{align}
where $H_{\mu}$ is the energy of the configuration with $\sigma_{i}= \mu$ while all other spins keep the old value. Single spin flips are proposed randomly.\\
\indent The system is evolved starting from the completely ordered state
\begin{align}
\sigma_{i} = 0\, ,\quad \forall i\in \{1,\ldots, L^{2}\}.
\end{align}  
Finite volume effects (FVE) are expected to be sizable when using large interaction ranges $R$ since the metastable correlation length $\xi_{s}^{ms}$ is expected to scale as $\xi_{s}^{ms}\sim R (h-h_{sp})^{-1/2}$. Thus the ratio $\xi_{s}^{ms}/L$ should be sufficiently small. After performing preliminary tests, the choice $L=200$ seemed a good compromise and corresponds to $L/\xi_{s}^{ms} > 6$ for all simulated points. \\
\indent Concerning the evaluation of the average nucleation time, its estimate is expected to be rather noisy: nucleation is indeed a stochastic Poisson process, and the probability to form a critical droplet at time $t_{nucl}$ is exponentially distributed via 
\begin{align}
P(t_{nucl}) =  \frac{1}{\tau}\exp{\left(-\frac{t_{nucl}}{\tau}\right)}\, ;
\end{align}  
thus $\text{Var}(t_{nucl})= \tau^{2}$. In order to control the statistical errors, we analyzed $2\cdot 10^{2}$ independent nucleation events for each value of $q,J,h$ and $R$ considered. Moreover, to determine the average nucleation time, we approximate $t_{nucl}$ with the time $t_{50\%}$ that takes to the system to lose $50\%$ of the original magnetization. Our definition of nucleation time contains some degree of arbitrariness, and does not average exactly to $\tau$. Intervention methods~\cite{Monette1992} are usually required in order to pinpoint the instant when the critical droplet appears in the system. However, for spinodal nucleation one expects that the time it takes to the critical droplet to grow and halve the magnetization is small, as compared to $t_{nucl}$. The mistake we commit by using the previous definition is thus negligible.\\
\indent On the contrary, a more precise determination of $t_{nucl}$ is required if one wants to measure observables associated to the critical droplet, such as its profile. The early stages of spinodal nucleation are characterized by an explosive growth that takes place on time scales of order $\mathcal{O}(10)$ Monte Carlo steps per spin ($t_{MC}/\text{spins}$). In this case, instead of using intervention methods, we evaluate $t_{nucl}$ on the basis of a stability analysis of the centers of mass (COM) of the largest clusters (see Sec.~[\ref{IV}]). As reported in Ref.~\cite{Monette1992}, in the case of the Ising model with LRI, the two methods provide similar determinations of the nucleation time.\\

\subsection{CK clusters}
\label{IIIa}
In this subsection we discuss how to properly define the droplets in the Potts model. A geometrical cluster definition, where the droplets are identified as connected trees of parallel interacting spins, is known to be inappropriate \cite{10.1063/1.4959235}. Indeed a proper droplet definition requires that its average length and size diverge respectively as the correlation length $\xi$ and susceptibility $\chi$ of the order parameter, when approaching a second order critical point. In the case of the three-dimensional Ising model with nearest neighbors interactions, it was shown that the geometrical clusters do not fulfill these properties: they were found to diverge at a temperature $T=0.945(5)T_{c}$~\cite{10.1016/0375-9601}. In two dimensions instead, the mean cluster size diverges with a critical exponent $\gamma_{p}= 91/48 \ne \gamma = 1.75$~\cite{10.1103/PhysRevLett.62.1067}.\\
\indent Geometrical clusters are indeed too large, and one has to build a definition of droplets capable of disentangling the spin-spin correlations effects from statistical fluctuations. This problem was solved for the Ising model by Coniglio and Klein in Ref.~\cite{10.1088/0305-4470/13/8/025}, and then extended to the Potts model by Coniglio and Peruggi in Ref.~\cite{10.1088/0305-4470/15/6/028}. By making use of the Kasteleyn-Fortuin theorems~\cite{1969PSJJS..26...11K,10.1016/0031-8914}, which map the partition function of the $q-$state  Potts model into that of a bond-correlated percolation problem (so-called random cluster model), they showed that a correct droplet definition is obtained introducing a fictitious bond between any pair of parallel interacting spins $(\sigma_{i}, \sigma_{j})$ with probability
\begin{align}
\label{bond_probability}
P_{bond} = 1- \exp{\{-2J_{ij}}\}\, .
\end{align}
A \emph{Potts droplet} is then defined as a maximal set of spins connected by bonds. The Coniglio-Klein droplet definition is particularly useful because a certain number of relations between the connectivity properties of the droplets and the distribution of the magnetization can be obtained analytically~\cite{10.1007/978-0-387-30440-3_104}. For instance at zero external field, one has
\begin{align}
\label{connectivity_rho}
&\langle \rho_{\infty} \rangle_{CK} = \langle |M|\rangle\, ,\\[8pt]
\label{connectivity_p}
&\langle p_{ij} \rangle_{CK} = \langle m_{i} m_{j}^{\dag}\rangle\, , 
\end{align}  
where $m_{i} = \sum_{\alpha=0}^{q-1} e^{i2\pi \alpha/q}\delta_{\sigma_{i} \alpha}$, $M=\frac{1}{V}\sum_{i}m_{i}$, $\rho_{\kappa}$ is the number density of droplets of any kind with size $\kappa$, and $\langle p_{ij}\rangle$ is the probability that the sites $i$ and $j$ belong to the same droplet. The average $\langle . \rangle_{CK}$ in the l.h.s. is performed over both spin and bond configurations. Using both  renormalization group techniques~\cite{10.1088/0305-4470/13/8/025,10.1088/0305-4470/15/6/028}, and Monte Carlo simulations~\cite{10.1088/0305-4470/15/12/008}, it has been extensively shown that droplet size and linear dimension diverge respectively with Potts critical exponents $\gamma$ and $\nu$. \\
\indent The previous approach was further extended to the Ising spinodal. In practice, one seeks a generalization of Eq.~[\ref{bond_probability}] such that droplet observables as measured in the metastable state, become critical when approaching the spinodal value of the magnetic field $h=h_{sp}$. Since the spinodal point truly exists only in the limit of infinite range interactions, one has to map the mean-field free energy into that of a percolation problem, enforcing the occurrence of the percolation transition at the spinodal point. We extended this calculation to the $q-$state  Potts model. The details are presented in App.~[\ref{DPM}] . Here, we only quote the main result, namely the probability of activating a bond between sites of parallel spins $\sigma$ is given by
\begin{align}
\label{bond_probability_spinodal}
P_{bond} =
\begin{cases}
1 - \exp{ \{-2qJ_{ij} n_{0}\}} & \sigma \in \left\{1,\ldots, q-1\right\} \\[8pt]
1 - \exp{ \{-2qJ_{ij} n_{\alpha>0}}\} &  \sigma = 0
\end{cases}\, .
\end{align}
For $q=2$ our formula coincides with the one obtained in Ref.~\cite{10.1142/S0217979294000646}. In all our simulations, we use Eq.~[\ref{bond_probability_spinodal}] to create our clusters. Throughout the paper we will denote with $\hat{s}_{i}$ the size of the largest cluster $\mathcal{C}^{max}_{i}$ made of spins $\sigma = i$, and with $\bar{r}_{i}^{2}$ its mean square radius
\begin{align}
\bar{r}_{i}^{2} = \frac{1}{\hat{s}_{i}}{\displaystyle \sum_{j \in \mathcal{C}_{i}^{max}}} \big |\vec{r}_{j}- \vec{r}_{i,com}\big|^{2}~,
\end{align}
where the sum is over the lattice sites belonging to $\mathcal{C}_{i}^{max}$, and $\vec{r}_{i,com}$ is the position of its center of mass. 
\section{Numerical results}
\label{IV}
\begin{figure*}
\includegraphics[scale=0.54]{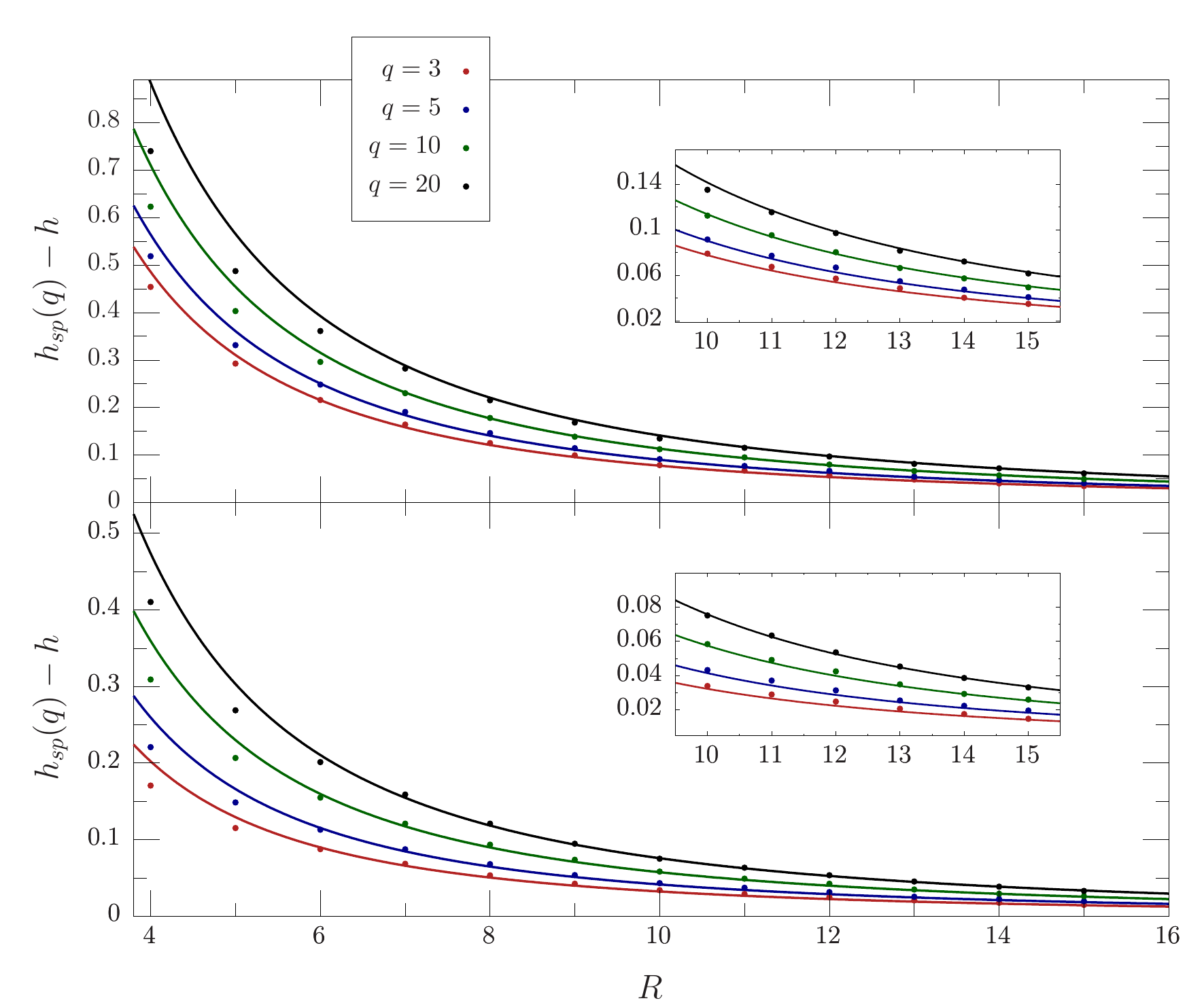}
\caption{\small \it Values of $h_{sp}(q)-h$ corresponding to a constant average nucleation time $\tau=10^3~\tau_{MC}/\textrm{spins}$ (filled markers) along with the theoretical prediction from Eq.~[\ref{eq:critical_droplet_energy}] (lines). The values are reported for $q=3,5,10$ and $20$ as a function of the interaction range. The upper and lower figures correspond respectively to $T=0.5T_c$ and $T=0.8T_c$. The error bars are smaller than the point size.}
\label{fig:Nucleation_times}
\end{figure*}
\begin{figure*}
\hspace{-6mm}
     \subfloat[\label{fig:n_iMTCtime_Q10_R10}]{%
       \includegraphics[scale=0.25]{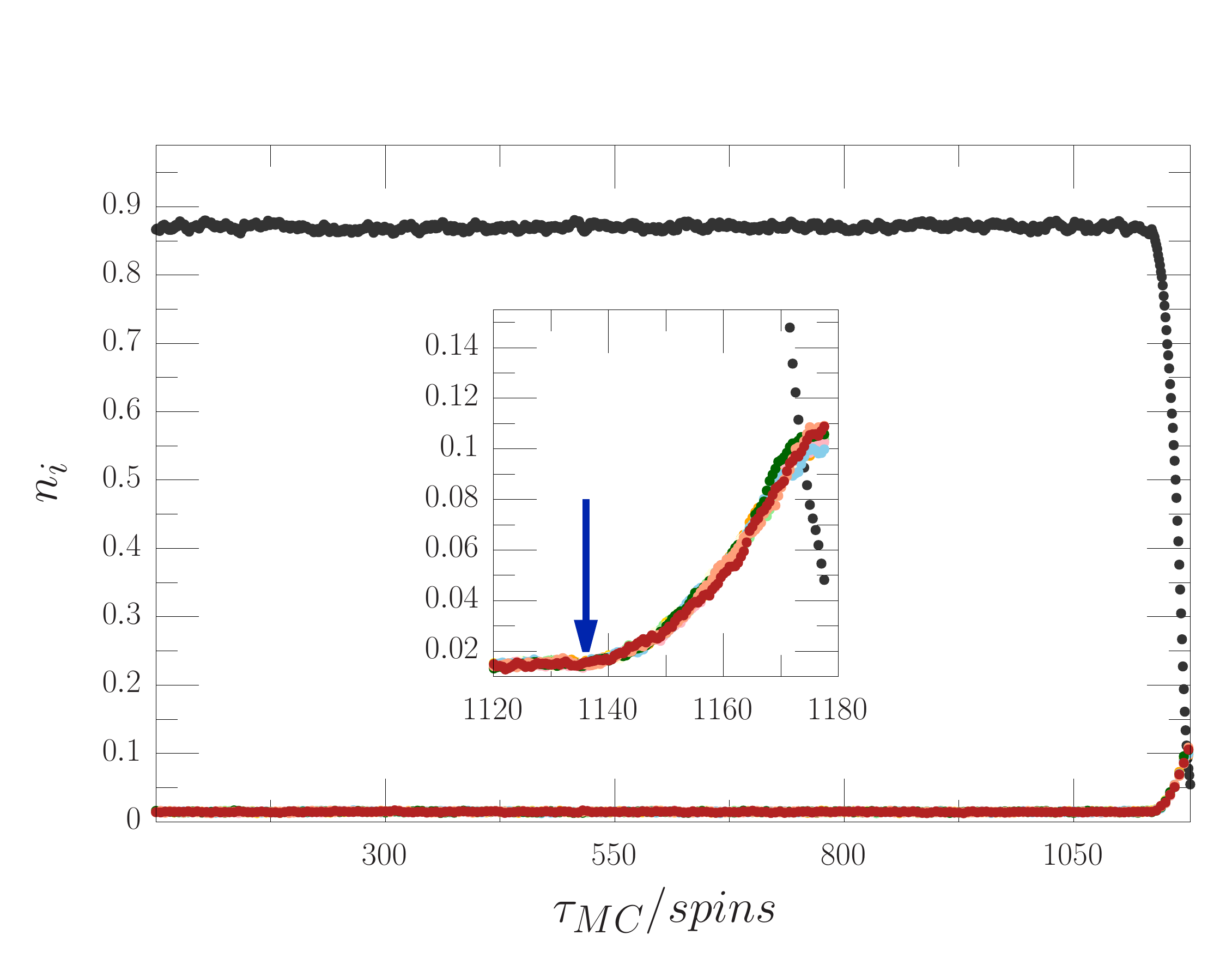}}
     \subfloat[\label{fig:clustermax_Q10_R10}]{%
       \includegraphics[scale=0.25]{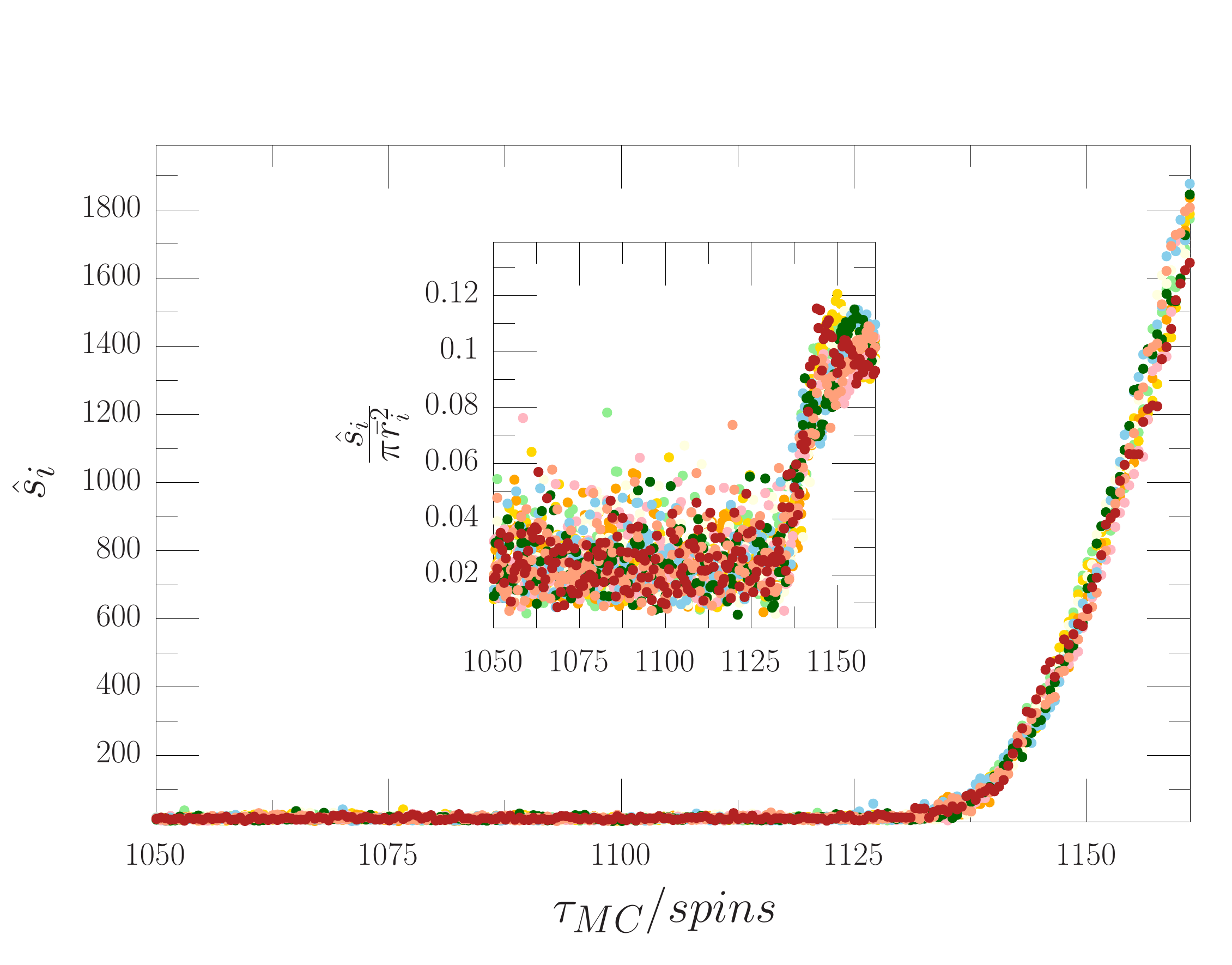}} \\
       \vspace{-9mm}
       \hspace{-6mm}
     \subfloat[\label{fig:COM_Q10_R10}]{%
       \includegraphics[scale=0.25]{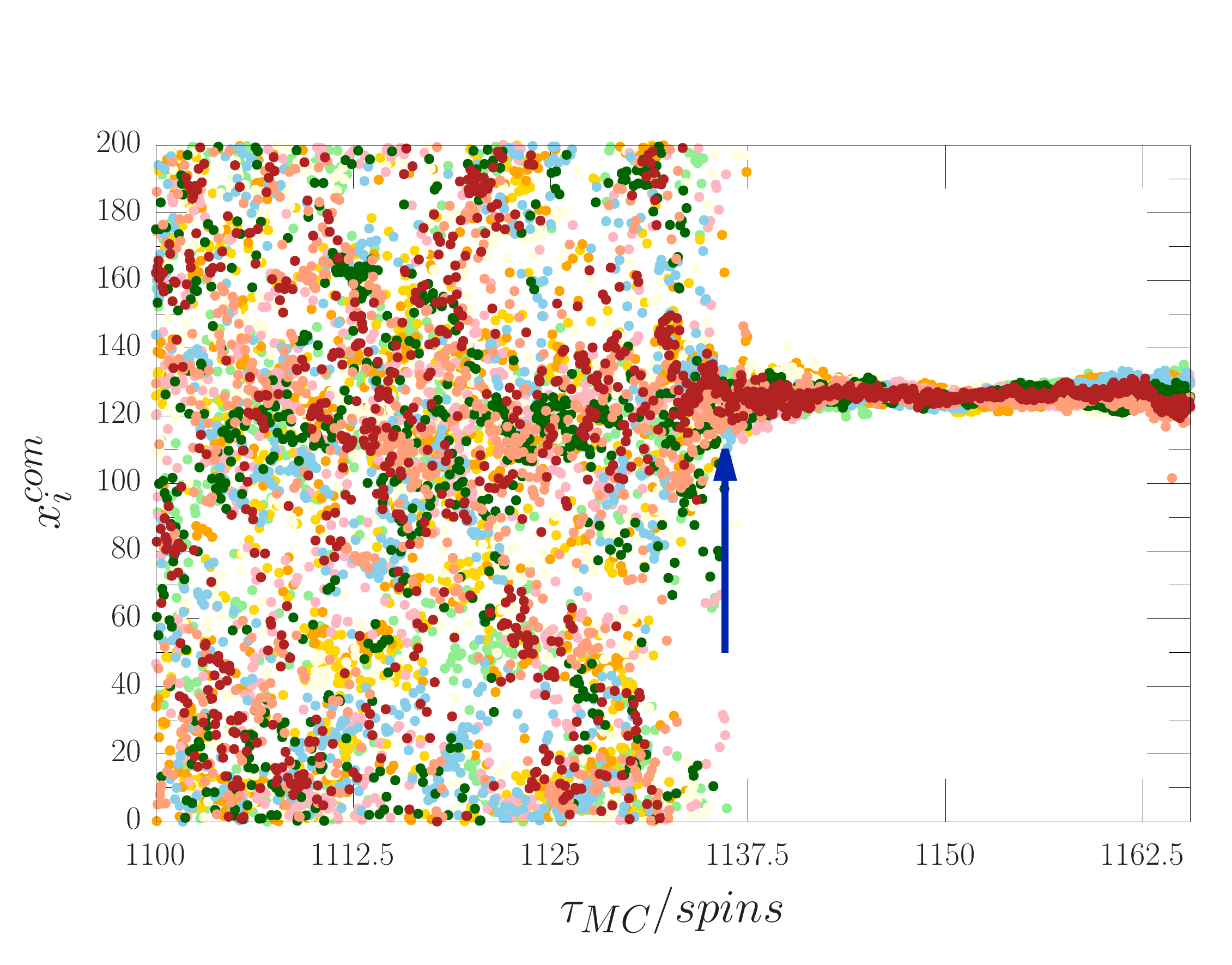}}
     \subfloat[\label{fig:gyration_Q10_R10}]{%
       \includegraphics[scale=0.25]{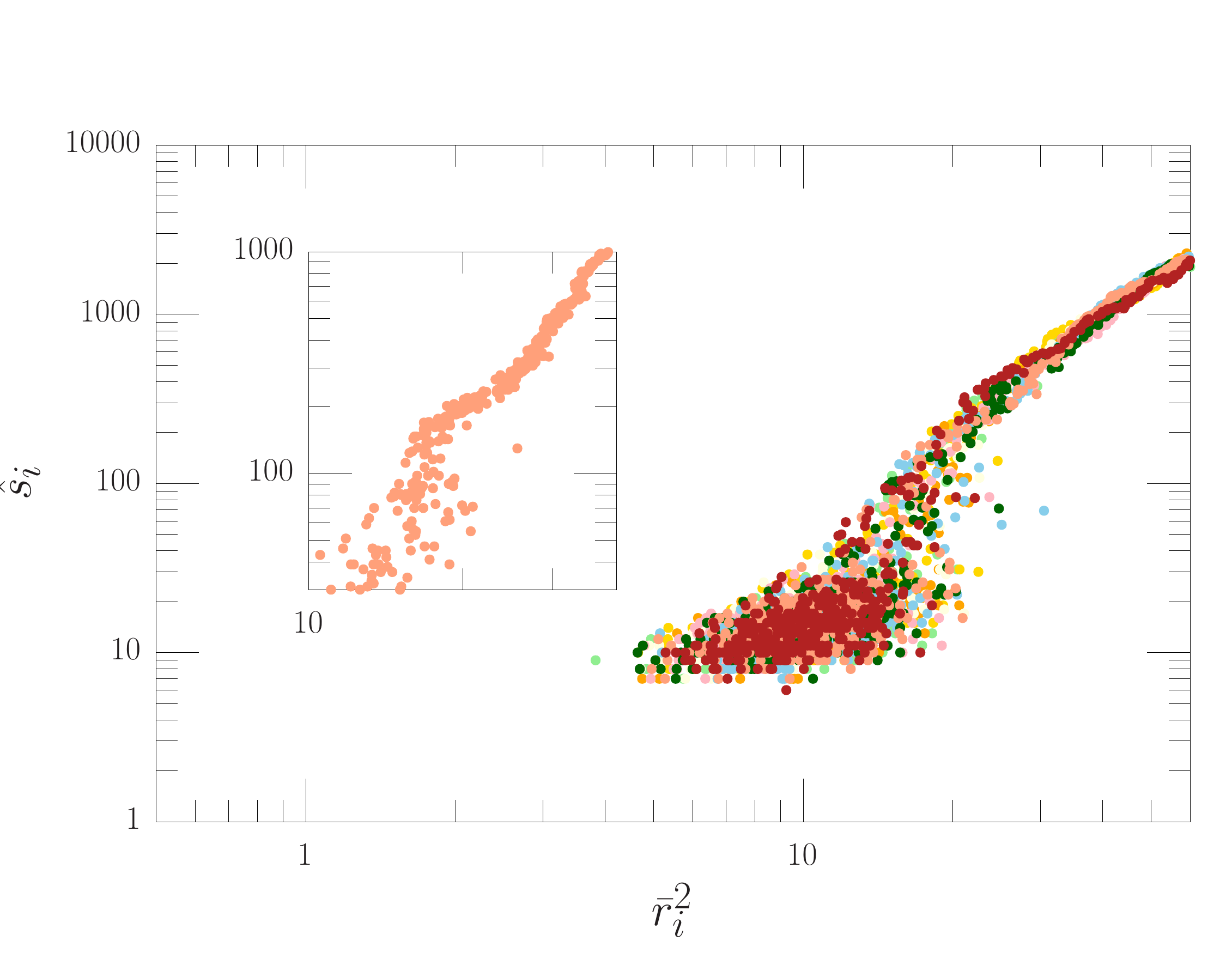}}       
\caption{\it \small Example of a typical spinodal nucleation process for $q=10,~T=0.8T_c,~h= 0.95\,h_{sp}$ and $R=10$. We show the time history of: a) the occupation numbers $n_i$, b) the sizes $\hat{s}_{i}$ of the clusters  $\mathcal{C}^{max}_{i>0}$ and c) the $x$ coordinate of their center of mass. The inset of figure b) shows the densities $\hat{s_i}/(\pi r^{2}_{i})$ for all clusters $\mathcal{C}^{max}_{i>0}$. Finally, in figure d) we show the sizes $\hat{s}_i$ as a function of the mean square radii $r^{2}_{i}$ in logarithmic scale. The blue arrows pinpoint the instant when the centers of mass collapse and nucleation occurs.}
\label{fig:Nucleation_process_Q10_R10}
\end{figure*}
\begin{figure*}%
\hspace{-6mm}
     \subfloat[\label{fig:n_iMTCtime_Q10_R2}]{%
       \includegraphics[scale=0.25]{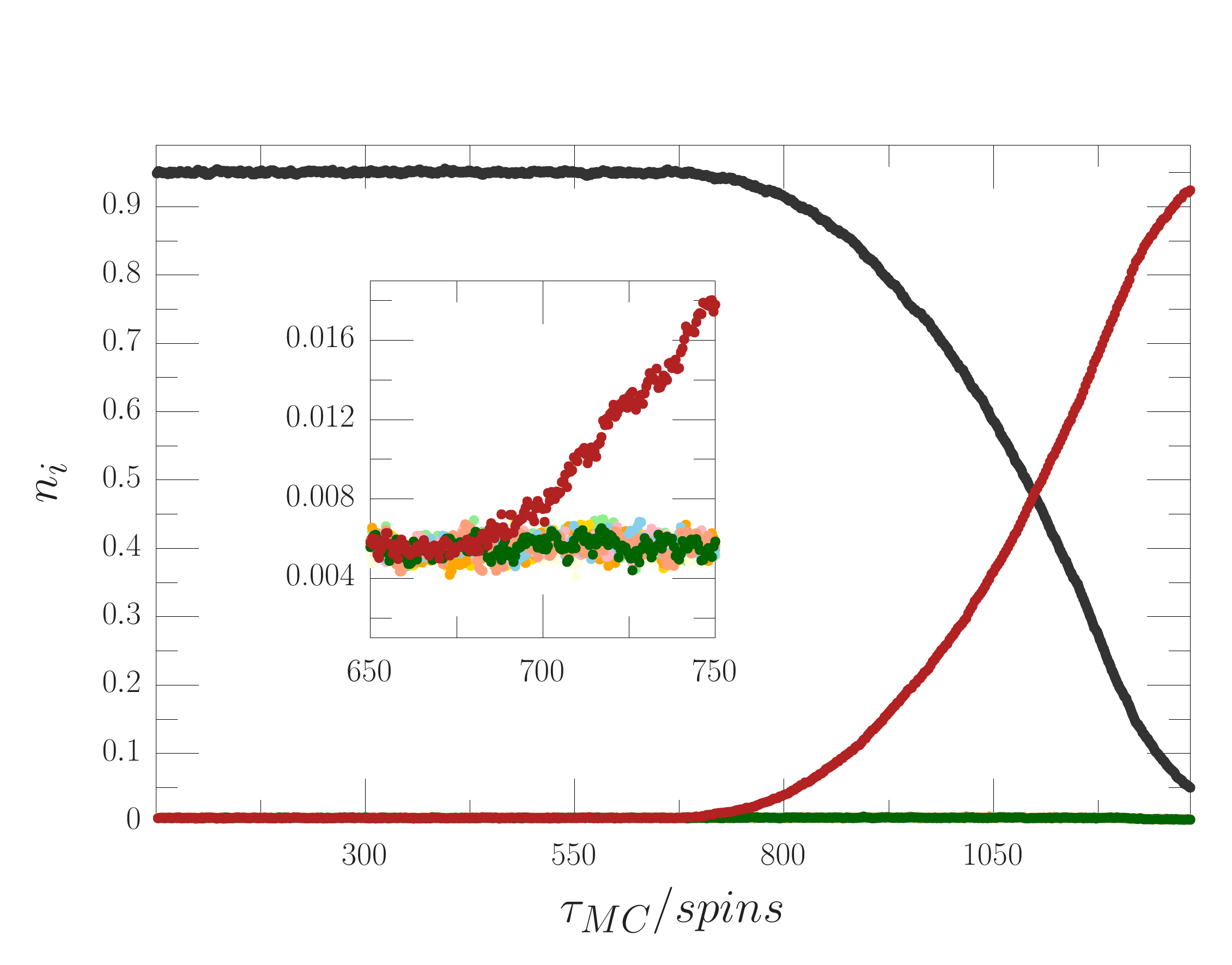}}
     \subfloat[\label{fig:clustermax_Q10_R2}]{%
       \includegraphics[scale=0.25]{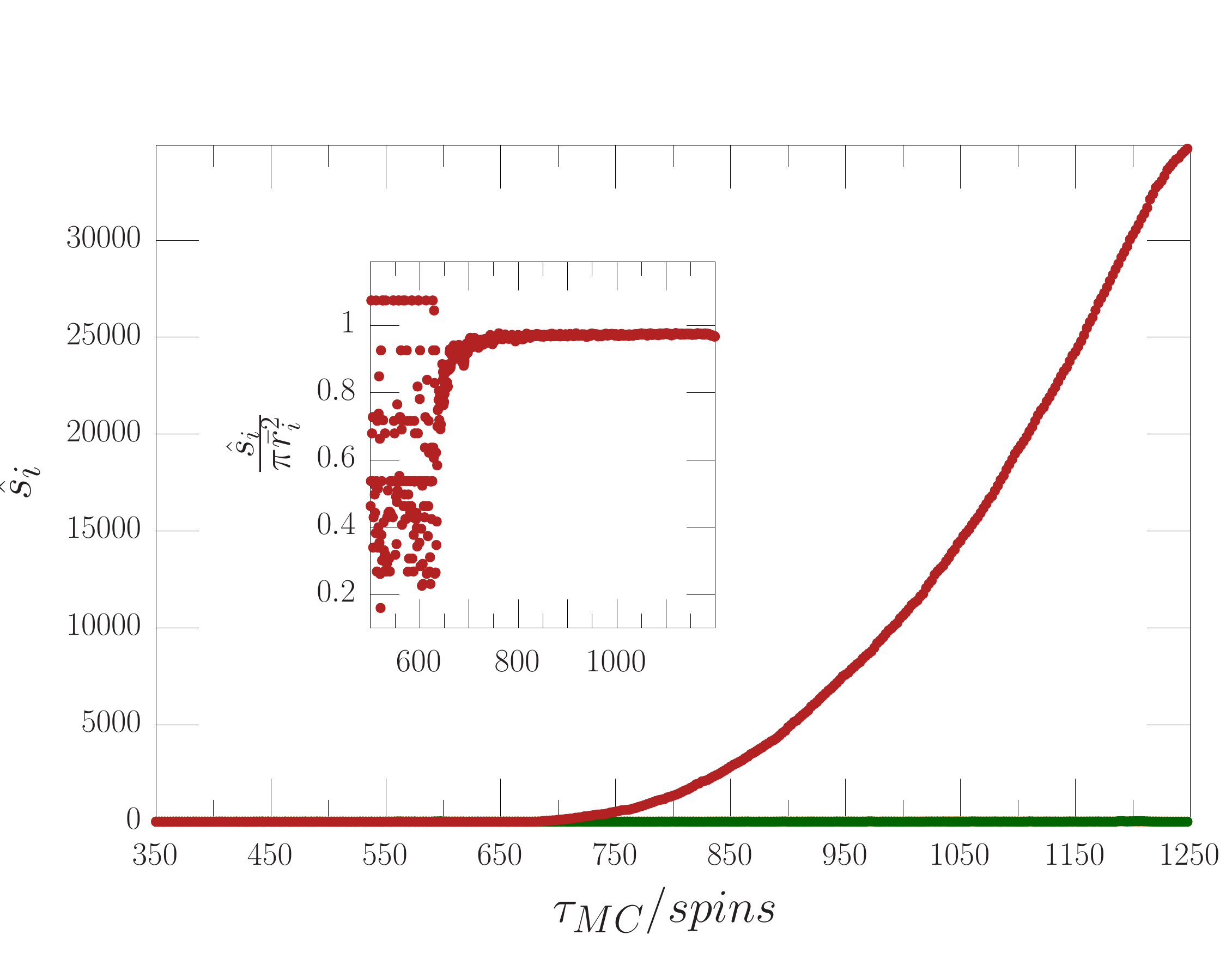}} \\
       \vspace{-9mm}
       \hspace{-6mm}
     \subfloat[\label{fig:COM_Q10_R2}]{%
       \includegraphics[scale=0.25]{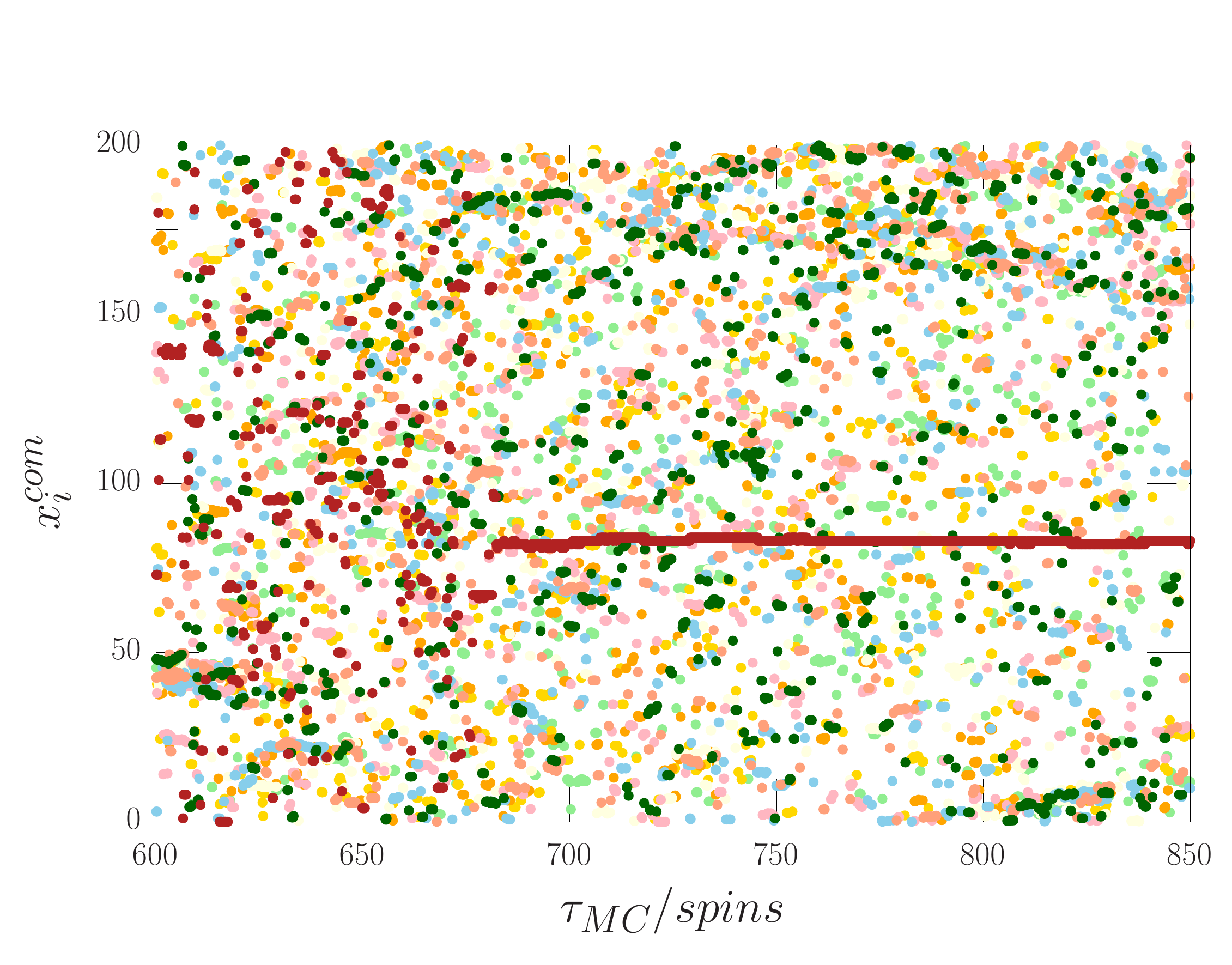}}
     \subfloat[\label{fig:gyration_Q10_R2}]{%
       \includegraphics[scale=0.25]{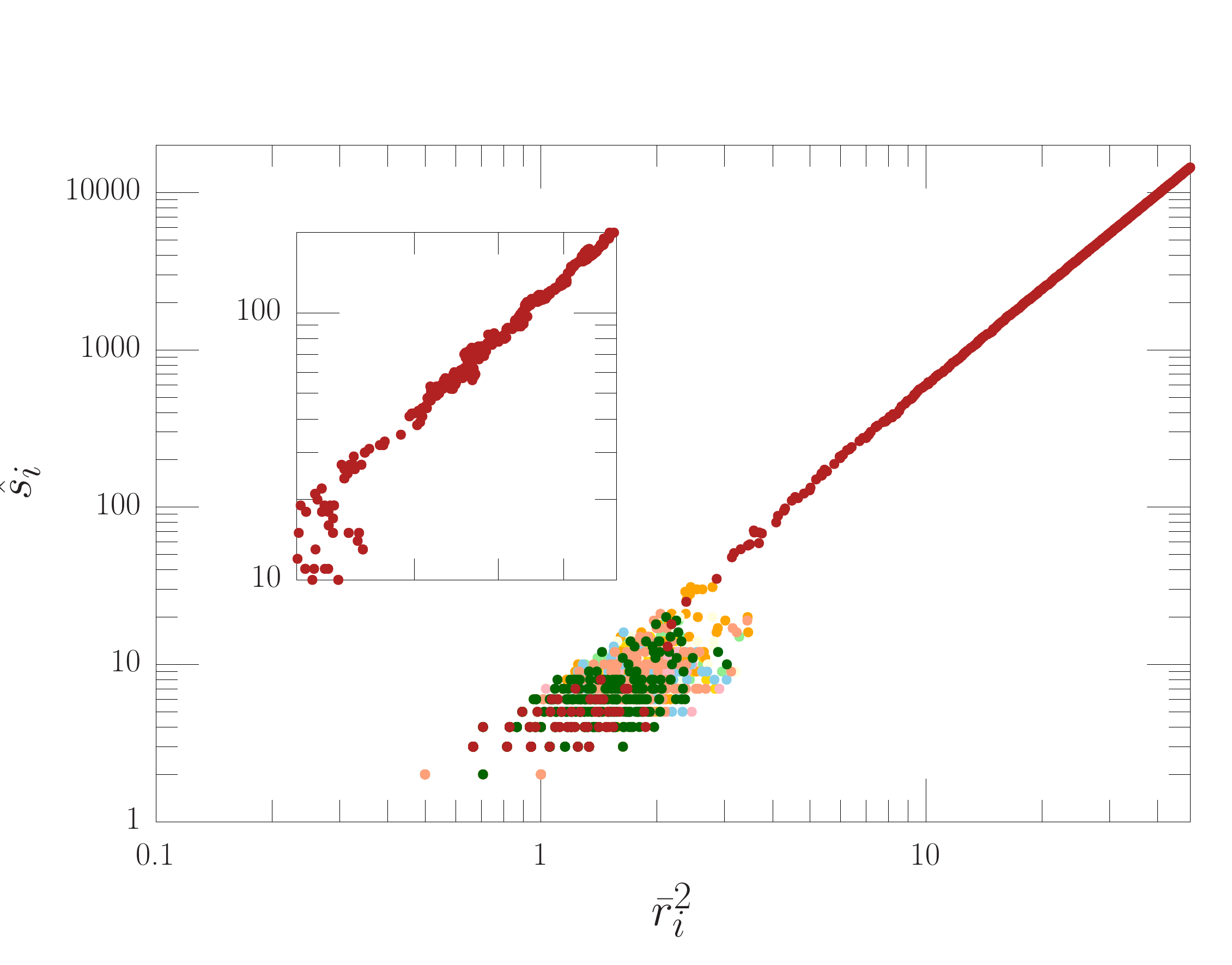}}       
\caption{\it \small Same quantities as in Fig.~[\ref{fig:Nucleation_process_Q10_R10}], for $q=10,~T=0.8T_c,~h= 0.43\,h_{sp}$ and $R=2$.  }
\label{fig:Nucleation_process_Q10_R2}
\end{figure*}
\begin{figure*}%
       \includegraphics[scale=0.50]{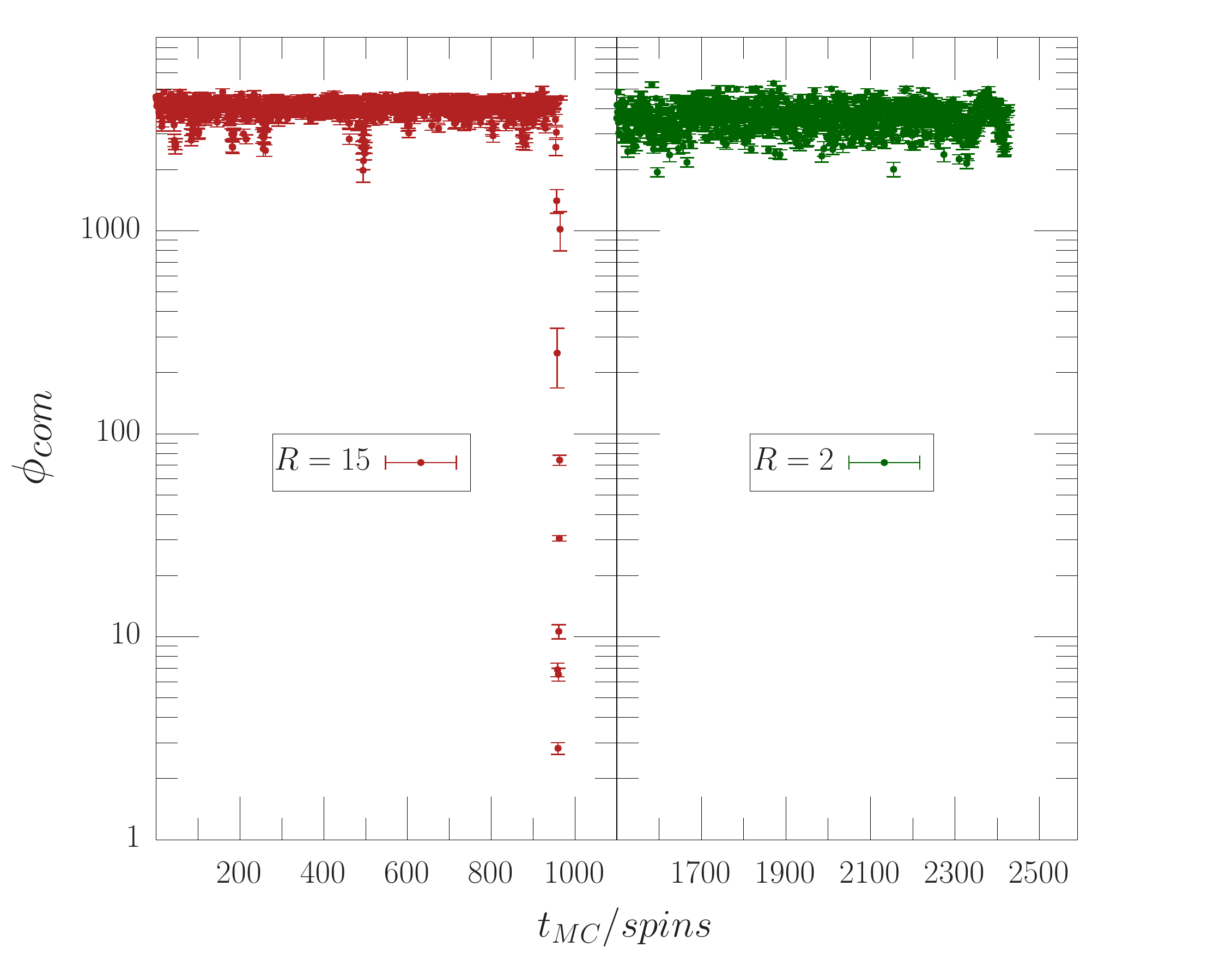}
\caption{ \small \it Time history of $\phi_{com}(t)$ from simulations of the ten-state Potts model at $T=0.8T_{c}$, and for two different values of the interaction range $R$. In both cases $\phi_{com}(t)$ has been measured every $1/20\cdot ~t_{MC}/\textrm{spins}$ and then averaged over a time interval $t_{ave}= 1\cdot ~t_{MC}/\textrm{spins}$.}
\label{fig:phi_com_history}
\end{figure*}
\begin{figure*}%
\hspace{-6mm}
     \subfloat[\label{fig:orderparam0.8Tc}]{%
       \includegraphics[scale=0.400]{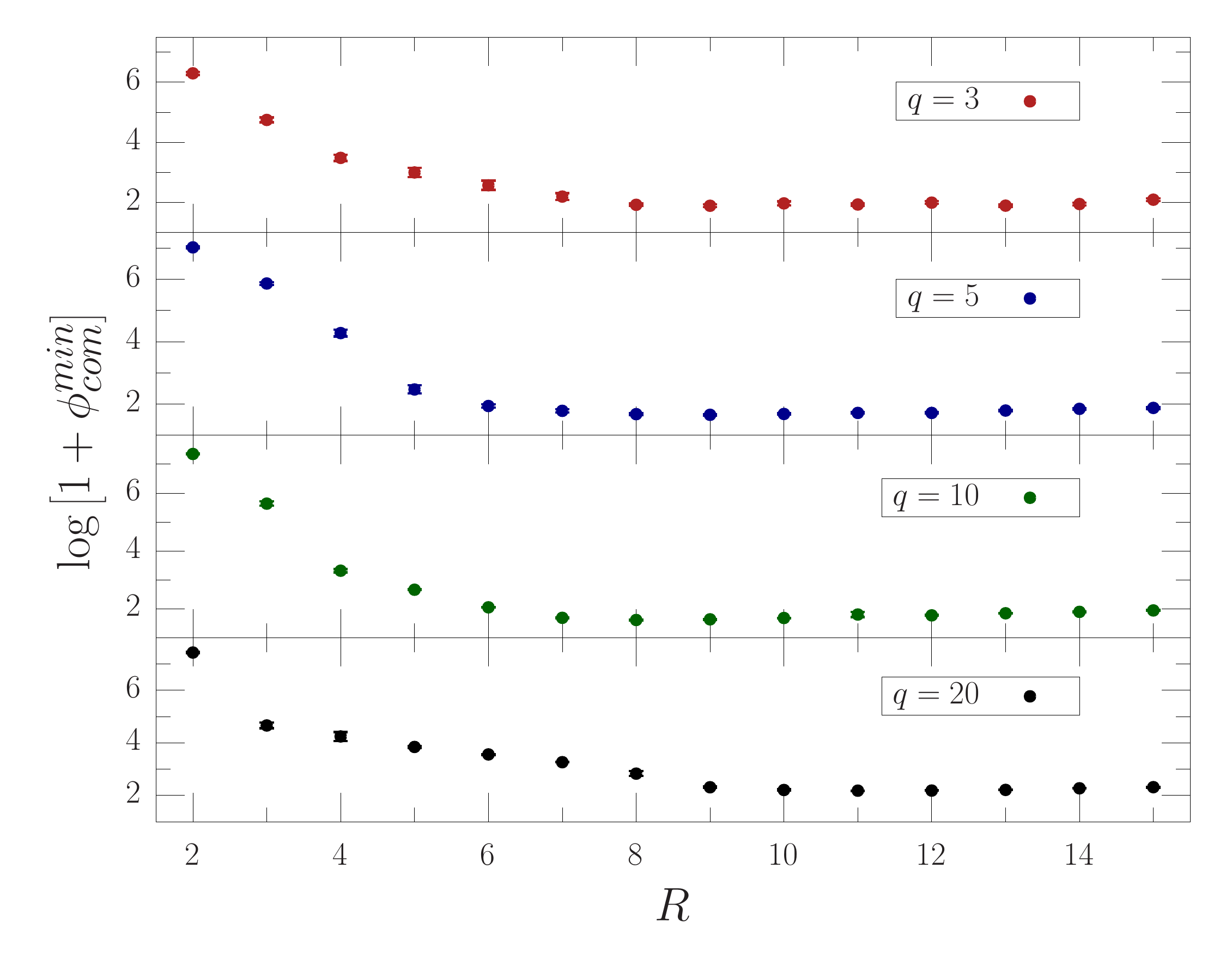}}
     \subfloat[\label{fig:orderparam0.5Tc}]{%
       \includegraphics[scale=0.400]{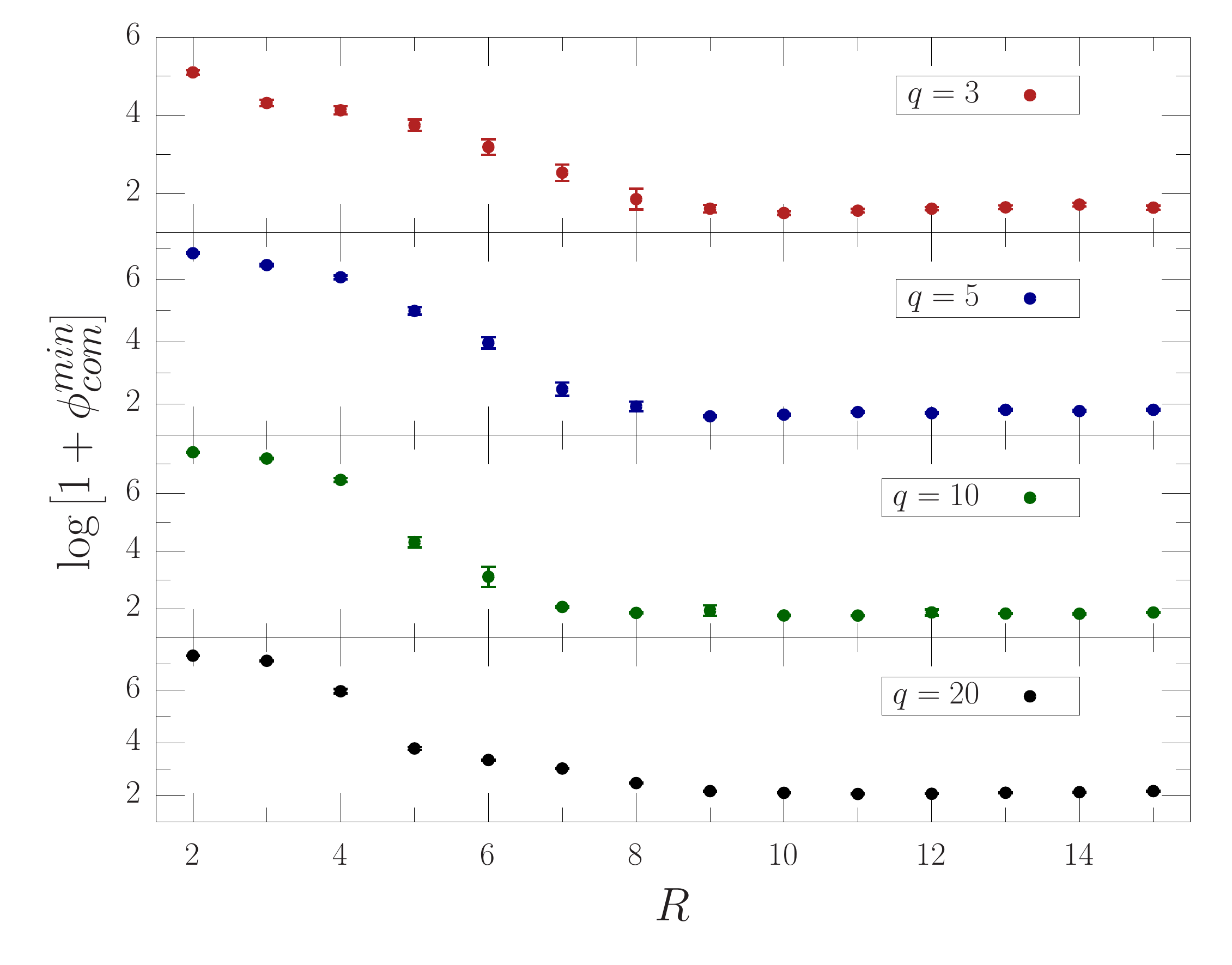}} 
\caption{ \small \it $\log[1+\phi^{min}_{com}]$ as a function of the interaction range $R$ for $T=0.8T_c$ (a) and $T=0.5T_c$ (b), for different $q$. The corresponding values of the magnetic field are the ones collected in Tab.~[\ref{tab:simulation_points}].}
\label{fig:orderparam}
\end{figure*}

We now present the numerical results obtained from Monte Carlo simulations of the two-dimensional $q-$state  Potts model. As already discussed in the previous sections, we make use of the heat bath dynamics to evolve the system from the completely ordered configuration, with all spins pointing in the $0-$th direction. The presence of a magnetic field that disfavours the $0-$th state quickly brings the system in a metastable state. During this phase, we measure the occupation numbers $n_{i}$, the susceptibility $\chi_s$ as well as various quantities related to the CK clusters. In particular we keep track of size, position and shape of the largest CK cluster for all the spins that are not coupled to the magnetic field. After nucleation has taken place, the magnetization of the $0-$th state decreases and the simulation is stopped when $n_{0} \simeq 0.05$.\\
\indent As shown in  Sec.~[\ref{III}], if close to the spinodal point the system shows a mean-field behavior, the structure of the critical droplet as well as the average nucleation time $\tau(q, J, h ,R)$ and the scalar susceptibility $\chi_{s}$ should display peculiar scaling properties (Eqs.~[\ref{eq:chi_xi_mf}],~[\ref{eq:tauexp}],~[\ref{eq:critical_droplet_energy}] and ~[\ref{eq:saddle_point}]). We begin describing the latter two, which can be thought as indirect probes of spinodal nucleation.  In order to facilitate the comparison between nucleation processes occurring at different values of $q,J,h,R$, we only considered quenches of the magnetic field leading to same average nucleation time $\tau$, i.e. constant free energy barrier, for all $q,J,R$. As shown in Sec.~[\ref{III}], under the assumption that the nucleation rate is dominated by the one-instanton contribution, this is equivalent to fix the Ginzburg parameter $G$  for all the runs. In this way, the simulation parameters can be changed and the spinodal point approached without altering the mean-field character of the process. We shall choose a $\tau$ which is large enough to ensure the existence of a well-defined metastable state; at the same time, it must be small enough to correspond to quenches of the magnetic field that are sufficiently close to the spinodal point, at least for the largest $R$ considered in this work. We choose $\tau=10^3\, t_{MC}/\textrm{spins}$. For all values of $q,J$ and $R$, we found the value of the magnetic field $h(q,J,R)$ that leads to $\tau \simeq10^3\, t_{MC}/\textrm{spins}$ employing a bisection algorithm. The resulting relative uncertainty on the value of $\tau$ is of order $\mathcal{O}\left(\frac{1}{\sqrt{n_r}}\right)$, where $n_r=10^{2}$ is the number of simulated nucleation events used for its determination. Since the dependence of $\tau$ on the magnetic field $h$ is essentially exponential (Eqs.~[\ref{eq:tauexp}],~[\ref{eq:critical_droplet_energy}]), the induced uncertainty on $h$ is small, and in our case it is typically of order $\mathcal{O}\left(0.01\%-0.1\%\right)$. The values of $h(q,J,R)$ obtained applying this procedure are collected in Tab.~[\ref{tab:simulation_points}]. \\
\begin{table*}[]
\begin{center}
\renewcommand{\arraystretch}{0.90}
\begin{tabular}{|c c | c c|}
\hline
\multicolumn{4}{|c|}{$q=3$ \rule{0pt}{2ex}} \\
\hline
\multicolumn{2}{|c|}{$T=0.8T_{c}$ \rule{0pt}{2ex}} & \multicolumn{2}{c|}{$T=0.5T_{c}$} \\
\hline
 $R$ \rule{0pt}{2ex} & $h$ & $R$  & $h$ \\
\hline
 $2$ \rule{0pt}{2ex} & $0.038268$ & $2$  & $0.702285$ \\
 $3$  & $0.134851$ & $3$  & $1.169240$ \\
 $4$  & $0.204291$ & $4$  & $1.418290$ \\
 $5$  & $0.259780$ & $5$  & $1.579550$ \\
 $6$  & $0.287406$ & $6$  & $1.656290$ \\
 $7$  & $0.306491$ & $7$  & $1.707800$ \\
 $8$  & $0.321512$ & $8$  & $1.746840$ \\
 $9$  & $0.332419$ & $9$  & $1.772710$ \\
 $10$ & $0.340930$ & $10$ & $1.793200$ \\
 $11$ & $0.345875$ & $11$ & $1.804790$ \\
 $12$ & $0.350050$ & $12$ & $1.815040$ \\
 $13$ & $0.354212$ & $13$ & $1.823540$ \\
 $14$ & $0.357338$ & $14$ & $1.831850$ \\
 $15$ & $0.360058$ & $15$ & $1.837190$ \\
\hline
\end{tabular}
\begin{tabular}{|c c | c c|}
\hline
\multicolumn{4}{|c|}{$q=5$ \rule{0pt}{2ex}} \\
\hline
\multicolumn{2}{|c|}{$T=0.8T_{c}$ \rule{0pt}{2ex}} & \multicolumn{2}{c|}{$T=0.5T_{c}$} \\
\hline
 $R$ \rule{0pt}{2ex} & $h$ & $R$  & $h$ \\
\hline
 $2$ \rule{0pt}{2ex} & $0.199860$ & $2$  & $1.421490$ \\
 $3$  & $0.380757$ & $3$  & $2.100510$ \\
 $4$  & $0.477496$ & $4$  & $2.387660$ \\
 $5$  & $0.549732$ & $5$  & $2.574710$ \\
 $6$  & $0.585312$ & $6$  & $2.657630$ \\
 $7$  & $0.610906$ & $7$  & $2.715320$ \\
 $8$  & $0.630217$ & $8$  & $2.759720$ \\
 $9$  & $0.644511$ & $9$  & $2.791430$ \\
 $10$ & $0.654977$ & $10$ & $2.814690$ \\
 $11$ & $0.661092$ & $11$ & $2.828930$ \\
 $12$ & $0.666821$ & $12$ & $2.839320$ \\
 $13$ & $0.672738$ & $13$ & $2.851310$ \\
 $14$ & $0.675869$ & $14$ & $2.858670$ \\
 $15$ & $0.678595$ & $15$ & $2.865320$ \\
\hline
\end{tabular}
\begin{tabular}{|c c | c c|}
\hline
\multicolumn{4}{|c|}{$q=10$ \rule{0pt}{2ex}} \\
\hline
\multicolumn{2}{|c|}{$T=0.8T_{c}$ \rule{0pt}{2ex} } & \multicolumn{2}{c|}{$T=0.5T_{c}$} \\
\hline
 $R$ \rule{0pt}{2ex} & $h$ & $R$  & $h$ \\
\hline
 $2$ \rule{0pt}{2ex} & $0.530407$ & $2$  & $2.582420$ \\
 $3$  & $0.785666$ & $3$  & $3.440290$ \\
 $4$  & $0.921141$ & $4$  & $3.771600$ \\
 $5$  & $1.023590$ & $5$  & $3.991320$ \\
 $6$  & $1.075340$ & $6$  & $4.098250$ \\
 $7$  & $1.109390$ & $7$  & $4.164140$ \\
 $8$  & $1.136840$ & $8$  & $4.216240$ \\
 $9$  & $1.156620$ & $9$  & $4.255780$ \\
 $10$ & $1.171870$ & $10$ & $4.282110$ \\
 $11$ & $1.181120$ & $11$ & $4.299280$ \\
 $12$ & $1.187740$ & $12$ & $4.314230$ \\
 $13$ & $1.195320$ & $13$ & $4.328090$ \\
 $14$ & $1.200920$ & $14$ & $4.337150$ \\
 $15$ & $1.204210$ & $15$ & $4.344970$ \\
\hline
\end{tabular}
\begin{tabular}{|c c | c c|}
\hline
\multicolumn{4}{|c|}{$q=20$ \rule{0pt}{2ex}}  \\
\hline
\multicolumn{2}{|c|}{$T=0.8T_{c}$ \rule{0pt}{2ex} }  & \multicolumn{2}{c|}{$T=0.5T_{c}$} \\
\hline
 $R$  \rule{0pt}{2ex} & $h$ & $R$  & $h$ \\
\hline
 $2$  \rule{0pt}{2ex} & $0.934506$ & $2$  & $3.921590$ \\
 $3$  & $1.266780$ & $3$  & $4.858650$ \\
 $4$  & $1.455080$ & $4$  & $5.259390$ \\
 $5$  & $1.596480$ & $5$  & $5.512000$ \\
 $6$  & $1.664050$ & $6$  & $5.638040$ \\
 $7$  & $1.706540$ & $7$  & $5.716890$ \\
 $8$  & $1.744560$ & $8$  & $5.783730$ \\
 $9$  & $1.770640$ & $9$  & $5.830700$ \\
 $10$ & $1.790280$ & $10$ & $5.864120$ \\
 $11$ & $1.801930$ & $11$ & $5.883840$ \\
 $12$ & $1.811790$ & $12$ & $5.902130$ \\
 $13$ & $1.820050$ & $13$ & $5.917510$ \\
 $14$ & $1.826740$ & $14$ & $5.927060$ \\
 $15$ & $1.832230$ & $15$ & $5.937520$ \\
\hline
\end{tabular}
\end{center}
\caption{\it \small The complete list of simulated points on the $L\times L$ lattice with $L=200$.}
\label{tab:simulation_points}
\end{table*}
\indent Neglecting the (slowly varying) contributions $f_{1-loop}$ and $f_{dyn}$ in Eq.~[\ref{eq:tauexp}] one has that the average nucleation time $\tau$ depends on the simulation parameters via
\begin{align}
\tau \propto \frac{1}{V}\exp{\big\{ E_{drop}(q,J,h,R) \big\}}~;
\end{align}
hence for constant $\tau$ (and volume $V$) one has $E_{drop}(q,J,h,R)=C$, for some constant $C$. This implies, making use of  Eq.~[\ref{eq:critical_droplet_energy}] that the distance between the magnetic field $h(q,J,R)$ giving $\tau= 10^{3}\,t_{MC}/\textrm{spins}$ and the corresponding spinodal point $h_{sp}(q,J)$,  has the following asymptotic expression
\begin{align}
h_{sp}(J,q)-h(J,q,R) = \frac{qC}{(q-1)\theta K 2\pi} \frac{a(q,J)}{R^{2}}\, .   
\label{eq:prediction_h}
\end{align}
In Fig.~[\ref{fig:Nucleation_times}] we compare the numerical values of $h_{sp}(q,J)-h(q,J,R)$ leading to $\tau = 10^{3}\,t_{MC}/\textrm{spins}$, against the theoretical prediction of Eq.~[\ref{eq:prediction_h}], setting $C=16$. As the figure shows, for both temperatures the simulation results quickly converge to the predicted values. In particular, the agreement with the theoretical curve starts around $R=6-8$ with a mild dependence on $q$ and $J$. We stress that the family of curves shown in Fig.~[\ref{fig:Nucleation_times}] only depends on the single parameter $C$. We also tried to fit our numerical data employing an Ansatz containing the logarithmic corrections to $\tau$ (Eq.~[\ref{eq:tauexp}]), but we could not detect the presence of such terms. \\
\begin{table*}[]
\begin{center}
\renewcommand{\arraystretch}{0.90}
\begin{tabular}{|c c |c c c | c c | c c c|}
\hline
\multicolumn{5}{|c|}{$T=0.5T_{c}$ \rule{0pt}{2ex}} & \multicolumn{5}{c|}{$T=0.8T_{c}$} \\
\hline
$q$ \rule{0pt}{2ex} & $R$ & $\emph{MC} $& $Eq.~[\ref{eq:chi_xi_mf}]$ & $\emph{m.f.}$ & $q$ & $R$ &  $\emph{MC}$ & $Eq.~[\ref{eq:chi_xi_mf}]$ & $\emph{m.f.}$ \\
\hline
$20$ \rule{0pt}{2ex} & $15$ & $0.2808(31) $ & $0.264 $ &$ 0.205$ & $20$ & $15$ & $0.755(11) $ &$0.622 $ &$0.516 $  \\
$10$ & $15$ & $0.4411(55) $ & $0.400 $ &$0.318 $ & $10$ & $15$ & $1.261(20) $ &$0.979 $ &$0.825 $ \\
$5$ & $15$ & $0.791(11) $ & $0.681 $ & $0.553 $ & $5$ & $15$ & $2.575(55) $ &$1.829 $ & $1.563 $ \\
$3$ & $15$ & $1.429(18) $ & $1.203 $ & $0.991 $ & $3$ & $15$ & $5.38(12) $ &$3.616 $ & $3.123 $ \\
\hline
\end{tabular}
\end{center}
\caption{\it \small Metastable susceptibility $\chi_{s}$ obtained from Monte Carlo simulations (MC) along with the prediction from the the mean-field asymptotic formula of Eq.~[\ref{eq:chi_xi_mf}], and the exact mean-field result (m.f.). In the table we only show the results obtained at the largest interaction range $R=15$. The values of the magnetic field are set to the ones of Table \ref{tab:simulation_points}.} 
\label{tab:susc_comparison}
\end{table*}
\begin{figure*}
\centering
     \subfloat[\label{fig:susc0.8Tc}]{%
       \includegraphics[scale=0.38]{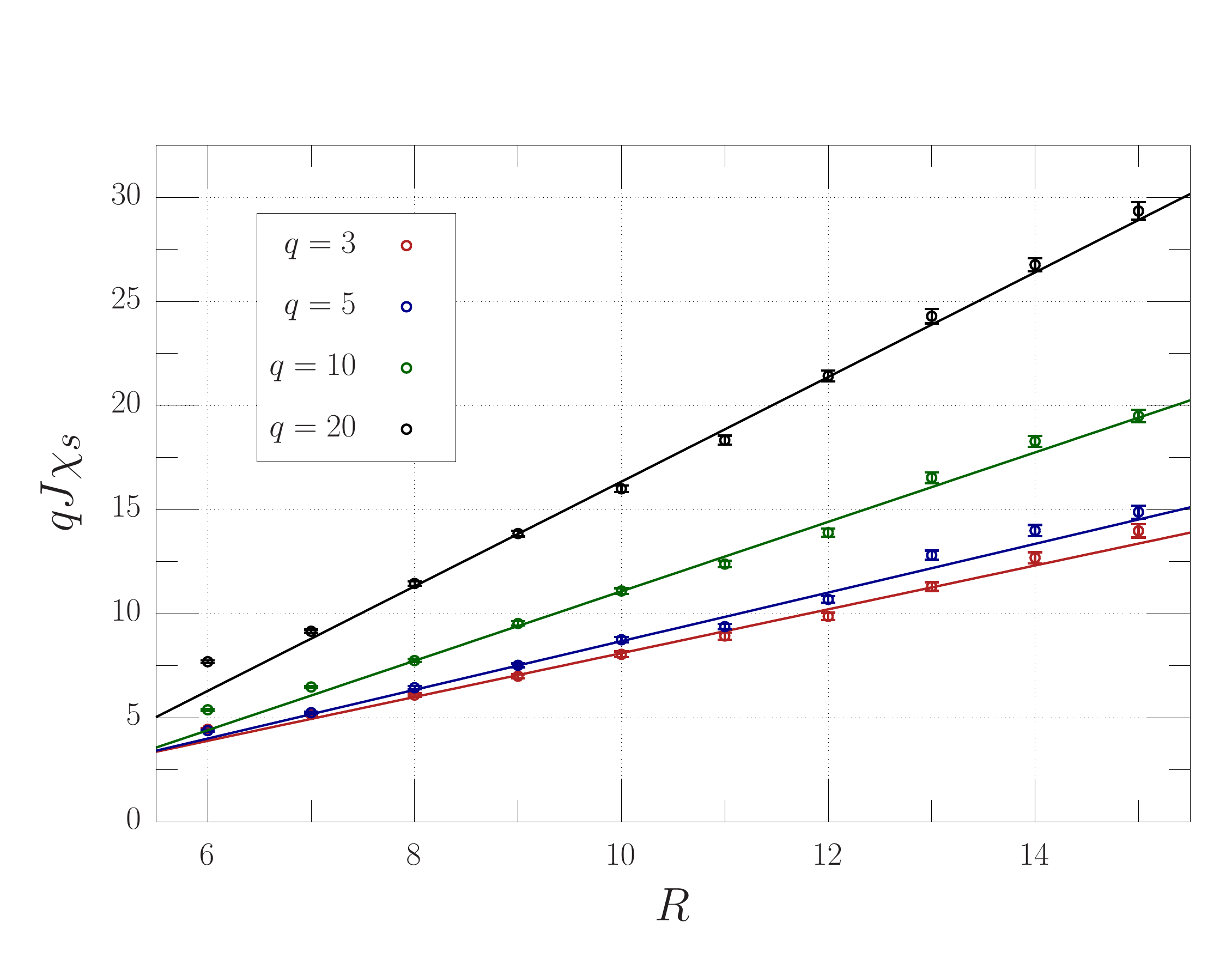}}
     \subfloat[\label{fig:susc0.5Tc}]{%
       \includegraphics[scale=0.38]{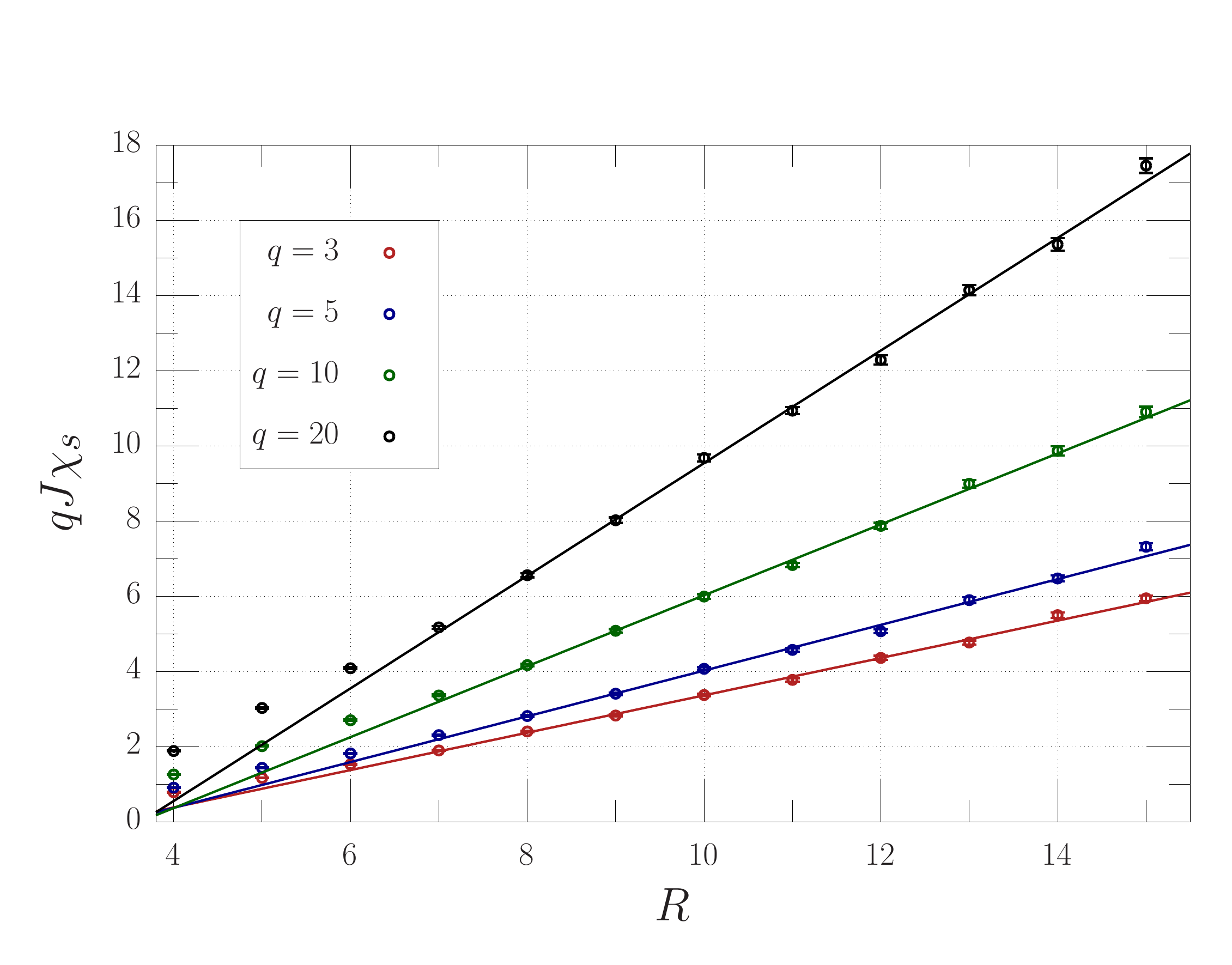}}
\caption{\small \it Metastable susceptibility $\chi_{s}$ as a function of the interaction range $R$ at (a) $T=0.8T_c$ and (b) $T=0.5T_c$. The solid lines are the results of a linear fit to the data in the interval $R\in [8,15]$.}
\label{fig:Susceptibility}
\end{figure*}
\indent As for the magnetic susceptibility, its value has been estimated during the metastable phase through
\begin{align}
\chi_{s}(q,J,h,R) =  V\left[\langle s^{2} \rangle_{ms} - \langle s\rangle ^{2}_{ms}\right]
\end{align}
where
\begin{align}
s = \frac{1}{V}\sum_{i=1}^{V} \frac{q}{q-1}(\delta_{0 \sigma_{i}}-\frac{1}{q})~.
\end{align}
For each simulation point we accumulated a total of $\mathcal{O}(10^{5})$ metastable configurations, and no measurement has  been taken in a given event, if nucleation occurred in less than $10^{3} \,t_{MC}/\textrm{spins}$. Mean values and standard errors were obtained by means of a standard Jackknife analysis. As the mean-field analysis of the previous section shows, one expects that close to the spinodal $\chi_{s} \sim (h-h_{sp})^{-1/2}$ for all $q$. However, if one moves towards the spinodal point at fixed $R$, eventually the system enters the regime where $G\lesssim 1$ and the mean-field approximation is no longer valid. Mean-field critical behavior is instead expected if one approaches the spinodal at constant Ginzburg parameter $G\gg 1$ (see Sec.~[\ref{sub:validity}]), as in the case of our numerical simulations. In turn, working at (almost) constant $G$ implies $b(q,h) \propto R^{-2}$, therefore $\chi_{s} \propto R$ ($R^{2d/6-d}$ in $d$ dimensions), as pointed out in Ref.~\cite{10.1063/1.4959235}. In Fig.~[\ref{fig:Susceptibility}] we show as a function of $R>3$, the metastable susceptibility $\chi_{s}$ for all $q=3,5,10,20$ and for $T=0.5T_c$ and $T=0.8T_{c}$. In the figure, $\chi_{s}$ has been rescaled by a factor $qJ$ for visualization purposes. The onset of the linear regime is visible at interaction ranges $R\sim 8$; small deviations from the fitting curve can be presumably attributed to statistical fluctuations. We reported in Tab.~[\ref{tab:susc_comparison}] the values of the susceptibility $\chi_{s}(q,J,h,R)$ for the largest interaction range we simulated ($R=15$), alongside the exact mean-field prediction, and its diverging part Eq.~[\ref{eq:xi_mf}]. It must be noted that the linear behavior of the susceptibility at fixed $\tau$ depends on whether the lines of constant nucleation time $\tau$, and the lines of constant Ginzubrg parameter $G$, coincide. 
Although this is true in the region of the parameter space we explored, by marching towards the spinodal point at fixed $G$, the logarithmic corrections to the nucleation times will no longer be small, and the two lines will depart from each other. In particular, since the logarithmic corrections increase the energy barrier that the system must overcome, the lines of constant $\tau$ will be closer to $h_{sp}(q)$, resulting in larger values of $\chi_{s}$ w.r.t. the observed linear behavior. \\
\begin{figure*}
\includegraphics[scale=0.45]{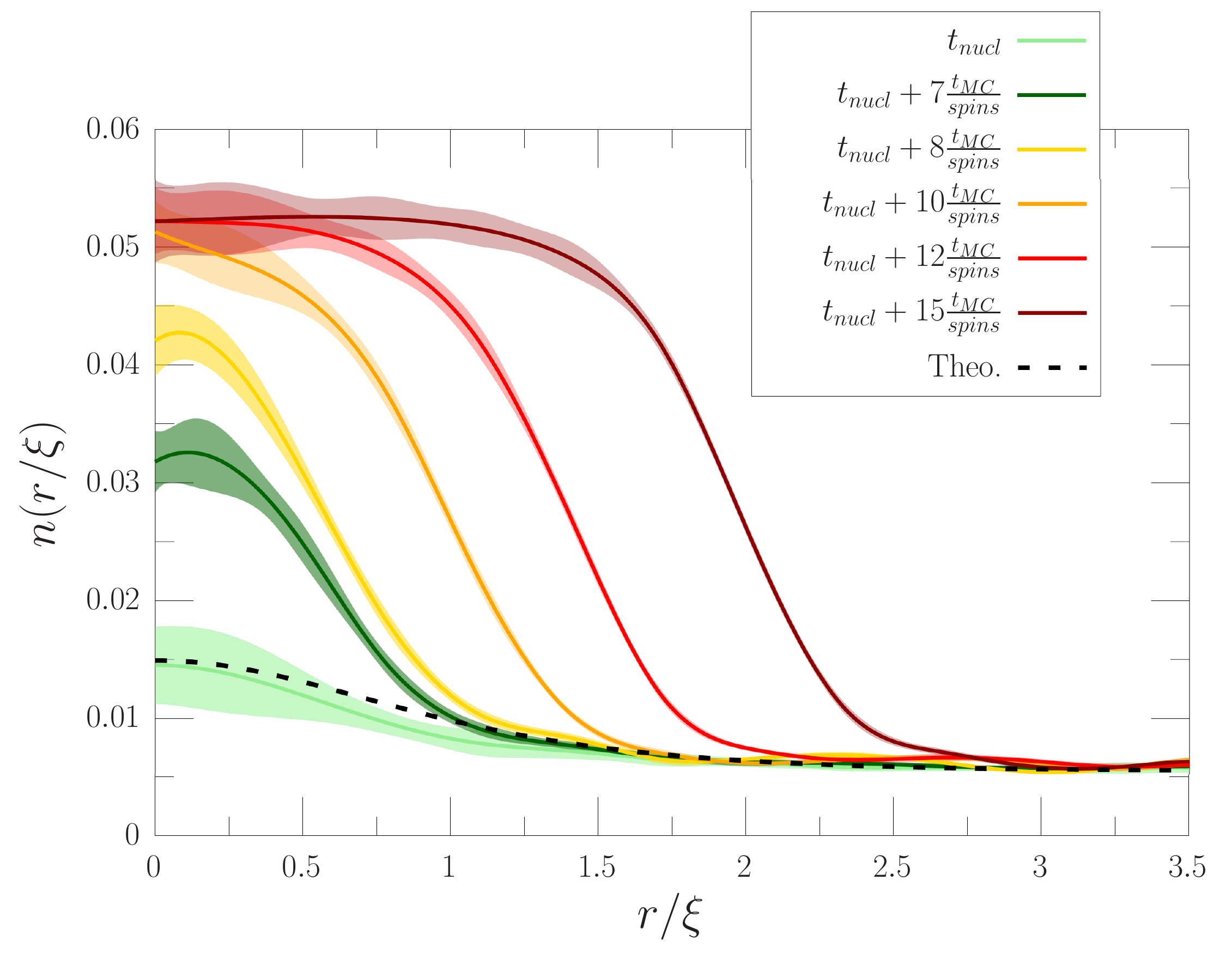}
\caption{\small \it Local magnetization $n(r/\xi)$ as a function of the distance from $(\bar{x}_{com}, \bar{y}_{com})$, and for different Monte Carlo times. Distances are expressed in units of the mean-field correlation length $\xi$ defined in Eq.~[\ref{eq:saddle_point}]. The data correspond to simulations of the twenty-state Potts model with $R=15$, and temperature $T=0.8T_{c}$. The semi-transparent bands stem from the statistical uncertainty on the determination of the profile, and for $t=t_{nucl}$ they also take into account the uncertainty in the exact determination of the nucleation time $t_{nucl}$. Finally, the dashed black line is the theoretical mean-field prediction obtained solving numerically Eq.~[\ref{eq:E-L_equation}].}
\label{fig:critical_droplet_profile}
\end{figure*}
\indent We now turn into the discussion of the structure of the nucleating droplet. As we discussed, the presence of a spinodal channel in the mean-field free energy for deep quenches of the magnetic field should allow the system to leave the metastable phase through the formation of a nucleating droplet consisting of $q-1$ spins. The new phase reached by the system can in turn be unstable or metastable, and a two-steps decay can be observed. In the case of shallow quenches close to the coexistence curve, the spinodal channel is disfavored and nucleation can only occur through the formation of a droplet of a single phase. To highlight the difference between the two regimes, we show in the panels of Fig.~[\ref{fig:Nucleation_process_Q10_R10}] and Fig.~[\ref{fig:Nucleation_process_Q10_R2}], two typical nucleation processes in the ten state Potts model at $T=0.8T_{c}$. The two examples correspond to two extreme cases $R=10$ and $R=2$. In the panels, we show several observables evaluated during Monte Carlo evolution: the occupation numbers of all spins (a), the sizes $\hat s_i$ of the largest clusters $\mathcal{C}_{i>0}^{max}$ (b), the lattice coordinates of their center of mass (c), and the $\hat s_i$ as a function of the mean square radius $\bar{r}_{i}^{2}$ (d). In the short range case nucleation proceeds ``classically", i.e. the nucleating droplet corresponds to a large localized fluctuation of a single spin, that expands by adding spins at its surface since $\bar{r}^{2} \propto \hat s$. The magnetization and the cluster sizes of the other $q-2$ spins remain small, while their centers of mass keep fluctuating randomly. In the long range case $R=10$, the emergence of spinodal-assisted nucleation is instead rather clear. The centers of mass of the clusters $\mathcal{C}_{i>0}^{max}$ simultaneously stabilize in a narrow lattice region, indicating that the system is leaving the metastable state along the symmetric channel. The clusters grow together as it is also evident from the evolution of their sizes and from the simultaneous bump in all the occupation numbers. The bump is slightly delayed with respect to the instant where the centers of mass stabilize because it takes some time for a small amplitude delocalized fluctuation to produce a visible effect in the total magnetization. The insets in Figs.~[\ref{fig:clustermax_Q10_R10}] and [\ref{fig:clustermax_Q10_R2}] also show the mean densities $\hat{s}_{i}/(\pi \bar{r}_{i}^{2})$, assuming that each cluster is a nearly spherical object with radius $\sqrt{\bar{r}_{i}^2}$. In the short range case, as expected, $\hat{s}_{i}/(\pi \bar{r}_{i}^{2}) = \delta_{i i_{nucl}}$ where $\mathcal{C}_{i_{nucl}}^{max}$ is the unique nucleating cluster, while in the long range case after an initial transient that corresponds to the compactification of the clusters, we observe $\hat{s}_{i}/(\pi \bar{r}_{i}^{2}) = 1/(q-1)$ for all $i\in \{1,\ldots, q-1\}$,  meaning that the clusters are all compenetrating. The evolution of $\bar{r}_{i}^2$ and $\hat{s}_{i}$  as a function of time shows that in the long range case the early stage of nucleation are characterized by a filling in of the clusters, since they grow at almost constant $\bar{r}_{i}^2$. With the heat bath algorithm this regime lasts for order $\mathcal{O}(10\,t_{MC}/\textrm{spins})$, and can be only identified by measuring cluster observables at fractional $t_{MC}/\textrm{spins}$. At larger times the growth is similar to the one observed for $R=2$, where $\bar{r}^2 \propto \hat{s}$.\\ 
\indent We also evaluated for $R=15$, $q=20$ and $T=0.8T_{c}$, the profile of the nucleating droplet from time $t=t_{nucl}$ up to $t= t_{nucl} + 15~t_{MC}/\textrm{spins}$ which roughly corresponds to the onset of compact growth. This is shown in Fig.~[\ref{fig:critical_droplet_profile}], where the average local magnetization $n(r) = \frac{1}{q-1}\sum_{i=1}^{q-1}n_{i}(r)$ is plotted as a function of the distance $r$ from the barycenter $(\bar{x}_{com}, \bar{y}_{com})$ of the centers of mass $(x_{i,com}, y_{i,com})$ of the clusters $\mathcal{C}_{i>0}^{max}$. Our estimate of the critical droplet profile, which takes into account the statistical errors associated to $n(r)$ as well as the systematics related to the exact determination of the nucleation time $t_{nucl}$, is compared with the numerical solution of Eq.~[\ref{eq:E-L_equation}], and we find substantial agreement between our determination and the theoretical prediction. The time evolution of the nucleating droplet displays qualitatively what it is expected from the spatial shape of the growth mode in Eq.~[\ref{eq:growth_mode}], and already indirectly observed in Fig.~[\ref{fig:Nucleation_process_Q10_R10}] : the density grows faster for $r\lesssim \xi$, until it reaches at the center its maximum value $n_{max}(0) = 1/(q-1)\sim 0.0526$, and the droplet grows by enlarging its surface. Such behavior is analogous to the one already observed in the Ising model~\cite{10.1103/PhysRevB.38.11607,10.1103/PhysRevLett.68.2336}. \\
\indent In order to characterize quantitatively the onset of spinodal nucleation as a function of $R$, we introduce a new observable $\phi_{com}^{min}$ defined as
\begin{align}
\label{eq:phi_com}
\phi_{com}(t) &= \frac{1}{(q-1)}\sum_{i=1}^{q-1}\left[ \big|x_{i, com}(t) - \bar{x}_{com}(t)\big|^{2} \right. \nonumber \\[8pt]
&+ \left.\big|y_{i, com}(t) - \bar{y}_{com}(t) \big|^{2}  \right]\, , \\[8pt]
\label{eq:com_min}
\phi_{com}^{min} &= \min\limits_{t}\left\{  \phi_{com}(t)\right\}\, .
\end{align}
The rationale behind the introduction of $\phi_{com}^{min}$  is that in the case of spinodal nucleation the appearance of the critical droplet should correspond to a drop in $\phi_{com}(t)$ since the clusters $\mathcal{C}_{i>0}^{max}$ are located in the same lattice region; hence $\phi_{com}^{min}$ can be used as a sort of order parameter to describe the crossover between SRI and LRI. During the metastable phase $\phi_{com}(t)$ highly fluctuates, since the spatial distribution of the centers of mass is almost uniform. We measure $\phi_{com}(t)$ every $1/20~ t_{MC}/\textrm{spins}$, and then to damp out the noise we average over all measurements taken within a time interval $t_{ave}$ of order $~1 t_{MC}/\textrm{spins}$, choosing a slightly smaller $t_{ave}$ for larger interaction ranges $R$. In Fig.~[\ref{fig:phi_com_history}], we show the time history of $\phi_{com}(t)$ for a  ``classical" and a spinodal nucleation process. The drop of $\phi_{com}(t)$ in presence of LRI is clearly visible and can be used to estimate the nucleation time $t_{nucl}$ with an associated uncertainty of few $t_{MC}/\textrm{spins}$. Our determination of $\phi_{com}^{min}$ is instead shown in Fig.~[\ref{fig:orderparam}] for both temperatures $T=0.5T_{c},~0.8T_{c}$ and for all simulated values of $q$, as a function of $R$. As it is clear from the figure, the behavior is almost independent from $q$ and $T$, with $\phi_{com}^{min}$ reaching a plateaux value at $R\gtrsim 7-9$. The drop is rather abrupt and hints of spinodal nucleation are visible at very small $R=4-5$.

\subsection{Multi-step or one-step nucleation?}
\label{sec:multi_step}
Soon after the first decay, the $q-1$ nucleating clusters start interacting and depending on $q$ and $T$ they can repel each other or continue to be interpenetrating. In the former case, random fluctuations may cause some of the $q-1$ spins to rapidly disappear from the system, while the remaining clusters keep growing in different regions of space. The occurrence of either the first or second scenario depends basically only on temperature and can be understood through the following argument: after decaying, the occupation number $n_{0}$ drops down to a negligible value and the system can be effectively considered as a $q-1$ state Potts model without external field. The system is prepared in a high-temperature phase given that along the symmetric channel the occupation numbers of the $q-1$ spins are locally equal. The fate of such state thus depends on whether the initial temperature $T < T_{c}(q) < T_{c}(q-1)$ allows for the existence of a disordered metastable state in the $q-1$ state Potts model. As argued in Sec.~[\ref{II}], the instability point of the high-temperature phase occurs, in the mean-field limit, at a spinodal temperature $ T_{c}^{-}(q-1)$, which corresponds to the intersection between the spinodal line $h_{sp}^{-}(q-1)$ and the horizontal axis  $h=0$ (see Fig.~[\ref{fig:Spinodal_lines}]). As a consequence, one expects that the new state is metastable only if $T(q) > T_{c}^{-}(q-1)$. The presence of a finite external field $h$ pushes the exact value  where a new metastable state can exist after the first decay, to slightly smaller values of $T$. In the mean-field limit such temperature can be found exactly from the free energy Eq.~[\ref{mean_field_free_energy}], and it is given by
\begin{align}
T_{sp}^{ms2} = \frac{2d}{q}(1-s^{ms2})~,
\end{align}
where $s^{ms2}<0$ is the value of the scalar order parameter $s$ in the global minimum of the free energy along the symmetric channel (see Fig.~[\ref{fig:free_energy_slices}]). In the limit $h\to \infty$, $s^{ms2} \to -1/(q-1)$ and $T_{sp}^{ms2} \to 2d/(q-1)  = T_{c}^{-}(q-1)$. In the case $q=3$, since the thermal transition of the Ising model is second order, there are no metastable states. For the other values of $q$ considered, we have
\begin{align}
0.8T_{c}(20) & >  \,\, 0.5T_{c}(20)> T_{sp}^{ms2}(20)~,\nonumber \\[8pt] 
0.8T_{c}(10) &> T_{sp}^{ms2}(10)\,\,>0.5T_{c}(10) ~,\nonumber \\[8pt] 
T_{sp}^{ms2}(5) &> \,\,\,0.8T_{c}(5) \,\, > 0.5T_{c}(5)~, 
\end{align}
for all simulated values of the magnetic field; thus for LRI we expect a new metastable state to form at $T=0.8T_{c}$ and $T=0.5T_{c}$ for $q=20$, and only at $T=0.8T_{c}$ for $q=10$. For $q=5$ instead, both values of $T$ are smaller than the spinodal temperature $T_{sp}^{ms2}$, and a mestastable state should never exist.\\
To determine how large $R$ must be for the mean-field predictions to hold true, we analyzed for all simulation points of Tab.~[\ref{tab:simulation_points}], $10^{2}$ nucleation events without stopping the simulation after the first decay. We then measured how many times the second state survived for a Monte Carlo time of at least $4\cdot 10^{2}~ t_{MC}/\textrm{spins}$, considering it to be decayed if
\begin{align}
\label{eq:second_ms_exit_cond}
\max{\left\{ n_{i}\right\}_{i=1,q-1}   } - \min{\left\{ n_{i}\right\}_{i=1,q-1}} \geq 0.05~.
\end{align}

\begin{figure*}
    \centering
    \includegraphics[scale=0.59]{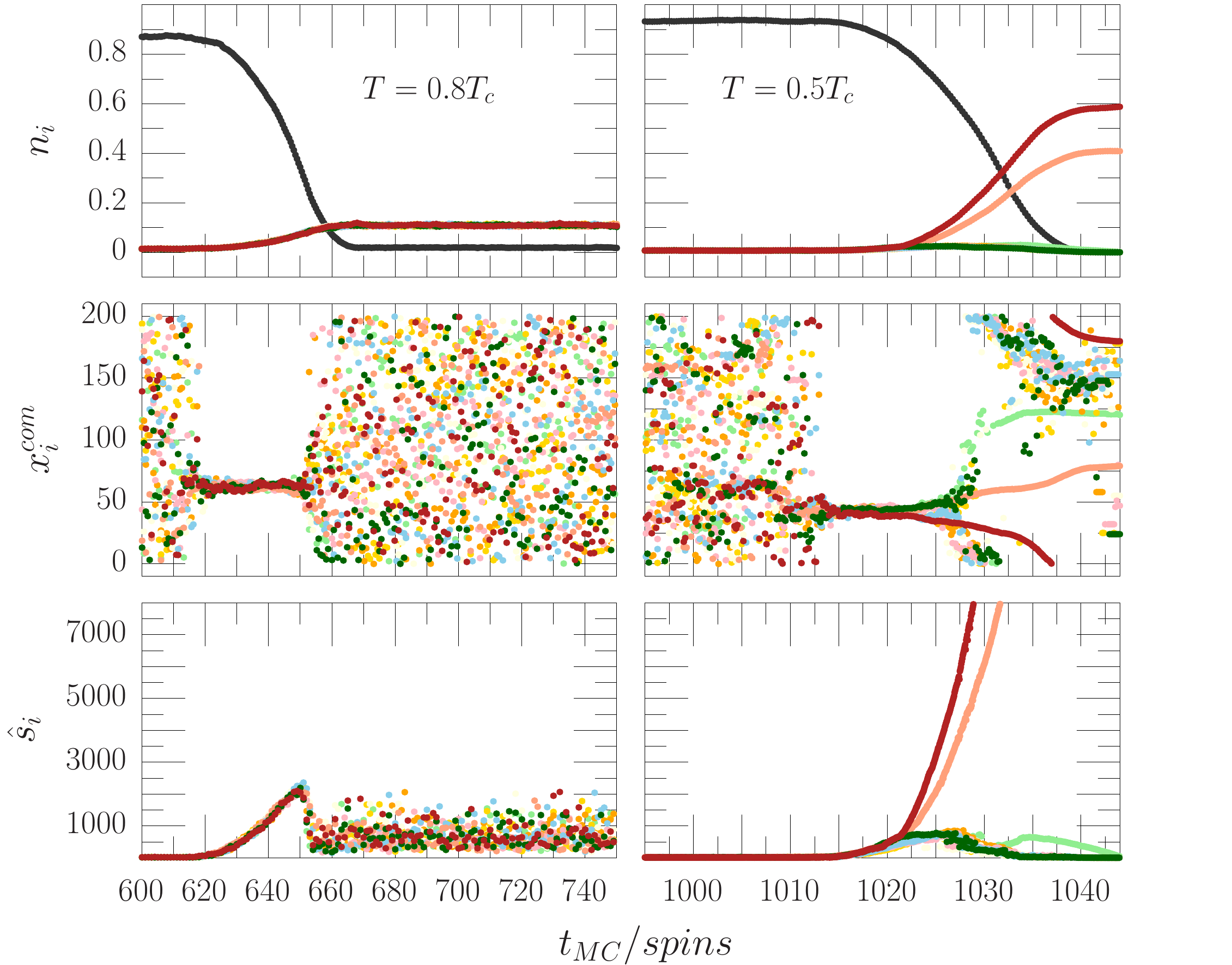}
    \caption{\small\it Time evolution of the occupation numbers $n_{i}$, of the $x$-coordinate of the centers of mass $x_{i}^{com}$ of the largest cluster $\mathcal{C}_{max}^{i}$ and of the corresponding sizes $\hat{s}_{i}$, during the first nucleation event and in the subsequent phase. The data correspond to simulations of the ten-state Potts model with $R=10$ and temperatures $T=0.8T_{c}$ (Left)  and $T=0.5T_{c}$ (Right).}
    \label{fig:metastable_2nd}
\end{figure*}

The results are collected in Tab.~[\ref{tab:percentage}]. For $q=10,20$ at $T=0.8T_{c}$ the system starts to be metastable for interaction ranges $R\geq 5$ ($q=20$) and $R\geq 8$ ($q=10$). At $T=0.5T_{c}$ only the twenty-state Potts model is metastable in the mean-field limit; Tab.~[\ref{tab:percentage}] indicates that signals of metastabilities are visible for $R=12$, while for $R=15$ no nucleation events have been observed.  For $q=3$ and $q=5$, nucleation occurred in all runs in less than 1 $t_{MC}/\textrm{spins}$; hence we did not report the corresponding values in Tab.~[\ref{tab:percentage}]. In Fig.~[\ref{fig:metastable_2nd}] we show for the ten-state Potts model with $R=10$ the time history of the occupations numbers, and of $x_{i}^{com}$ and $\hat{s}_{i}$ for the clusters $\mathcal{C}_{i>0}^{max}$. The figure illustrates the different dynamics associated to the two cases $T>T_{sp}^{ms2}$ and $T<T_{sp}^{ms2}$. For $T=0.5T_{c}$, the second state is not metastable and soon after the first nucleation event the clusters start to interact, with few of them becoming increasingly larger than the others. The formation of large domains, spatially well separated, indicates the relaxation towards the equilibrium state, even though it might take a very long time until one prevails. Instead, for $T=0.8T_{c}$ the occupation numbers continue to be the same for long time (Fig.~[\ref{fig:metastable_2nd}] Left). The system is effectively locked in a high-temperature metastable state. The clusters $\mathcal{C}_{i>0}^{max}$ are located in the same region of space, and they keep growing at the same rate. However, when $n_{0} \ll n_{i>0}$ the bond probability $P_{bond}= 1 -\exp{\left\{-2qJ_{ij}n_{0}\right\}}$ in Eq.~[\ref{bond_probability_spinodal}] is no longer adequate since the first metastable state decayed, and we have to switch to $P_{bond} = 1- \exp{\left\{-2J_{ij}\right\}}$. The cluster sizes must therefore drop, as they do, until eventually a new critical droplet forms and thermalization is reached.

\begin{table*}[]
\renewcommand{\arraystretch}{1.19}
    \centering
    \begin{tabular}{|c|c|c|c|c|c|c|c|c|}
    \hline
    & \multicolumn{4}{c|}{$q=10$} & \multicolumn{4}{c|}{$q=20$} \\
    \cline{2-9}
   $R$ & \multicolumn{2}{c|}{$T=0.5T_{c}$} & \multicolumn{2}{c|}{$T=0.8T_{c}$} & \multicolumn{2}{c|}{$T=0.5T_{c}$} & \multicolumn{2}{c|}{$T=0.8T_{c}$} \\
    \cline{2-9}
   & perc. & $\bar{\tau}$ & perc. & $\bar{\tau}$ &perc. & $\bar{\tau}$ & perc. & $\bar{\tau}$ \\
   \hline
    2 & $0\%$ & $<1$ & $0\%$ & $<1$  & $0\%$ & $<1$  & $0\%$ & $<1$   \\
    3 & $0\%$ & $<1$ & $0\%$ & $<1$ &  $0\%$ & $<1$ & $0\%$ & $<1$  \\
    4 & $0\%$ & $<1$ & $0\%$  & $<1$ &  $0\%$ & $<1$  & $0\%$ & $34$  \\
    5 & $0\%$ & $<1$ & $0\%$  &  $<1$ &  $0\%$ & $<1$ &$100\%$ & $>400$ \\
    6 & $0\%$ & $<1$ & $0\%$  & $3$ &  $0\%$& $<1$ & $100\%$ & $>400$ \\
    7 & $0\%$ & $<1$ & $0\%$  & $56$ &  $0\%$ & $<1$ & $100\%$ & $>400$ \\
    8 & $0\%$ & $<1$ & $78\%$  & $>400$ &  $0\%$ & $<1$ & $100\%$ & $>400$  \\
    9 & $0\%$ & $<1$ &$97\%$  & $>400$ &  $0\%$ & $2$ & $100\%$ & $>400$ \\
    10 & $0\%$ & $<1$ & $100\%$  & $>400$ &   $0\%$ & $11$ & $100\%$ & $>400$ \\
    11 & $0\%$ & $<1$ & $100\%$ & $>400$  &   $0\%$ & $31$ & $100\%$ & $>400$ \\
    12 & $0\%$ & $<1$ & $100\%$ & $>400$  &   $6\%$ & $136$ & $100\%$ & $>400$ \\
    13 & $0\%$ & $<1$ & $100\%$ & $>400$  &   $68\%$ & $>400$  & $100\%$ & $>400$ \\
    14 & $0\%$ & $<1$ & $100\%$ & $>400$  &   $95\%$ & $>400$ & $100\%$ & $>400$ \\
    15 & $0\%$ & $<1$ & $100\%$ & $>400$  &   $100\%$ & $>400$ & $100\%$ & $>400$  \\
    \hline
    \end{tabular}
    \caption{\small \it Average nucleation time $\bar{\tau}$  and percentage of events for which the second metastable state lasts at least $4\cdot 10^{2}~t_{MC}/\textrm{spins}$, as a function of $R$, $q$ and $T$. For each simulation point the magnetic field is set to the corresponding value in Tab.~[\ref{tab:simulation_points}].}
\label{tab:percentage}
\end{table*}

\section{Discussion and conclusion}
\label{V}
In this work we studied by means of Monte Carlo simulations on a $L=200$  square lattice, the mechanism of spinodal-assisted nucleation in the two-dimensional $q-$state  Potts model for $q=3,5,10,20$. Focusing on the low temperature phase $T<T_{c}(q)$, we prepared the system in a state with all spins aligned in a given direction ($0-$th direction), then monitored the evolution of the metastable state formed after an instantaneous quench of the magnetic field that disfavors the $0-$th state. Depending on the tunable parameters of the model, such as temperature, strength of the magnetic quench, and length scale $R$ of the spin-spin coupling, different nucleation patterns arise. We concentrated on the quantitative description of the crossover between the nucleation regime in presence of SRI ($R~\sim 1$) and LRI ($R\gg 1$), where different types of critical droplets allow the system to leave the metastable state. In the short range case the critical droplet consists of a large-amplitude localized fluctuation of a single spin, and thermalization is achieved in a single step. By increasing the length scale of the ferromagnetic coupling between the spins, the system starts to gradually display mean-field features, and nucleation can take place in a reasonable computer time, only for magnetic fields $h$ within a narrow interval below a spinodal point $h_{sp}(q)$. By means of a semi-classical calculation, we showed that in this case the critical droplet consists of a collective small-amplitude fluctuation of all $q-1$
spins orthogonal to the magnetic field, and computed the associated decay rate as a function of the external parameters, using the same formalism developed in Refs.~\cite{10.1103/PhysRevB.29.2698, 10.1103/PhysRevB.31.6127, 10.1103/PhysRevB.28.445} . The field theoretical predictions have been compared with the result of Monte Carlo simulations, including the asymptotic expression for nucleation times, metastable susceptibility and critical droplet profile. We find that quantitative agreement is obtained for interaction ranges $R\sim \mathcal{O}(10)$, with a mild dependence on temperature and simulated values of $q$, while hints of spinodal nucleation are already present for remarkably small values of $R \sim 4-5$. In order to compare nucleation processes at different values of $q$ and temperatures, we considered only magnetic quenches leading to approximately same average nucleation time. In the region of the parameter space where the logarithmic corrections to the decay rates are negligible, this corresponds to keep constant the spinodal Ginzburg parameter $G$. In this way the mean-field character of the system is not altered getting closer to the spinodal point. We provided evidence that the metastable susceptibility diverges linearly with the interaction range $R$ for all $q$, if one approaches the spinodal value of the magnetic field at constant $G \propto R^{2}(h-h_{sp}(q))$. \\ 
\indent Moreover, spinodal nucleation brings the system in a disordered state, still far from equilibrium, where almost only the $q-1$ spins that do not couple to the magnetic field are present. This state can be metastable or unstable depending basically only on temperature, and a multi-step decay process can be thus observed. In the limit of large $R$, the condition for the metastability of such phase can be again inferred from a mean-field calculation: a sharp spinodal point at $T=T_{sp}^{ms2}$ appears close to the instability point of the high-temperature phase in the $(q-1)$-state Potts model at zero external field. Our numerical findings become fully consistent with the mean-field predictions for interaction ranges $R$ of order $\mathcal{O}(10)$, where the occurrence of multi-step nucleation is observed only if $T>T_{sp}^{ms2}$.

\acknowledgments
We thank Dr. Nicola Bonini for the kind support. We dedicate this work to the memory of our friend Federico Tonielli.
The simulations were performed using the Cirrus UK National Tier-2 HPC Service at EPCC  (http://www.cirrus.ac.uk) funded by the University of Edinburgh and EPSRC (EP/P020267/1). G.G. is supported by MIUR (Italy) under grant PRIN
20172LNEEZ. 
\appendix 
\section{The site-bond Potts correlated polychromatic percolation model}
\label{DPM}
The Hamiltonian of the site-bond Potts correlated polychromatic percolation model is given by
\begin{align}
\label{dilute_hamiltonian}
H^{\textrm{DPM}} &= -\sum_{\alpha=0}^{q-1}\sum_{i j}\,\, (J_{b_{\alpha}})_{ij}\cdot\left(\delta_{\rho_{i}  \rho_{j}} -1\right)\,\delta_{\sigma_{i} \alpha}\,\delta_{\sigma_{j} \alpha} \nonumber \\
& \underbrace{-\sum_{i j} J_{ij}~\delta_{\sigma_{i} \sigma_{j}} + h\sum_{i} \delta_{\sigma_{i} 0}}_{H_{\textrm{Potts}}}\, ,
\end{align}
where $\rho_{i} \in \{ 0,\ldots, s-1\}$ is an additional Potts field. This model is particularly useful due to a generalization of the Kasteleyn and Fortruin theorem~\cite{10.1016/0031-8914,1969PSJJS..26...11K} obtained by Coniglio and Peruggi in Ref.~\cite{10.1088/0305-4470/15/6/028}. The theorem states that
\begin{align}
\label{eq:Kasteleyn}
\frac{\partial f^{\textrm{DPM}}(J_{b_{\alpha}}, J,h, s)}{\partial s}\bigg|_{s=1} = \frac{\langle n_{cl}\rangle}{N}\, ,  
\end{align}
where $f^{\textrm{DPM}}(J_{b_{\alpha}}, J,h, s)$  is the free energy density corresponding to the Hamiltonian Eq.~[\ref{dilute_hamiltonian}], $N$ is the total number of spins, while $\langle n_{cl} \rangle$ is the average total number of clusters in the percolation model where the Potts spins $\sigma$ are distributed according to $H_{\textrm{Potts}}$, and the probability of activating a bond between any pair of interacting parallel spins $\sigma_{i}= \sigma_{j}=\alpha$ is 
\begin{align}
P_{bond} = 1- \exp{\left(-2(J_{b_{\alpha}})_{ij}\right)}\, . 
\end{align}
Our goal is to solve exactly the model Eq.~[\ref{dilute_hamiltonian}] in the limit of infinite range interactions
\begin{align}
\left(J_{b_{\alpha}}\right)_{ij} = d\frac{J_{b_{\alpha}}}{N}~,\quad J_{ij} = d\frac{J}{N}~,
\end{align} 
then finding the critical values $J_{b_{\alpha}}^{cr}\left(J, h_{sp}(J,q)\right)$ such that $\langle n_{cl}\rangle$ becomes critical exactly at the spinodal point. To do so, we introduce the occupation numbers $n_{\alpha}^{\beta}$ with $\alpha \in \{0,\ldots, q-1\}$ and $\beta \in \{ 0,\ldots,s-1\}$. $n_{\alpha}^{\beta}$ corresponds to the density of spins $\sigma_{i} = \alpha$ of a given configuration that occupy lattice sites where $\rho_{i}= \beta$; thus the following constraint applies
\begin{align}
\sum_{\beta=0}^{s-1}\, n_{\alpha}^{\beta} = n_{\alpha} \implies \sum_{\alpha=0}^{q-1}\,\sum_{\beta=0}^{s-1}\, n_{\alpha}^{\beta} = 1\, . 
\end{align}
In terms of the occupation numbers, the Hamiltonian Eq.~[\ref{dilute_hamiltonian}] can be rewritten as
\begin{align}
\frac{H^{\textrm{DPM}}}{N} &= -2d\sum_{\alpha=0}^{q-1}\,J_{b_{\alpha}}\, \, \left[- \frac{n_{\alpha}^{2}}{2} +\sum_{\beta=0}^{s-1}\frac{\left(n_{\alpha}^{\beta}\right)^{2}}{2}\right] \nonumber \\
&-2dJ\sum_{\alpha=0}^{q-1} \frac{n_{\alpha}^{2}}{2} + h n_{0}+ \mathcal{O}(\frac{1}{N}) \, . 
\end{align}
In turn, a simple combinatorial calculation shows that
\begin{align}
\mathcal{Z}^{\textrm{DPM}} &= \sum_{\{\rho\}}\sum_{\{\sigma\}} e^{-H^{\textrm{DPM}}} \nonumber \\
&= \sum_{\{n_{\alpha}\}}\sum_{\{n_{\alpha}^{\beta} | n_{\alpha} \}}\!\!\! V! \,\,{\displaystyle \prod_{\alpha,\beta}}\,\,\frac{1}{n_{\alpha}^{\beta}!}\, e^{- H^{\textrm{DPM}}(\{n_{\alpha}^{\beta}\})}\, ;
\end{align}
hence the mean-field free energy density $f^{\textrm{DPM}}$ reads
\begin{align}
\label{eq:free_energy_DPM}
f^{\textrm{DPM}} = \sum_{\alpha,\beta} \, n_{\alpha}^{\beta}\, \log{(n_{\alpha}^{\beta})} - H^{\textrm{DPM}}(\{n_{\alpha}^{\beta}\})\, .
\end{align}
It is now natural to impose on the $n_{\alpha}^{\beta}$ the same symmetry breaking pattern that led to Eq.~[\ref{eq:occupations}], i.e. we assume that the occupations minimizing $f^{\textrm{DPM}}$ are of the form
\begin{align}
\label{eq:occupations_DPM}
\frac{n_{0}^{\beta}}{V} &= \begin{cases}
\frac{1}{qs}\left[ (1 + (q-1)\phi)\cdot (1 + (s-1)\psi_{0})\right] & \beta= 0 \\[8pt]
\frac{1}{qs}\left[ (1 + (q-1)\phi)\cdot ( 1 - \psi_{0})\right] & \beta > 0
\end{cases} \, ,
\nonumber \\[15pt]
\frac{n_{\alpha > 0}^{\beta}}{V} &= \begin{cases}
\frac{1}{qs}\left[ (1 -\phi)\cdot ( 1 + (s-1)\psi_{\alpha})\right] & \,\,\,\,\,\qquad\beta= 0 \\[8pt]
\frac{1}{qs}\left[ (1 -\phi)\cdot ( 1 - \psi_{\alpha})\right] & \,\,\,\,\,\qquad\beta > 0
\end{cases} \, ,
\end{align}
where the $\psi_{\alpha}$ are percolating fields describing the $q$ clusters made up of spins $\sigma= \alpha$. Substituting back Eq.~[\ref{eq:occupations_DPM}] into Eq.~[\ref{eq:free_energy_DPM}], and expanding the $\psi_{\alpha}$ up to cubic order we get
\begin{widetext}
\begin{align}
f^{\textrm{DPM}}[\psi_{\alpha}, \phi] &= \textrm{const}. + \mathcal{F}[\phi]+\frac{1}{2q^{2}s}\left[ (s-1)(1 + (q-1)\phi)(qs - 2dJ_{b}( (q-1)\phi -1))\psi_{0}^{2} \right. \nonumber \\[8pt]
&+ \left. \frac{q}{3}s(s-1)(s-2)( (q-1)\phi -1)\psi_{0}^{3}\right] +\frac{1}{2 q^{2} s}\sum_{\alpha=1}^{s}\left[(s-1)(1-\phi)(qs - 2dJ_{b}(1-\phi))\psi_{\alpha}^{2}\right. \nonumber \\[8pt]
&+ \left.\frac{q}{3}s(s-1)(s-2)(\phi-1)\psi_{\alpha}^{3}\right]\, , \\[8pt]
\mathcal{F}[\phi] &= \frac{1}{q}\left[1+(q-1)\phi\right]\log{\left[\frac{1}{q}\left[1+(q-1)\phi\right]\right]} +(q-1)\frac{1}{q}(1-\phi)\log{\left[\frac{1}{q}(1-\phi)\right]} -\frac{dJ}{q}(q-1)\phi^2\nonumber \\[8pt]
&  +\frac{dJ_{b_{0}}}{q^{2}}\left( 1+ (q-1)\phi\right)^{2} +\frac{d}{q^{2}}\left(\sum_{\alpha=1}^{q-1}J_{b_{\alpha}}\right)(1-\phi)^{2}         +\frac{h}{q}(q-1)\phi \, .
\end{align}
\end{widetext}
By evaluating the derivative $\frac{\partial f}{\partial s}\big|_{s=1}$, and making use of Eq.~[\ref{eq:Kasteleyn}], each of the percolating fields $\psi_{\alpha}$ undergoes a second order phase transition when the coefficient of the $\psi_{\alpha}^{2}$ term vanishes. This yields
\begin{align}
J_{b_{0}}^{cr} = \frac{q}{2d[1+(q-1)\phi]}\, , \qquad J_{b_{\alpha > 0}}^{cr}= \frac{q}{2d[1-\phi]}\, .
\end{align} 
Finally, we need to impose that the transition occurs exactly when $\phi = s_{sp}$. Thus, making use of Eq.~[\ref{eq:Pottspinodals}] and noticing that at the spinodal point
\begin{align}
\frac{q}{(1-s_{sp})(1+ (q-1)s_{sp})} = 2dJ 
\end{align}
we get the final result
\begin{align}
\label{eq:critical_cluster_probability}
J_{b_{0}}^{cr} = q J n_{\alpha > 0}\, ,\qquad J_{b_{\alpha > 0}}^{cr} = q J n_{0}\,  \,\,\,\,\,\,\qquad \square .
\end{align}
In the Ising limit $q=2$, our formula coincides with the one determined in Ref.~\cite{10.1142/S0217979294000646}. 

\bibliography{biblio_mache_doi.bib} 

\begin{thebibliography}{36}%
\makeatletter
\providecommand \@ifxundefined [1]{%
 \@ifx{#1\undefined}
}%
\providecommand \@ifnum [1]{%
 \ifnum #1\expandafter \@firstoftwo
 \else \expandafter \@secondoftwo
 \fi
}%
\providecommand \@ifx [1]{%
 \ifx #1\expandafter \@firstoftwo
 \else \expandafter \@secondoftwo
 \fi
}%
\providecommand \natexlab [1]{#1}%
\providecommand \enquote  [1]{``#1''}%
\providecommand \bibnamefont  [1]{#1}%
\providecommand \bibfnamefont [1]{#1}%
\providecommand \citenamefont [1]{#1}%
\providecommand \href@noop [0]{\@secondoftwo}%
\providecommand \href [0]{\begingroup \@sanitize@url \@href}%
\providecommand \@href[1]{\@@startlink{#1}\@@href}%
\providecommand \@@href[1]{\endgroup#1\@@endlink}%
\providecommand \@sanitize@url [0]{\catcode `\\12\catcode `\$12\catcode
  `\&12\catcode `\#12\catcode `\^12\catcode `\_12\catcode `\%12\relax}%
\providecommand \@@startlink[1]{}%
\providecommand \@@endlink[0]{}%
\providecommand \url  [0]{\begingroup\@sanitize@url \@url }%
\providecommand \@url [1]{\endgroup\@href {#1}{\urlprefix }}%
\providecommand \urlprefix  [0]{URL }%
\providecommand \Eprint [0]{\href }%
\providecommand \doibase [0]{http://dx.doi.org/}%
\providecommand \selectlanguage [0]{\@gobble}%
\providecommand \bibinfo  [0]{\@secondoftwo}%
\providecommand \bibfield  [0]{\@secondoftwo}%
\providecommand \translation [1]{[#1]}%
\providecommand \BibitemOpen [0]{}%
\providecommand \bibitemStop [0]{}%
\providecommand \bibitemNoStop [0]{.\EOS\space}%
\providecommand \EOS [0]{\spacefactor3000\relax}%
\providecommand \BibitemShut  [1]{\csname bibitem#1\endcsname}%
\let\auto@bib@innerbib\@empty
\bibitem [{\citenamefont {Becker}\ and\ \citenamefont
  {D{\"o}ring}(1935)}]{10.1002/andp.19354160806}%
  \BibitemOpen
  \bibfield  {author} {\bibinfo {author} {\bibfnamefont {R.}~\bibnamefont
  {Becker}}\ and\ \bibinfo {author} {\bibfnamefont {W.}~\bibnamefont
  {D{\"o}ring}},\ }\href {\doibase 10.1002/andp.19354160806} {\bibfield
  {journal} {\bibinfo  {journal} {Annalen der Physik}\ }\textbf {\bibinfo
  {volume} {416}},\ \bibinfo {pages} {719} (\bibinfo {year}
  {1935})}\BibitemShut {NoStop}%
\bibitem [{\citenamefont {Gunton}(1983)}]{10008767133}%
  \BibitemOpen
  \bibfield  {author} {\bibinfo {author} {\bibfnamefont {D.}~\bibnamefont
  {Gunton}},\ }\href {https://ci.nii.ac.jp/naid/10008767133/en/} {\bibfield
  {journal} {\bibinfo  {journal} {Phase transitions and critical phenomena}\
  }\textbf {\bibinfo {volume} {8}},\ \bibinfo {pages} {267} (\bibinfo {year}
  {1983})}\BibitemShut {NoStop}%
\bibitem [{\citenamefont {Debenedetti}(1996)}]{9780691085951}%
  \BibitemOpen
  \bibfield  {author} {\bibinfo {author} {\bibfnamefont {P.}~\bibnamefont
  {Debenedetti}},\ }\href {http://www.jstor.org/stable/j.ctv10crfs5} {\emph
  {\bibinfo {title} {Metastable Liquids: Concepts and Principles}}},\
  Vol.~\bibinfo {volume} {1}\ (\bibinfo  {publisher} {Princeton University
  Press},\ \bibinfo {year} {1996})\BibitemShut {NoStop}%
\bibitem [{\citenamefont {Debenedetti}\ and\ \citenamefont
  {Stillinger}(2001)}]{10.1038/35065704}%
  \BibitemOpen
  \bibfield  {author} {\bibinfo {author} {\bibfnamefont {P.}~\bibnamefont
  {Debenedetti}}\ and\ \bibinfo {author} {\bibfnamefont {F.}~\bibnamefont
  {Stillinger}},\ }\href {\doibase 10.1038/35065704} {\bibfield  {journal}
  {\bibinfo  {journal} {Nature}\ }\textbf {\bibinfo {volume} {410}},\ \bibinfo
  {pages} {259} (\bibinfo {year} {2001})}\BibitemShut {NoStop}%
\bibitem [{\citenamefont {Callan}\ and\ \citenamefont
  {Coleman}(1977)}]{10.1103/PhysRevD.16.1762}%
  \BibitemOpen
  \bibfield  {author} {\bibinfo {author} {\bibfnamefont {C.~G.}\ \bibnamefont
  {Callan}}\ and\ \bibinfo {author} {\bibfnamefont {S.}~\bibnamefont
  {Coleman}},\ }\href {\doibase 10.1103/PhysRevD.16.1762} {\bibfield  {journal}
  {\bibinfo  {journal} {Phys. Rev. D}\ }\textbf {\bibinfo {volume} {16}},\
  \bibinfo {pages} {1762} (\bibinfo {year} {1977})}\BibitemShut {NoStop}%
\bibitem [{\citenamefont {Kosterlitz}\ and\ \citenamefont
  {Thouless}(1972)}]{10.1088/0022-3719/5/11/002}%
  \BibitemOpen
  \bibfield  {author} {\bibinfo {author} {\bibfnamefont {J.~M.}\ \bibnamefont
  {Kosterlitz}}\ and\ \bibinfo {author} {\bibfnamefont {D.~J.}\ \bibnamefont
  {Thouless}},\ }\href {\doibase 10.1088/0022-3719/5/11/002} {\bibfield
  {journal} {\bibinfo  {journal} {Journal of Physics C: Solid State Physics}\
  }\textbf {\bibinfo {volume} {5}},\ \bibinfo {pages} {L124} (\bibinfo {year}
  {1972})}\BibitemShut {NoStop}%
\bibitem [{\citenamefont {Binder}(1987)}]{10.1088/0034-4885/50/7/001}%
  \BibitemOpen
  \bibfield  {author} {\bibinfo {author} {\bibfnamefont {K.}~\bibnamefont
  {Binder}},\ }\href {\doibase 10.1088/0034-4885/50/7/001} {\bibfield
  {journal} {\bibinfo  {journal} {Reports on Progress in Physics}\ }\textbf
  {\bibinfo {volume} {50}},\ \bibinfo {pages} {783} (\bibinfo {year}
  {1987})}\BibitemShut {NoStop}%
\bibitem [{\citenamefont {Binder}\ and\ \citenamefont
  {Virnau}(2016)}]{10.1063/1.4959235}%
  \BibitemOpen
  \bibfield  {author} {\bibinfo {author} {\bibfnamefont {K.}~\bibnamefont
  {Binder}}\ and\ \bibinfo {author} {\bibfnamefont {P.}~\bibnamefont
  {Virnau}},\ }\href {\doibase 10.1063/1.4959235} {\bibfield  {journal}
  {\bibinfo  {journal} {The Journal of Chemical Physics}\ }\textbf {\bibinfo
  {volume} {145}},\ \bibinfo {pages} {211701} (\bibinfo {year} {2016})},\
  \Eprint {http://arxiv.org/abs/https://doi.org/10.1063/1.4959235}
  {https://doi.org/10.1063/1.4959235} \BibitemShut {NoStop}%
\bibitem [{\citenamefont {Herrmann}\ \emph {et~al.}(1982)\citenamefont
  {Herrmann}, \citenamefont {Klein},\ and\ \citenamefont
  {Stauffer}}]{10.1103/PhysRevLett.49.1262}%
  \BibitemOpen
  \bibfield  {author} {\bibinfo {author} {\bibfnamefont {D.~W.}\ \bibnamefont
  {Herrmann}}, \bibinfo {author} {\bibfnamefont {W.}~\bibnamefont {Klein}}, \
  and\ \bibinfo {author} {\bibfnamefont {D.}~\bibnamefont {Stauffer}},\ }\href
  {\doibase 10.1103/PhysRevLett.49.1262} {\bibfield  {journal} {\bibinfo
  {journal} {Phys. Rev. Lett.}\ }\textbf {\bibinfo {volume} {49}},\ \bibinfo
  {pages} {1262} (\bibinfo {year} {1982})}\BibitemShut {NoStop}%
\bibitem [{\citenamefont {Monette}\ \emph {et~al.}(1989)\citenamefont
  {Monette}, \citenamefont {Klein}, \citenamefont {Zuckermann}, \citenamefont
  {Khadir},\ and\ \citenamefont {Harris}}]{10.1103/PhysRevB.38.11607}%
  \BibitemOpen
  \bibfield  {author} {\bibinfo {author} {\bibfnamefont {L.}~\bibnamefont
  {Monette}}, \bibinfo {author} {\bibfnamefont {W.}~\bibnamefont {Klein}},
  \bibinfo {author} {\bibfnamefont {M.}~\bibnamefont {Zuckermann}}, \bibinfo
  {author} {\bibfnamefont {A.}~\bibnamefont {Khadir}}, \ and\ \bibinfo {author}
  {\bibfnamefont {R.}~\bibnamefont {Harris}},\ }\href {\doibase
  10.1103/PhysRevB.38.11607} {\bibfield  {journal} {\bibinfo  {journal}
  {Physical review. B, Condensed matter}\ }\textbf {\bibinfo {volume} {38}},\
  \bibinfo {pages} {11607} (\bibinfo {year} {1989})}\BibitemShut {NoStop}%
\bibitem [{\citenamefont {Monette}\ \emph {et~al.}(1992)\citenamefont
  {Monette}, \citenamefont {Klein},\ and\ \citenamefont
  {Zuckermann}}]{Monette1992}%
  \BibitemOpen
  \bibfield  {author} {\bibinfo {author} {\bibfnamefont {L.}~\bibnamefont
  {Monette}}, \bibinfo {author} {\bibfnamefont {W.}~\bibnamefont {Klein}}, \
  and\ \bibinfo {author} {\bibfnamefont {M.}~\bibnamefont {Zuckermann}},\
  }\href {\doibase 10.1007/BF01060062} {\bibfield  {journal} {\bibinfo
  {journal} {Journal of Statistical Physics}\ }\textbf {\bibinfo {volume}
  {66}},\ \bibinfo {pages} {117} (\bibinfo {year} {1992})}\BibitemShut
  {NoStop}%
\bibitem [{\citenamefont {Unger}\ and\ \citenamefont
  {Klein}(1984)}]{10.1103/PhysRevB.29.2698}%
  \BibitemOpen
  \bibfield  {author} {\bibinfo {author} {\bibfnamefont {C.}~\bibnamefont
  {Unger}}\ and\ \bibinfo {author} {\bibfnamefont {W.}~\bibnamefont {Klein}},\
  }\href {\doibase 10.1103/PhysRevB.29.2698} {\bibfield  {journal} {\bibinfo
  {journal} {Phys. Rev. B}\ }\textbf {\bibinfo {volume} {29}},\ \bibinfo
  {pages} {2698} (\bibinfo {year} {1984})}\BibitemShut {NoStop}%
\bibitem [{\citenamefont {Unger}\ and\ \citenamefont
  {Klein}(1985)}]{10.1103/PhysRevB.31.6127}%
  \BibitemOpen
  \bibfield  {author} {\bibinfo {author} {\bibfnamefont {C.}~\bibnamefont
  {Unger}}\ and\ \bibinfo {author} {\bibfnamefont {W.}~\bibnamefont {Klein}},\
  }\href {\doibase 10.1103/PhysRevB.31.6127} {\bibfield  {journal} {\bibinfo
  {journal} {Phys. Rev. B}\ }\textbf {\bibinfo {volume} {31}},\ \bibinfo
  {pages} {6127} (\bibinfo {year} {1985})}\BibitemShut {NoStop}%
\bibitem [{\citenamefont {Klein}\ and\ \citenamefont
  {Unger}(1983)}]{10.1103/PhysRevB.28.445}%
  \BibitemOpen
  \bibfield  {author} {\bibinfo {author} {\bibfnamefont {W.}~\bibnamefont
  {Klein}}\ and\ \bibinfo {author} {\bibfnamefont {C.}~\bibnamefont {Unger}},\
  }\href {\doibase 10.1103/PhysRevB.28.445} {\bibfield  {journal} {\bibinfo
  {journal} {Phys. Rev. B}\ }\textbf {\bibinfo {volume} {28}},\ \bibinfo
  {pages} {445} (\bibinfo {year} {1983})}\BibitemShut {NoStop}%
\bibitem [{\citenamefont {Langer}(1967)}]{10.1016/0003-4916(67)90200-X}%
  \BibitemOpen
  \bibfield  {author} {\bibinfo {author} {\bibfnamefont {J.}~\bibnamefont
  {Langer}},\ }\href {\doibase https://doi.org/10.1016/0003-4916(67)90200-X}
  {\bibfield  {journal} {\bibinfo  {journal} {Annals of Physics}\ }\textbf
  {\bibinfo {volume} {41}},\ \bibinfo {pages} {108 } (\bibinfo {year}
  {1967})}\BibitemShut {NoStop}%
\bibitem [{\citenamefont {Gunther}\ \emph {et~al.}(1980)\citenamefont
  {Gunther}, \citenamefont {Wallace},\ and\ \citenamefont
  {Nicole}}]{10.1088/0305-4470/13/5/034}%
  \BibitemOpen
  \bibfield  {author} {\bibinfo {author} {\bibfnamefont {N.~J.}\ \bibnamefont
  {Gunther}}, \bibinfo {author} {\bibfnamefont {D.~J.}\ \bibnamefont
  {Wallace}}, \ and\ \bibinfo {author} {\bibfnamefont {D.~A.}\ \bibnamefont
  {Nicole}},\ }\href {\doibase 10.1088/0305-4470/13/5/034} {\bibfield
  {journal} {\bibinfo  {journal} {Journal of Physics A: Mathematical and
  General}\ }\textbf {\bibinfo {volume} {13}},\ \bibinfo {pages} {1755}
  (\bibinfo {year} {1980})}\BibitemShut {NoStop}%
\bibitem [{\citenamefont {Ostwald}(1897)}]{ostwald1897file}%
  \BibitemOpen
  \bibfield  {author} {\bibinfo {author} {\bibfnamefont {W.}~\bibnamefont
  {Ostwald}},\ }\href@noop {} {\bibfield  {journal} {\bibinfo  {journal}
  {Zeitschrift f{\"u}r physikalische Chemie}\ }\textbf {\bibinfo {volume}
  {22}},\ \bibinfo {pages} {289} (\bibinfo {year} {1897})}\BibitemShut
  {NoStop}%
\bibitem [{\citenamefont {Rutkevich}(2002)}]{10.1142/S0129183102003255}%
  \BibitemOpen
  \bibfield  {author} {\bibinfo {author} {\bibfnamefont {S.}~\bibnamefont
  {Rutkevich}},\ }\href {\doibase 10.1142/S0129183102003255} {\bibfield
  {journal} {\bibinfo  {journal} {International Journal of Modern Physics C}\
  }\textbf {\bibinfo {volume} {13}},\ \bibinfo {pages} {495} (\bibinfo {year}
  {2002})},\ \Eprint
  {http://arxiv.org/abs/https://doi.org/10.1142/S0129183102003255}
  {https://doi.org/10.1142/S0129183102003255} \BibitemShut {NoStop}%
\bibitem [{\citenamefont {Corberi}\ \emph {et~al.}(2021)\citenamefont
  {Corberi}, \citenamefont {Cugliandolo}, \citenamefont {Esposito},
  \citenamefont {Mazzarisi},\ and\ \citenamefont {Picco}}]{corberi2021phases}%
  \BibitemOpen
  \bibfield  {author} {\bibinfo {author} {\bibfnamefont {F.}~\bibnamefont
  {Corberi}}, \bibinfo {author} {\bibfnamefont {L.~F.}\ \bibnamefont
  {Cugliandolo}}, \bibinfo {author} {\bibfnamefont {M.}~\bibnamefont
  {Esposito}}, \bibinfo {author} {\bibfnamefont {O.}~\bibnamefont {Mazzarisi}},
  \ and\ \bibinfo {author} {\bibfnamefont {M.}~\bibnamefont {Picco}},\
  }\href@noop {} {\enquote {\bibinfo {title} {How many phases nucleate in the
  bidimensional potts model?}}\ } (\bibinfo {year} {2021}),\ \Eprint
  {http://arxiv.org/abs/2102.01003} {arXiv:2102.01003 [cond-mat.stat-mech]}
  \BibitemShut {NoStop}%
\bibitem [{\citenamefont {Coniglio}\ and\ \citenamefont
  {Klein}(1980)}]{10.1088/0305-4470/13/8/025}%
  \BibitemOpen
  \bibfield  {author} {\bibinfo {author} {\bibfnamefont {A.}~\bibnamefont
  {Coniglio}}\ and\ \bibinfo {author} {\bibfnamefont {W.}~\bibnamefont
  {Klein}},\ }\href {\doibase 10.1088/0305-4470/13/8/025} {\bibfield  {journal}
  {\bibinfo  {journal} {Journal of Physics A: Mathematical and General}\
  }\textbf {\bibinfo {volume} {13}},\ \bibinfo {pages} {2775} (\bibinfo {year}
  {1980})}\BibitemShut {NoStop}%
\bibitem [{\citenamefont {Coniglio}\ and\ \citenamefont
  {Fierro}(2009)}]{10.1007/978-0-387-30440-3_104}%
  \BibitemOpen
  \bibfield  {author} {\bibinfo {author} {\bibfnamefont {A.}~\bibnamefont
  {Coniglio}}\ and\ \bibinfo {author} {\bibfnamefont {A.}~\bibnamefont
  {Fierro}},\ }\enquote {\bibinfo {title} {Correlated percolation},}\ in\ \href
  {\doibase 10.1007/978-0-387-30440-3_104} {\emph {\bibinfo {booktitle}
  {Encyclopedia of Complexity and Systems Science}}},\ \bibinfo {editor}
  {edited by\ \bibinfo {editor} {\bibfnamefont {R.~A.}\ \bibnamefont
  {Meyers}}}\ (\bibinfo  {publisher} {Springer New York},\ \bibinfo {address}
  {New York, NY},\ \bibinfo {year} {2009})\ pp.\ \bibinfo {pages}
  {1596--1615}\BibitemShut {NoStop}%
\bibitem [{\citenamefont {Wu}(1982)}]{10.1103/RevModPhys.54.235}%
  \BibitemOpen
  \bibfield  {author} {\bibinfo {author} {\bibfnamefont {F.~Y.}\ \bibnamefont
  {Wu}},\ }\href {\doibase 10.1103/RevModPhys.54.235} {\bibfield  {journal}
  {\bibinfo  {journal} {Rev. Mod. Phys.}\ }\textbf {\bibinfo {volume} {54}},\
  \bibinfo {pages} {235} (\bibinfo {year} {1982})}\BibitemShut {NoStop}%
\bibitem [{\citenamefont {Hubbard}(1959)}]{PhysRevLett.3.77}%
  \BibitemOpen
  \bibfield  {author} {\bibinfo {author} {\bibfnamefont {J.}~\bibnamefont
  {Hubbard}},\ }\href {\doibase 10.1103/PhysRevLett.3.77} {\bibfield  {journal}
  {\bibinfo  {journal} {Phys. Rev. Lett.}\ }\textbf {\bibinfo {volume} {3}},\
  \bibinfo {pages} {77} (\bibinfo {year} {1959})}\BibitemShut {NoStop}%
\bibitem [{\citenamefont {{Stratonovich}}(1957)}]{1957SPhD....2..416S}%
  \BibitemOpen
  \bibfield  {author} {\bibinfo {author} {\bibfnamefont {R.~L.}\ \bibnamefont
  {{Stratonovich}}},\ }\href@noop {} {\bibfield  {journal} {\bibinfo  {journal}
  {Soviet Physics Doklady}\ }\textbf {\bibinfo {volume} {2}},\ \bibinfo {pages}
  {416} (\bibinfo {year} {1957})}\BibitemShut {NoStop}%
\bibitem [{\citenamefont {Langer}(1969)}]{10.1016/0003-4916(69)90153-5}%
  \BibitemOpen
  \bibfield  {author} {\bibinfo {author} {\bibfnamefont {J.}~\bibnamefont
  {Langer}},\ }\href {\doibase https://doi.org/10.1016/0003-4916(69)90153-5}
  {\bibfield  {journal} {\bibinfo  {journal} {Annals of Physics}\ }\textbf
  {\bibinfo {volume} {54}},\ \bibinfo {pages} {258 } (\bibinfo {year}
  {1969})}\BibitemShut {NoStop}%
\bibitem [{\citenamefont {Monette}\ and\ \citenamefont
  {Klein}(1992)}]{10.1103/PhysRevLett.68.2336}%
  \BibitemOpen
  \bibfield  {author} {\bibinfo {author} {\bibfnamefont {L.}~\bibnamefont
  {Monette}}\ and\ \bibinfo {author} {\bibfnamefont {W.}~\bibnamefont
  {Klein}},\ }\href {\doibase 10.1103/PhysRevLett.68.2336} {\bibfield
  {journal} {\bibinfo  {journal} {Phys. Rev. Lett.}\ }\textbf {\bibinfo
  {volume} {68}},\ \bibinfo {pages} {2336} (\bibinfo {year}
  {1992})}\BibitemShut {NoStop}%
\bibitem [{\citenamefont {Binder}(1984)}]{10.1103/PhysRevA.29.341}%
  \BibitemOpen
  \bibfield  {author} {\bibinfo {author} {\bibfnamefont {K.}~\bibnamefont
  {Binder}},\ }\href {\doibase 10.1103/PhysRevA.29.341} {\bibfield  {journal}
  {\bibinfo  {journal} {Phys. Rev. A}\ }\textbf {\bibinfo {volume} {29}},\
  \bibinfo {pages} {341} (\bibinfo {year} {1984})}\BibitemShut {NoStop}%
\bibitem [{\citenamefont {Gunton}\ and\ \citenamefont
  {Yalabik}(1978)}]{10.1103/PhysRevB.18.6199}%
  \BibitemOpen
  \bibfield  {author} {\bibinfo {author} {\bibfnamefont {J.~D.}\ \bibnamefont
  {Gunton}}\ and\ \bibinfo {author} {\bibfnamefont {M.~C.}\ \bibnamefont
  {Yalabik}},\ }\href {\doibase 10.1103/PhysRevB.18.6199} {\bibfield  {journal}
  {\bibinfo  {journal} {Phys. Rev. B}\ }\textbf {\bibinfo {volume} {18}},\
  \bibinfo {pages} {6199} (\bibinfo {year} {1978})}\BibitemShut {NoStop}%
\bibitem [{\citenamefont {Cardy}\ \emph {et~al.}(1996)\citenamefont {Cardy},
  \citenamefont {Goddard},\ and\ \citenamefont {Yeomans}}]{9780521499590}%
  \BibitemOpen
  \bibfield  {author} {\bibinfo {author} {\bibfnamefont {J.}~\bibnamefont
  {Cardy}}, \bibinfo {author} {\bibfnamefont {P.}~\bibnamefont {Goddard}}, \
  and\ \bibinfo {author} {\bibfnamefont {J.}~\bibnamefont {Yeomans}},\ }\href
  {https://books.google.it/books?id=Wt804S9FjyAC} {\emph {\bibinfo {title}
  {Scaling and Renormalization in Statistical Physics}}},\ Cambridge Lecture
  Notes in Physics\ (\bibinfo  {publisher} {Cambridge University Press},\
  \bibinfo {year} {1996})\BibitemShut {NoStop}%
\bibitem [{\citenamefont {Müller-Krumbhaar}(1974)}]{10.1016/0375-9601}%
  \BibitemOpen
  \bibfield  {author} {\bibinfo {author} {\bibfnamefont {H.}~\bibnamefont
  {Müller-Krumbhaar}},\ }\href {\doibase
  https://doi.org/10.1016/0375-9601(74)90337-5} {\bibfield  {journal} {\bibinfo
   {journal} {Physics Letters A}\ }\textbf {\bibinfo {volume} {50}},\ \bibinfo
  {pages} {27 } (\bibinfo {year} {1974})}\BibitemShut {NoStop}%
\bibitem [{\citenamefont {Stella}\ and\ \citenamefont
  {Vanderzande}(1989)}]{10.1103/PhysRevLett.62.1067}%
  \BibitemOpen
  \bibfield  {author} {\bibinfo {author} {\bibfnamefont {A.~L.}\ \bibnamefont
  {Stella}}\ and\ \bibinfo {author} {\bibfnamefont {C.}~\bibnamefont
  {Vanderzande}},\ }\href {\doibase 10.1103/PhysRevLett.62.1067} {\bibfield
  {journal} {\bibinfo  {journal} {Phys. Rev. Lett.}\ }\textbf {\bibinfo
  {volume} {62}},\ \bibinfo {pages} {1067} (\bibinfo {year}
  {1989})}\BibitemShut {NoStop}%
\bibitem [{\citenamefont {Coniglio}\ and\ \citenamefont
  {Peruggi}(1982)}]{10.1088/0305-4470/15/6/028}%
  \BibitemOpen
  \bibfield  {author} {\bibinfo {author} {\bibfnamefont {A.}~\bibnamefont
  {Coniglio}}\ and\ \bibinfo {author} {\bibfnamefont {F.}~\bibnamefont
  {Peruggi}},\ }\href {\doibase 10.1088/0305-4470/15/6/028} {\bibfield
  {journal} {\bibinfo  {journal} {J. Phys. A}\ }\textbf {\bibinfo {volume}
  {15}},\ \bibinfo {pages} {1873} (\bibinfo {year} {1982})}\BibitemShut
  {NoStop}%
\bibitem [{\citenamefont {{Kasteleyn}}\ and\ \citenamefont
  {{Fortuin}}(1969)}]{1969PSJJS..26...11K}%
  \BibitemOpen
  \bibfield  {author} {\bibinfo {author} {\bibfnamefont {P.~W.}\ \bibnamefont
  {{Kasteleyn}}}\ and\ \bibinfo {author} {\bibfnamefont {C.~M.}\ \bibnamefont
  {{Fortuin}}},\ }\href@noop {} {\bibfield  {journal} {\bibinfo  {journal}
  {Physical Society of Japan Journal Supplement}\ }\textbf {\bibinfo {volume}
  {26}},\ \bibinfo {pages} {11} (\bibinfo {year} {1969})}\BibitemShut {NoStop}%
\bibitem [{\citenamefont {Fortuin}\ and\ \citenamefont
  {Kasteleyn}(1972)}]{10.1016/0031-8914}%
  \BibitemOpen
  \bibfield  {author} {\bibinfo {author} {\bibfnamefont {C.}~\bibnamefont
  {Fortuin}}\ and\ \bibinfo {author} {\bibfnamefont {P.}~\bibnamefont
  {Kasteleyn}},\ }\href {\doibase https://doi.org/10.1016/0031-8914(72)90045-6}
  {\bibfield  {journal} {\bibinfo  {journal} {Physica}\ }\textbf {\bibinfo
  {volume} {57}},\ \bibinfo {pages} {536 } (\bibinfo {year}
  {1972})}\BibitemShut {NoStop}%
\bibitem [{\citenamefont {Jan}\ \emph {et~al.}(1982)\citenamefont {Jan},
  \citenamefont {Coniglio},\ and\ \citenamefont
  {Stauffer}}]{10.1088/0305-4470/15/12/008}%
  \BibitemOpen
  \bibfield  {author} {\bibinfo {author} {\bibfnamefont {N.}~\bibnamefont
  {Jan}}, \bibinfo {author} {\bibfnamefont {A.}~\bibnamefont {Coniglio}}, \
  and\ \bibinfo {author} {\bibfnamefont {D.}~\bibnamefont {Stauffer}},\ }\href
  {\doibase 10.1088/0305-4470/15/12/008} {\bibfield  {journal} {\bibinfo
  {journal} {Journal of Physics A: Mathematical and General}\ }\textbf
  {\bibinfo {volume} {15}},\ \bibinfo {pages} {L699} (\bibinfo {year}
  {1982})}\BibitemShut {NoStop}%
\bibitem [{\citenamefont {Monette}(1994)}]{10.1142/S0217979294000646}%
  \BibitemOpen
  \bibfield  {author} {\bibinfo {author} {\bibfnamefont {L.}~\bibnamefont
  {Monette}},\ }\href {\doibase 10.1142/S0217979294000646} {\bibfield
  {journal} {\bibinfo  {journal} {International Journal of Modern Physics B}\
  }\textbf {\bibinfo {volume} {08}},\ \bibinfo {pages} {1417} (\bibinfo {year}
  {1994})},\ \Eprint
  {http://arxiv.org/abs/https://doi.org/10.1142/S0217979294000646}
  {https://doi.org/10.1142/S0217979294000646} \BibitemShut {NoStop}%
\end{thebibliography}%
\bibliographystyle{apsrev4-1}

\end{document}